\documentclass[12pt,letterpaper]{article}
\usepackage{eurosym}
\usepackage{amssymb}
\usepackage{amsmath}
\usepackage{amsfonts}
\usepackage{verbatim}
\usepackage{geometry}
\usepackage[scaled]{helvet}
\usepackage{graphicx}
\usepackage{subfig}
\usepackage{amssymb}
\usepackage{amsmath}
\usepackage{amsfonts}
\usepackage{scalefnt}
\usepackage{pgflibraryarrows}
\usepackage{pgflibrarysnakes}
\usepackage{xcolor}
\usepackage{caption}
\usepackage[onehalfspacing]{setspace}
\usepackage{lipsum}
\usepackage[disable]{todonotes}
\usepackage{threeparttable}
\usepackage{longtable}
\usepackage{float}
\usepackage{natbib}
\usepackage{enumitem}
\usepackage{placeins}
\usepackage{booktabs}
\usepackage{caption}
\usepackage{comment}
\usepackage[colorlinks=true, linkcolor=red!50!black, citecolor=blue!70!black, urlcolor=blue]{hyperref}

\newtheorem{theorem}{Theorem}

\newtheorem{corollary}{Corollary}

\newtheorem{definition}{Definition}

\newtheorem{lemma}{Lemma}

\newtheorem{proposition}{Proposition}
\newtheorem{remark}{Remark}

\newenvironment{proof}[1][Proof]{\noindent\textbf{#1.} }{\ \rule{0.5em}{0.5em}}
\input{tcilatex}
\DeclareMathOperator\supp{supp}
\DeclareMathOperator\cl{cl}
\DeclareMathOperator*{\argmax}{argmax}

\newcommand{\editYS}[1]{#1}

\begin{document}

\title{\textbf{Optimal Wage Band for Job Matching with Signaling\thanks{
We thank Maxim Ivanov, Jiangtao Li, Takashi Kunimoto, Gabor Virag, and seminar participants
at Singapore Management University, Theory and Experiment at McMaster (TEAM) Workshop,
the 35th Stony Brook International Conference on Game Theory, the fall 2024 Midwest Economic
Theory Conference. Han, Sam, and Shin gratefully acknowledge financial support from the Social Sciences
and Humanities Research Council of Canada. }}}
\author{ Seungjin Han\thanks{\fontsize{10}{13}\selectfont Dept.\ of
Economics, McMaster University, Canada. Email: hansj@mcmaster.ca} \and Alex
Sam\thanks{\fontsize{10}{13}\selectfont Dept.\ of Economics, University of
Calgary, Canada. Email: alex.sam1@ucalgary.ca} \and Youngki Shin\thanks{
\fontsize{10}{13}\selectfont Dept.\ of Economics, McMaster University,
Canada. Email: shiny11@mcmaster.ca} }
\date{\today }
\maketitle

\begin{abstract}
We study an optimal wage band problem in a competitive matching labor market where education signals worker ability. We prove uniqueness of the competitive signaling equilibrium under a general class of utility and profit functions and show that the optimal wage band problem is isomorphic to a simpler optimal ability threshold problem. Using a parametric model, we analyze how wage bands improve welfare relative to no intervention. Our results highlight novel mechanisms driven by asymmetric information, contrasting with existing literature. The framework is broadly applicable to settings where agents invest in costly signals under asymmetric information in competitive matching environments.
\end{abstract}
%\newpage

%\tableofcontents

%\newpage

\section{Introduction}
Since the seminal work of \citet{spence1973jobmarket}, a growing body of
research has highlighted the signaling role of education, particularly when schooling serves as a costly signal of worker ability. In such environments, individuals may pursue excessive education
to gain informational rents, leading to socially suboptimal outcomes
\citep[e.g.,][]{riley1979informational, Altonji1998, kaymak2025quantifying}. Separately, there has been growing interest in
efficiency gains from minimum wages, driven by factors such as monopsonistic wage-setting,
search frictions, and efficient rationing \citep*[e.g.,][]{lee2012optimal,
flinn2006minimum, berger2025minimum}.
In this paper, we study an optimal wage band problem in which a policymaker
sets both minimum and maximum wages prior to a competitive matching labor market.
To this end, we develop a welfare analysis framework that simultaneously
addresses inefficiency arising from
excessive investment in education and the ripple effects caused by the minimum
wage.
The key feature of our approach is the consideration of asymmetric
information about worker ability and education investments made before matching.

In Spencean signaling models, posterior beliefs about a worker's
unobservable type (e.g., ability) emerge endogenously through the worker's costly
acquisition of signals. \editYS{Typically, the models feature} highly nonlinear environments where types have
common value. For example, workers invest in education to signal their
inherent ability to firms in the job market. \editYS{In equilibrium}, they \editYS{tend} to choose
inefficiently high levels of costly education to obtain information rents.
These rents are further amplified in matching markets, \editYS{since} higher perceived
ability enables a worker to match with a more productive firm type.

This raises the question: Can a policymaker influence endogenous posterior
beliefs in job matching markets with signaling? We study the policymaker's
optimal wage band problem. \editYS{A wage band can be implemented economy-wide or targeted to specific sectors such as technology, finance,
large corporations, the public sector (e.g., government, education, healthcare), or unionized industries (e.g., manufacturing, transit). In practice, a wage band are used for talent management, market-aligned compensation, consistency in pay decisions,
transparency, and budget control.} \editYS{However, little is known about} the welfare
impact of a wage band when \editYS{education serves} as a signaling mechanism prior to job matching.

In this optimal wage band problem, we consider the following sequence of
decision-making procedures.
First, the policymaker sets a wage band.
Heterogeneous workers then choose their education levels as signals. Subsequently,
heterogeneous firms choose wages within the feasible band, resulting in one-to-one matching
between a continuum of heterogeneous workers and firms. Our solution
concept is the monotone competitive signaling equilibrium (CSE) in the
stronger set order, as proposed by \citet*{HAN_et_al}.

From a methodological perspective, this paper makes the following
contributions.
First, we extend the uniqueness result in \citet{HAN_et_al} to a much
broader setting. Specifically, we show that, for any given wage band, the monotone CSE is unique, provided the
utility and profit functions satisfy monotonicity, supermodularity, and
other mild regularity conditions, even in the absence of quasilinearity.
Second, we show that the policymaker's optimal wage band problem is isomorphic
to choosing an optimal pair of ability thresholds. This reformulation improves the
tractability of the original problem.
As a result, a typical monotone CSE for a given wage band is
\editYS{characterized} by two ability thresholds: low-ability workers opt out, mid-ability
workers separate through education, and high-ability workers pool at the same
education level.
The separating region in the middle is determined by a system of coupled
nonlinear ordinary differential equations (ODEs).

Building on this theoretical foundation, we construct a flexible parametric
model and analyze a novel mechanism through which an optimal wage band affects
the outcomes of the monotone CSE.
The optimal wage policy literature largely overlooks the role of workers'
education investments made prior to job matching.
For example, the assignment model \citep[e.g.,][]{becker1973theory,
tinbergen1975substitution, sattinger1975comparative} does not address pre-match investments.
By explicitly accounting for them, we are able to derive far-reaching and interesting
implications of the optimal wage band policy.

First, we demonstrate that under monotone, supermodular, and concave utility
and profit functions, the surplus efficiency frontier is convex, reflecting decreasing
returns to scale. This convexity ensures that both optimal ability thresholds rise continuously with
the weight assigned to worker surplus in the social welfare objective.

Second, we show that an optimal minimum wage raises equilibrium wages for all
workers, reflecting the ripple effect documented in the empirical literature
\citep[e.g.,][]{engbom2022earnings, phelan2019hedonic}.
A key insight of our analysis is that it reveals a detailed mechanism underlying this ripple effect,
which operates through workers' education investments and their resulting equilibrium utilities.
Compared to the baseline separating CSE without a minimum wage, low-skilled workers near the
minimum wage experience lower welfare, as additional education
investments required are not fully compensated by increased wages. This utility
loss is reversed for high-skilled workers, whose wage gains are sufficient
to offset the higher education costs. The resulting utility ripple effects lead to more
nuanced and asymmetric welfare outcomes in contrast to the redistribution argument
supported in models without signaling \citep*[e.g.,][]{lee2012optimal,
 berger2025minimum}.

Third, we also show that the maximum wage may improve the welfare by generating
two opposing forces. On one hand, it introduces inefficiency due to random matching occurred in the
upper pooling region. On the other hand, it saves education costs for
high-skilled workers who no longer need to invest excessively in costly
signaling. The optimal level of the
maximum wage depends on the trade-off between these two effects--matching
inefficiency and education cost savings. Within the
pooling region, the random matching structure implies that top-skilled workers
experience lower utility compared to the baseline separating CSE with a
no-intervention scenario.

Finally, we conduct comparative statics over key model parameters. For example,
examining the effect of firm heterogeneity on the optimal wage band
reveals that increases in the mean and variance of firm
productivity raise both the minimum and maximum thresholds. This result, combined with
the empirical findings of \citet{poschke2018firm}--who shows that firm size
distributions tend to be larger in \editYS{richer} countries and have increased over time in the U.S.--suggests a potential
mechanism behind the long-run decline in top income tax rates. In 1942, the U.S.
briefly proposed a 100\% marginal tax rate on income above \$25,000, and
implemented a top marginal rate of 93\% via the Revenue Act. In contrast, the
current U.S.\ top federal income tax rate is 37\%. Our results imply that
increased firm heterogeneity may reduce the effectiveness of maximum wage
instruments, helping to explain the erosion of top-end income taxation over
time.

\paragraph{Related Literature}
The assignment model (e.g., \citet{becker1973theory}, \citet{tinbergen1975substitution}, \citet{sattinger1975comparative}) study job matching
between heterogeneous firms and workers given fixed characteristics distributions, so it does not
allow us to study the impact of the wage band on workers' education choices prior to job matching.

Our model relates to \citet{kurlat2021signalling}, who also introduce two-sided heterogeneity in a competitive signaling environment.
In their model, firm type does not directly affect profits but instead reflects the firm's ability to acquire a direct signal about
a worker’s type—for example, through interviews or other screening mechanisms. This ability is valuable
because a worker’s signal is considered a pure waste: it does not enhance her productivity.

In contrast, in our matching model, firm type directly influences profit, creating a direct payoff complementarity between
firm type and worker type (and education). We show that this complementarity drives a worker’s education decision in a general setting,
\editYS{allowing for both productivity-related and purely signaling roles of education.}
%where education may or may not affect the worker’s productivity.
In this sense, our model can be \editYS{viewed} as an extension of the pre-match investment competition through a worker-firm framework.

Pre-match investment
competition studies whether pre-match competition to match with a better
partner can solve the hold-up problem of non-contractible pre-match
investment that prevails when a match is considered in isolation
\citep[e.g.,][]{grossman1986costs, williamson1986economic}.
\citet{cole1995incorporating}, \citet{rege2008people}, and \citet*{hoppe2009theory}
consider pre-match investment with incomplete information and
non-transferable utility without monetary transfers (i.e., no wage choice by
a firm). Therefore, the worker-firm framework does not apply. Pre-match
investment with incomplete information in \citet{hopkins2012job} includes the
transferable-utility case but only with no restrictions on transfers. A
separating equilibrium is the focus in \citet{hopkins2012job}.

Our solution concept is a monotone CSE in the stronger set order for job matching with signaling.
Monotone equilibrium analysis has been studied mostly in a model without
signaling \citep[e.g.,][]{athey2001single, mcadams2003isotone,
reny2004existence, reny2011existence, van2007monotone}. While early literature on signaling shows a glimpse of
monotone equilibrium analysis, it is formally analyzed only recently. With
the monotone-supermodular property of the worker's utility, the monotonicity
of the worker's signaling choice is established in a two-period signaling
game between one worker and one firm \citep{liu2020monotone} and a dynamic
incomplete information games with multiple players \citep{mensch2020existence}.

\citet{HAN_et_al} extends the signaling model to matching markets
with two-sided heterogeneity and provide a full monotone equilibrium
analysis: Suppose that the worker utility is monotone-supermodular and the
firm's utility is weakly monotone-supermodular. Then, the worker's action
function, the firm's wage function, the matching function and the belief in
a signaling equilibrium are all monotone in the stronger set order if and
only if it passes Criterion D1 (\citet*{cho1987signaling}, \citet*{%
banks1987equilibrium}).

Our paper is related to the literature on delegation
\citep[e.g.,][]{holmstrom1978incentives,holmstrom1984,melumad1991communication,
alonso2008optimal} in that the policymaker
lets firms and workers decide their wages in a wage band. Delegation
problems have been studied in various problems such as tariff caps
\citep{amador2013}, delegated project choice \citep{armstrong2010},
consumption-saving problems \citep*{amador2006}, min-max optimal mechanisms
\citep{frankel2014aligned}, monetary policy \citep*{athey2005}, veto-based
delegation \citep{mylovanov2008veto}, dynamic
delegation \citep{escobar2021, grenadier2016, guo2016dynamic, lipnowski2020}.

Most of works on optimal delegation have focused on a single principal and a
single agent. Notably, our paper brings a delegation problem as the design
of markets for job matching with signaling to develop a new method to study how
a \editYS{policymaker} can affect endogenous beliefs in matching markets. By doing
so, we provide a noble insight into the policymaker's willingness to
delegate and the optimal delegation in matching markets.

\section{Preliminaries}

\label{section: preliminaries}

We consider an economy consisting of a unit measure of heterogeneous workers
and firms. Workers have unobserved abilities distributed over $\mathcal{Z}=[%
\underline{z},\overline{z}]\subset \mathbb{R}$ according to a probability
measure $G$, and firms have productivities distributed over $\mathcal{X}=[%
\underline{x},\bar{x}]\subset \mathbb{R}$ according to a probability measure
$H$. While $G$ and $H$ are publicly known measures, a worker's ability is
\emph{private} information known only to the worker.
% We use $G$ and $H$ as probability measures not exactly CDFs across the paper.

We first discuss the sequence of decisions and the associated equilibrium
concept.
%The policymaker sets a wage band $T=[t_{\ell },t_{h}] \subset [\underline{t},\overline{t}]$, where $t_{\ell}$ and $t_{h}$ represent to the minimum and the maximum wages, respectively.
The policymaker sets a wage band $T=[t_{\ell },t_{h}]$, where $t_{\ell }$
and $t_{h}$ represent the minimum and the maximum wages, respectively. Let $%
\underline{t}\geq 0$ denote the lowest possible subsistence level of wage
for workers. While $\underline{t}=0$ in most cases, it can be strictly
positive. Let
\begin{equation}
\mathcal{T}:=\left\{ \left[ t_{\ell },t_{h}\right] :\underline{t}\leq
t_{\ell }\leq t_{h}\leq \infty \right\}  \label{feasible_bands}
\end{equation}%
be the set of all possible wage bands that the policymaker can choose. %
\todo{typo: added `policymaker'}

Given a wage band $T=[t_{\ell },t_{h}]$, each worker then chooses an
education level $s\in S=%
%TCIMACRO{\U{211d} }%
%BeginExpansion
\mathbb{R}
%EndExpansion
_{+}$. After observing education levels $s$, each firm chooses a wage $t\in
T $ when matching with a worker in the market. Workers and firms have the
option not to participate in the market, which we normalize as $s=0$ and $%
t=0 $, with corresponding utility and profit set to zero.

We apply the notion of the competitive signaling equilibrium (CSE)
introduced by \citet{HAN_et_al} to this model. In this competitive market,
both workers and firms take the market wage function $\tau _{T}:S^{\ast
}\rightarrow T$ as given, where $S^{\ast }$ denotes the set of education
levels chosen by workers who participate in the market. Thus, a firm offers
a wage $\tau _{T}(s)$ if it wishes to match with a worker whose education
level is $s\in S^{\ast }$. In equilibrium, workers and firms share a common
posterior belief function $\mu _{T}:S\rightarrow \Delta (\mathcal{Z})$,
which specifies the conditional distribution of $z$ given $s$. The
dependence of the market wage function and \editYS{the} posterior belief function on the
wage band $T$ announced by the policymaker is explicitly indicated by the
subscript $T$.

\subsection{Competitive Signaling Equilibrium given $T$}

\label{subsection:eq}Fix a wage band $T=[t_{\ell },t_{h}]\in \mathcal{T}$
announced by the policymaker. The utility for a worker with unobserved
ability $z$, education level $s$, and wage $t$ is denoted by $u(t,s,z).$ The
profit for a firm with productivity level $x$ is denoted by $g(t,x,s,z)$
when it is matched with a worker of ability $z$ and education level $s$ at
wage $t$. When we establish the existence of a unique CSE given any wage
band and show the equivalence between the optimal wage band problem in
Section \ref{section:optimal_wage_band}, we only impose
monotone-supermodular assumptions (Assumptions A and B in Appendix B),
concavity assumption (Assumption C) and other mild technical assumptions
(Assumptions E and G).

When we conduct comparative analysis in the optimal wage band and policy
analysis in Section \ref{secnumerical_analysis}, we adopt a parametric
approach. In Section \ref{secnumerical_analysis}, the worker's utility
function is assumed to be separable:
\begin{equation}
u(t,s,z)=h(t)-c(s,z),  \label{eq: work's utility}
\end{equation}%
where $h(t)$ denotes the utility from wage $t$ and $c(s,z)$ is the cost of
acquiring signal $s.$ Specifically, Section \ref{secnumerical_analysis}
considers an iso-elastic specification in wage $t$ and a convex cost
(disutility) in
education $s$:
\begin{align}
h(t)& :=%
\begin{cases}
\dfrac{t^{1-\rho }-1}{1-\rho } & \text{if }\rho \neq 1 \\
\ln t & \text{if }\rho =1%
\end{cases}
\label{eq: iso-utility} \\
c(s,z)& :=%
\begin{cases}
\beta \dfrac{s^{b}}{z} & \text{if }s\neq 0 \\
0 & \text{if }s=0,
\end{cases}
\label{eq: cost fn}
\end{align}%
where $\beta>0$, $b\geq1$, and $\rho\geq0$. The parameter $\rho$ represents the elasticity of the marginal utility with respect to wage
$t$ (or the measure of relative risk aversion).
\begin{comment}
Note that $\lim_{\rho \rightarrow 1}(t^{1-\rho }-1)/(1-\rho )=\ln t$. Estimates for the
plausible range of $\rho $ can be found in Evans (2005), Layard, Mayraz, and
Nickell (2008), and Gandelman and Hern\'{a}ndez-Murillo (2015), which
indicate cross-country variation. While it is typically above zero in macro
studies, the range of $\rho $ varies a lot (See, for example, Chetty
(2006)). % -1 is omitted and the limit is different, then.
% as $\rho \rightarrow 1,$ $\frac{t^{1-\rho }}{1-\rho }\rightarrow \ln t$.
For the case where the receiver is a firm and the sender is a worker, $t$
represents the sender's wage.%
\end{comment}

\todo{Changed from `income' to emphasize a
	labor income.}

Given the market wage function $\tau _{T}:S^{\ast
}\rightarrow T$, an education choice function $\sigma :\mathcal{Z}%
\rightarrow S$ is optimal for workers if (i) for all $z\in \mathcal{Z}$,
\begin{equation*}
\sigma _{T}(z):=\left\{
\begin{array}{cc}
\argmax_{s\in S}u(\tau _{T}(s),s,z) & \text{if positive} \\
0 & \text{otherwise,}%
\end{array}%
\right.
\end{equation*}%
and (ii) there is no profitable worker deviation to an off-path education in
equilibrium. Roughly speaking, no profitable worker deviation to an off-path
education implies that there does not exist an education level $s^{\prime
}\notin \sigma _{T}(\mathcal{Z}):=\left\{ \sigma _{T}(z):z\in \mathcal{Z}%
\right\} $ that would allow for a worker and a firm to be better off by
forming a new match and transferring a wage to the worker.

%(See Appendix \ref{App_def_CSE} for the formal definition).

In Section \ref{secnumerical_analysis}, the firm's profit function is given
by
\begin{equation}
g(t,x,s,z):=v(x,s,z)-t, \label{eq: firm_profit}
\end{equation}%
where $v(x,s,z)$ is the matching production function:
\begin{equation}
v(x,s,z):=A\left( s^{a}+1\right) xz+1, \label{eq: firm_production}
\end{equation}%
where $A>0$ and $0\leq a<1$. Note that the parameter $a$ captures the contribution of education to
production. When $a=0$, education is purely a signal and has no effect on
production.

Given the market wage function $\tau _{T}:S^{\ast }\rightarrow T$, a hiring
choice function $\xi _{T}:\mathcal{X}\rightarrow S^{\ast }\cup \{0\}$ is
optimal for firms if (i) for all $x\in \mathcal{X}$,
\begin{equation*}
\xi _{T}(x):=\left\{
\begin{array}{cc}
\argmax_{s\in S^{\ast }}\mathbb{E}_{\mu _{T}(s)}\left[ g(\tau _{T}(s),x,s,z)%
\right] & \text{if positive } \\
0 & \text{otherwise,}%
\end{array}%
\right.
\end{equation*}%
where $\mathbb{E}_{\mu _{T}}$ denotes expectation over $z$ with respect to
the measure $\mu_{T}$. \todo{\small Changed $\mu$ to $\mu_T$} If $\xi
_{T}(x)\in S^{\ast }$, then a firm with productivity $x$ wants to match with
a worker with education $\xi _{T}(x)$ at wage $\tau _{T}(\xi _{T}(x))$.
Similar to $S^{\ast }$, we denote by $\mathcal{X}^{\ast }$ the set of firms
that participate in the market.

Let $\mathcal{B}(A)$ be the Borel sigma-algebra on a set $A$. For each $s\in
S^{\ast }$, let $m_{T}(s)\in \mathcal{B}(\mathcal{X}^{\ast })$\ denote the
set of firms matched with a worker whose education level is $s$. Thus, $%
m_{T}:S^{\ast }\rightarrow \mathcal{B}(\mathcal{X}^{\ast })$\ is a
set-valued matching function. With slight abuse of notation, let $m_{T}^{-1}(x)\in S^{\ast }$ denote the
education level chosen by a worker hired by firm $x\in \mathcal{X}^{\ast }$.

Given a wage band $T=[t_{\ell },t_{h}]$, the tuple $c_{T}:=\{\sigma _{T},\mu
_{T},\tau _{T},m_{T}\}$\ constitutes \emph{a competitive signaling
equilibrium} \editYS{(CSE)} if (i) $\sigma _{T}$ is optimal for workers, (ii) $\mu
_{T}$ is Bayesian consistent, (iii) given $\{\sigma _{T},\mu _{T}\}$, a pair
$\{\tau _{T},m_{T}\}$ is an equilibrium matching outcome\footnote{%
In an equilibrium matching outcome, if there is a positive measure of
workers who choose some education $s\in S^{\ast }$ (i.e.$G\left[ \left\{
z|\sigma _{T}\left( z\right) =s\right\} \right] >0$), then there must be the
same measure of firms who wants those workers at wage $\tau _{T}(s)$ to
clear the market (i.e., $H\left[ \left\{ x|x\in m_{T}(\xi _{T}(x))\text{, }%
\xi _{T}(x)=s\right\} \right] =G\left[ \left\{ z|\sigma _{T}\left( z\right)
=s\right\} \right] ).$ This is part of the definition of an equilibrium
matching outcome.}. A more detailed treatment of CSE is available in
Appendix \ref{App_def_CSE}.

\subsection{Policy maker's optimal wage band}

The policymaker believes that a monotone CSE will emerge, in the economy
given a wage band $T=[t_{\ell },t_{h}]$. A CSE $c_{T}=\left\{ \sigma
_{T},\mu _{T},\tau _{T},m_{T}\right\} $ is said to be monotone if $\sigma
_{T},$ $\mu _{T},$ $\tau _{T},$ and $m_{T}$ are all non-decreasing in the
\emph{stronger} set order.\footnote{%
We say that a (set-valued) function $M:K\rightarrow P(Y)$ with a partially
ordered set $K$ with a given relation $\geq $ is non-decreasing in the
stronger set order if $k^{\prime }\leq k$ implies that $M(k^{\prime })\leq
_{c}M(k)$. Consider a posterior belief function $\mu _{T}:S\rightarrow
\Delta (Z)$. The monotonicity of a posterior belief function is defined by
the stronger set order on the supports of the probability distributions. A
posterior belief function is non-decreasing in the stronger set order if $%
s^{\prime }\leq s$ implies supp $\mu _{T}(s^{\prime })\leq _{c}$ supp $\mu
_{T}(s)$. A matching function $m_{T}:S^{\ast }\rightarrow B(X^{\ast })$ is
non-decreasing in the stronger set order if $s^{\prime }\leq s$ implies $%
m(s^{\prime })\leq _{c}m(s)$. Note $\sigma $ and $\tau $ are single-valued
functions, so the stronger set order is identical to the strong set order
for $\sigma _{T}$ and $\tau _{T}$.} Let $\mathcal{C}_{T}$ be the set of all
monotone CSEs given a wage band $T=[t_{\ell },t_{h}]$.

The policymaker is concerned about the weighted sum of firm surplus and
worker surplus in a monotone CSE $c_{T}\in \mathcal{C}_{T}.$ Given a
monotone CSE $c_{T}\in \mathcal{C}_{T}$, firm surplus and worker surplus, $%
R(c_{T})$ and $S(c_{T})$ are defined respectively as follows.
\begin{gather*}
R(c_{T}):=\int_{x\in \mathcal{X}^{\ast }}\mathbb{E}_{\mu _{T}(s)}\left[
g(\tau _{T}(\xi _{T}(x)),x,\xi _{T}(x),z)\right] dH(x), \\
S(c_{T}):=\int_{z\in \left\{ z\in \mathcal{Z}:\sigma _{T}(z)\in S^{\ast
}\right\} }u(\tau _{T}(\sigma _{T}(z)),\sigma _{T}(z),z)dG(z).
\end{gather*}%
A \emph{social welfare function} is defined as the weighted sum of the firm
and worker surpluses:
\begin{equation*}
W(c_{T})=\omega R(c_{T})+(1-\omega )S(c_{T}),
\end{equation*}%
where $\omega \in \lbrack 0,1]$ represents the weight placed on the firm's
surplus. A business-friendly policymaker will have a high value of $\omega ,
$ whereas a labour-friendly policymaker will have a low value of $\omega $.
\begin{comment}
As common agency literature suggests (e.g., \citet*{dixit1997common}), the
value of $\omega $ may be determined endogenously by lobbying from special
interest groups such as unions, business associations, or big individual
firms. We do not address this in our paper.
\end{comment}
We study the policymaker's optimal wage band problem given $\omega $.

\begin{definition}[Optimal Wage Band Problem]\label{def:optimal_wage_band}
The policymaker's \emph{optimal wage band problem} is to find a solution
for
\begin{equation}
\max_{T\in \mathcal{T}}\left[ \max_{c_{T}\in \mathcal{C}_{T}}W(c_{T})\right]
.  \label{wage_band_problem}
\end{equation}
\end{definition}

Recall that $\mathcal{C}_T$ denotes the set of all monotone CSEs. \citet{%
HAN_et_al} show that, under some regularity conditions,
%if the worker's utility function is monotone-supermodular and the firm's profit function is weakly monotone-supermodular,
a CSE is monotone in the stronger set order if and only if it passes
Criterion D1 in \citet*{cho1987signaling} and \citet*{banks1987equilibrium}.
% To our knowledge, this is the first monotone signaling equilibrium result for a competitive singlaing equilibrium in matching markets.
However, uniqueness of equilibrium is generally not guaranteed, raising the
question of which equilibrium to select among multiple candidates. Following the standard mechanism
design literature, we define the optimal wage band problem as choosing $T$
to maximize the social welfare under the best equilibrium outcome as formulated in Definition \ref{def:optimal_wage_band}.

\section{Optimal Wage Band Problem\label{section:optimal_wage_band}}

In this section, we establish the existence and uniqueness of a monotone CSE
under a broad class of utility and profit functions. This result
significantly generalizes the earlier work by \citet{HAN_et_al}, who proved
uniqueness under the assumption of quasilinear utility and profit functions.
Notice that quasilinear payoff functions impose restrictions such as risk
neutrality and separability, so we generalize the result to allow more general functions such as the
iso-utility function in \eqref{eq:
iso-utility} used in our policy analysis. As shown in the next subsection, our uniqueness
result under a broader class of functions accommodates general utility and profit functions, enabling analysis under rich heterogeneity in the elasticities of marginal utility of consumption and functional forms.

In subsection~\ref{section:unique_CSE} we show that for any wage band $T\in
\mathcal{T}$, a unique monotone CSE exists under mild regularity conditions%
\^{a}\euro ''without requiring separability or risk neutrality. This
uniqueness result simplifies the optimal wage band problem:
\begin{equation*}
\max_{T\in \mathcal{T}}\left[ \max_{c_{T}\in \mathcal{C}_{T}}W(c_{T})\right]
=\max_{T\in \mathcal{T}}W(c_{T}).
\end{equation*}%
Borrowing the terminology in the implementation literature, we can say that
the uniqueness of a monotone CSE given any wage band $T$ ensures the \emph{%
full implementation} of $\left\{ \sigma _{T},\mu _{T},\tau
_{T},m_{T}\right\} $ in monotone CSE.

In subsection \ref{section_welfare_max} we show that the policymaker's problem
can be converted to the optimization over two ability thresholds for workers: the
low ability threshold for market participation and the high ability threshold
for a pooled education choice.

\begin{remark}
For notational simplicity, we omit the subscript $T$ in the monotone CSE $%
\left\{ \sigma _{T},\mu _{T},\tau _{T},m_{T}\right\} $  unless needed for
clarity.
\end{remark}

\subsection{Unique monotone CSE given $T$\label{section:unique_CSE}}

\todo[inline]{Changed general terminology, action/reaction, sender/receiver,
    signal etc.
to labor application terms.}

We fix a wage band $T\in \mathcal{T}$ in this subsection and shows that there
exists a unique monotone CSE under mild assumptions described in Appendix %
\ref{App_separating_CSE}. They are (i) monotone-supermodular assumptions on
worker and firm's utility and profit functions (Assumptions A and B), (ii)
concavity assumptions (Assumption C), and other technical assumptions
(Assumptions D - F). These assumptions are mild so that it allows for
non-separable functions, power functions, logarithmic functions, etc, which
increases the applicability in applied works.

Note that if the market wage function $\tau $ induces a positive measure of
workers who choose a pooled education level $s\in S^{\ast },$ then $m(s)$
has the same measure for the market clearing condition. Furthermore, $m(s)$
is a connected interval when $m$ is non-decreasing in the stronger set
order. Therefore, pooling or semi-pooling is allowed in a monotone CSE in
the stronger set order.

We say that a monotone CSE is \emph{well-behaved} if it is characterized by
two ability thresholds, $z_{\ell }$ and $z_{h}$ such that every worker
below $z_{\ell }$ stays out of the market, every worker in $[z_{\ell
},z_{h}) $ differentiates themselves with their unique education choice,
 and every worker in $[z_{h},\overline{z}]$ pools themselves with the
same education choice. If $z_{\ell }<z_{h}=\bar{z},$ then a well-behaved CSE
is separating. If $z_{\ell }=z_{h},$ then a well-behaved CSE is pooling. If $%
z_{\ell }<z_{h}<\bar{z},$ then it is strictly well-behaved with both
separating and pooling regions in the equilibrium. We shall show that a
monotone CSE exists and is unique and well-behaved.

\subsubsection{Separating CSE}

We first study a separating monotone CSE.%
\todo[inline]{\small Can we
drop this, or move to the later part? From the beginning, `see appendix for
details' does not deliver much information on the contents in this section and in the appendix.}
Fix a wage band $T=[t_{\ell },t_{h}].$ Assume that $t_{h}=\infty $ for now.
Let $n(z)$ be the firm productivity which has the same percentile on the firm side as
the worker productivity $z$ has on the worker side, i.e., $H(n(z))=G(z)$. In a
separating monotone CSE, $n(z)$ is the productivity of the firm that hires a worker with ability $z$.

The minimum wage $t_{\ell }$ in a wage band determines the worker ability and the firm productivity
types, denoted by $z_{\ell }$ and $x_{\ell }:=n(z_{\ell })$, in the
equilibrium bottom match and worker $z_{\ell }$'s education level and the
wage paid to worker $z_{\ell }$. If the minimum wage is low enough to induce
all workers and firms to enter the market, the bottom match consists of
worker $\underline{z}$ and firms $\underline{x}.$ Further, worker $%
\underline{z}$'s education level and wage paid to her are a bilaterally
efficient pair, denoted by , $(s^{\ast }(\underline{z}),t^{\ast }(\underline{%
z}))$, in the bottom match as there in no information rent for worker $%
\underline{z}$.

In order to see the level of $t_{\ell }$ that induces all workers and firms to enter the
market, we define a bilaterally efficient pair of education and wage $%
(s^{\ast }(z),t^{\ast }(z))$ between worker $z$ and firm $n(z)$ as
a solution to the following maximization problem:

\begin{equation}
\max_{(s,t)\in S\times
%TCIMACRO{\U{211d} }%
%BeginExpansion
\mathbb{R}
%EndExpansion
_{+}}g(t,n(z),s,z)\text{ s.t. }(s,t)\in D(z),  \label{maximization}
\end{equation}%
where $D(z):=\{(s,t)\in S\times
%TCIMACRO{\U{211d} }%
%BeginExpansion
\mathbb{R}
%EndExpansion
_{+}:u(t,s,z)\geq 0\}$. The constraint is binding at a solution $(s^{\ast
}(z),t^{\ast }(z))$. We normalize
\begin{equation}
g(t^{\ast }(\underline{z}),n(\underline{z}),s^{\ast }(\underline{z}),%
\underline{z})=0
\end{equation}%
and assume that $\underline{t}\leq t^{\ast }(\underline{z})$. If $t_{\ell
}\in \lbrack \underline{t},t^{\ast }(\underline{z})]$, all workers and firms
enter the market and hence the bottom match is formed between $z_{\ell }=%
\underline{z}$ and $x_{\ell }=\underline{x}=n(\underline{z})$ with $s^{\ast
}(\underline{z})$ and $t^{\ast }(\underline{z})$ as their education and wage
choices.

If the minimum wage $t_{\ell }$ in a wage band exceeds $t^{\ast }(\underline{%
z})$, then the worker's ability $z_{\ell }$ in the bottom match is strictly
greater than $\underline{z}$. For the case of $t^{\ast }(\underline{z}%
)<t_{\ell }$, define an education level, $\kappa (t_{\ell },z)$ that
satisfies
\begin{equation}
u(t_{\ell },\kappa (t_{\ell },z),z)=0,  \label{lemmaAe_1}
\end{equation}%
given $(t_{\ell },z)$. Let us define $\hat{t}$ as the value of wage that
satisfies
\begin{equation*}
g(\hat{t},n(\bar{z}),\kappa (\hat{t},\overline{z}),\overline{z})=0.
\end{equation*}%
If $t_{\ell }>\hat{t},$ no positive measure of workers and firms will enter
the market. If $t_{\ell }\in (t^{\ast }(\underline{z}),\hat{t}),$ then, we
have that $z_{\ell }\in (\underline{z},\overline{z})$ and the wage chosen by
firm $n(z_{\ell })$ is $t_{\ell }.$ Furthermore, given $t_{\ell }\in
(t^{\ast }(\underline{z}),\hat{t})$, the worker's ability $z_{\ell }$ and
her education level $s_{\ell }$ are simultaneously determined so that they
make worker type $z_{\ell }$ and her employee, firm type $n(z_{\ell })$
indifferent between entering the market forming the match and staying out of
the market:
\begin{align}
g\left( t_{\ell },n\left( z_{\ell }\right) ,s_{\ell },z_{\ell }\right) & =0,
\label{lem1e} \\
u\left( t_{\ell },s_{\ell },z_{\ell }\right) & =0.  \label{lem2e}
\end{align}%
Lemma \ref{lemma_bottom_types_e} in Appendix \ref{App_lem_bottom_match}
establishes the existence of a unique $\left( z_{\ell },s_{\ell }\right) $
given any $t_{\ell }\in \lbrack \underline{t},\hat{t}).$

In any separating monotone CSE, the worker's optimal education choice
satisfies that for all $z\in (z_{\ell },\bar{z})$,
\begin{equation}
u_{t}\left( \tau (s),s,z\right) \tau ^{\prime }(s)=-u_{s}\left( \tau
(s),s,z\right) .  \label{worker_optimal_e}
\end{equation}%
On the other hand, the optimal wage choice $\tau (s)$ by a firm who choose
to match with a worker with education $s$ satisfies
\begin{equation}
g_{t}\left( \tau (s),s,\mu (s),x\right) \tau ^{\prime }(s)+g_{z}\left( \tau
(s),s,\mu (s),x\right) \mu ^{\prime }(s)=-g_{s}\left( \tau (s),s,\mu
(s),x\right) .  \label{receiver_optimal_e}
\end{equation}

(\ref{worker_optimal_e}) and (\ref{receiver_optimal_e}) yield the couple
system of first-order nonlinear ODEs:
\begin{equation}
\begin{bmatrix}
\tau ^{\prime }(s) & \mu ^{\prime }(s)%
\end{bmatrix}
^{\mathsf{T}}=\Phi (s,\tau (s),\mu (s)),  \label{system_diff_eq_e}
\end{equation}
where
\begin{equation*}
\Phi (s,\tau (s),\mu (s)):=
\begin{bmatrix}
-\dfrac{u_{s}}{u_{t}} & \dfrac{g_{t}\times u_{s}-g_{s}\times u_{t}}{
u_{t}\times g_{z}}%
\end{bmatrix}
^{\mathsf{T}}.
\end{equation*}
$\tau $ and $\mu $ can be derived by solving it with the initial condition $%
(s_{\ell },t_{\ell },z_{\ell })$ that is unique and exists according to
Lemma \ref{lemma_bottom_types_e}. Let $\{\tilde{\sigma},\tilde{\mu},\tilde{
\tau},\tilde{m}\}$ denote a separating monotone CSE.

\begin{theorem}[Existence of separating CSE]
\label{thm_unique_separating_eq_e}If $g$\ and $u$\ are such that $\Phi $\
defined in (\ref{system_diff_eq_e}) is uniformly Lipshitz continuous, then
the solution exists and it is unique. Subsequently, a unique separating
monotone CSE $\{\tilde{\sigma},\tilde{\mu},\tilde{\tau},\tilde{m}\}$ exists.
\end{theorem}

The proof of Theorem \ref{thm_unique_separating_eq_e} is standard. It
follows from the Picard-Lindeloff existence and uniqueness theorem. Given
the unique solution $(\tilde{\mu},\tilde{\tau})$ for the coupled system of
first-order non-linear ODE, $\tilde{\sigma}$ and $\tilde{m}$ are derived as
follows:

\begin{enumerate}
\item $\tilde{\sigma}(\mu (s))=s$ for all $\mu (s)\geq z_{\ell }$ and $%
\tilde{\sigma}(z)=0$ for all $z<z_{\ell },$

\item $\tilde{m}(s)=n(\mu (s))$ for all $s\in S^{\ast }=\{\tilde{\sigma}
(z):z\geq z_{\ell }\}.$
\end{enumerate}

Let $Z(s)$ denote the set of the types of workers who choose education $s$.

\begin{lemma}
\label{lemma_no_bottom_bunching_e}If $Z(s)$ has a positive measure in a
monotone CSE, then it is an interval with $\max Z(s)=\overline{z}.$
\end{lemma}

Because the (stronger) monotone $\mu $ is equivalent to Criterion D1, we can
apply Criterion D1 to Lemma \ref{lemma_no_bottom_bunching_e} to show that a
unique monotone CSE does not involve pooling when $\tilde{\tau}(\tilde{\sigma%
}(\bar{z}))\leq t_{h}.$

\begin{theorem}
\label{thm_uniqueSME_e}Suppose that $T=[t_{\ell },t_{h}]$ satisfies
$\underline{t}\leq t_{\ell }<\tilde{\tau}(\tilde{\sigma}(\bar{z}))\leq t_{h}$.
Then, a monotone CSE is unique and it is separating.
\end{theorem}

\paragraph{Ripple Effect}
(\ref{lem1e}) and (\ref{lem2e}) are crucial to understand the ripple effect of the \emph{minimum wage}
on the equilibrium profits and equilibrium utilities net of education costs. To see this, first
note that in the baseline separating CSE with no minimum wage, all firms and workers enter
the market, the worst firm and the worst worker
receive their reservation profit and net utility, and equilibrium profit and net utility
are increasing in the firm productivity and the worker ability respectively thanks to the envelope theorem.

Suppose that the policymaker introduces a minimum wage that is greater than $t^{*}(\underline{z})$.
This results in a separating CSE $\{\tilde{\sigma},\tilde{\mu},\tilde{\tau},\tilde{m}\}$,
where low productivity firms and low ability workers stay out of the market
(i.e., those below $n(z_{\ell})$ and $z_{\ell}$) and hence makes them worse off than they are in the baseline CSE with no minimum wage.
In addition, (\ref{lem1e}) and (\ref{lem2e}) imply that firm  $n(z_{\ell})$ and worker $z_{\ell}$ receive their reservation profit and net utility. This, in conjunction with the continuity of the separating region of the CSE, implies that those firms and workers whose productivity and ability are higher than but close to $n(z_{\ell})$ and $z_{\ell}$ are also worse off than they are in the baseline separating CSE.

Our policy analysis in Section \ref{secnumerical_analysis} demonstrates that all firms are in fact worse off
but workers whose ability are close to the top ability are better off. This shows a stark contrast to the ripple effect of the minimum wage
on equilibrium wages \citep{engbom2022earnings, phelan2019hedonic} in that the minimum wage makes low skilled workers worse off, while benefiting only high skilled workers.

\subsubsection{Strictly well-behaved CSE}

If $t_{h}<\tilde{\tau}(\tilde{\sigma}(\bar{z}))$, then we have no separating
CSE. Given Lemma \ref{lemma_no_bottom_bunching_e}, Lemma \ref{lemma_binding_upper_bound_e} shows that if there is pooling on the top of
the worker side, the wage to those senders pooled at the top must be the
maximum wage $t_{h}.$

\begin{lemma}
\label{lemma_binding_upper_bound_e}If $Z(s)$ has a positive measure in a
monotone CSE, then $t_{h}$ is the wage to the workers in $Z(s)$.
\end{lemma}

If $t_{h}<\tilde{\tau}(\tilde{\sigma}(\bar{z}))$, there are only two types
of non-separating monotone CSEs as shown in Lemma \ref{theorem_all_eq_w/o_separating_e}. The reason is that pooling can happen
only among workers in an interval with $\bar{z}$ being its maximum due to
Lemma \ref{lemma_no_bottom_bunching_e}.

\begin{lemma}
\label{theorem_all_eq_w/o_separating_e}If $\underline{t}\leq t_{h}<\tilde{\tau}(\tilde{\sigma}(\bar{z}))$, then, there are two possible monotone CSEs:
(i) a strictly well-behaved monotone CSE with both separating and pooling
regions and (ii) a monotone pooling CSE.
\end{lemma}

Lemma \ref{theorem_all_eq_w/o_separating_e} follows Lemmas \ref%
{lemma_no_bottom_bunching_e} and \ref{lemma_binding_upper_bound_e}. When $
t_{\ell }<t_{h}<\tilde{\tau}(\tilde{\sigma}(\bar{z}))$, we have a strictly
well-behaved CSE $\left\{ \hat{\sigma},\hat{\mu},\hat{\tau},\hat{m}\right\} $
with both separating and pooling regions. The upper ability threshold $%
z_{h}$ and the pooled education $s_{h}$ are determined by the following
system of equations:
\begin{gather}
u(t_{h},s,z)=u(\tilde{\tau}\left( \tilde{\sigma}\left( z\right) \right) ,
\tilde{\sigma}\left( z\right) ,z)  \label{jumping_seller_e} \\
\mathbb{E}[g(t_{h},s,z^{\prime },n\left( z\right) )|z^{\prime }\geq z]=g(%
\tilde{\tau}\left( \tilde{\sigma}\left( z\right) \right) ,\tilde{\sigma}%
\left( z\right) ,z,n\left( z\right) ).  \label{jumping_buyer_e}
\end{gather}

Let $(s_{h},z_{h})$ denote a solution of (\ref{jumping_seller_e}) and (\ref%
{jumping_buyer_e}). Note that (\ref{jumping_seller_e}) makes worker $z_{h}$ indifferent
between choosing $s_{h}$ for $t_{h}$ and $\tilde{\sigma}%
\left( z_{h}\right) $ for $\tilde{\tau}\left( \tilde{\sigma}\left(
z_{h}\right) \right) .$ The expression on the left hand side of (\ref%
{jumping_seller_e}) is the equilibrium utility for worker $z_{h}$. The
expression on the left hand side of (\ref{jumping_buyer_e}) is the profit
for firm $n\left( z_{h}\right) $ who chooses a worker with education $%
s_{h}$ as his partner by choosing $t_{h}$ for her. This is the equilibrium
profit for firm type $n(z_{h})$. The expression on the right-hand side is
his profit if he chooses a worker type $z_{h}$ with education $\tilde{\sigma}%
\left( z_{h}\right) $ as his partner by paying the wage $\tilde{\tau}\left(
\tilde{\sigma}\left( z_{h}\right) \right) $.

\begin{lemma}
\label{lemma_well_behaved_e}Given $\underline{t}\leq t_{\ell }<t_{h}<\tilde{
\tau}(\tilde{\sigma}(\bar{z}))$, there exists a unique $\left(
s_{h},z_{h}\right) \in (s_{\ell },\tilde{\sigma}(\bar{z}))\times \left(
z_{\ell },\overline{z}\right) $ that satisfies (\ref{jumping_seller_e}) and
( \ref{jumping_buyer_e}).
\end{lemma}

The strictly well-behaved monotone CSE $\left\{ \hat{\sigma},\hat{\mu},\hat{%
\tau},\hat{m}\right\} $ follows the baseline separating monotone CSE $\{%
\tilde{\sigma},\tilde{\mu},\tilde{\tau},\tilde{m}\}$ up until the upper
ability threshold $z_{h}.$ Jumping occurs at $z_{h}$ and all workers
above $z_{h}$ chooses the pooled education $s_{h}$ strictly higher
than $\tilde{\sigma}\left( z_{h}\right) $ for $t_{h},$ which is also
strictly higher than $\tilde{\tau}\left( \tilde{\sigma}\left( z_{h}\right)
\right) $:

\begin{lemma}
\label{lemma2_e}We have that $\tilde{\tau}\left( \tilde{\sigma}\left(
z_{h}\right) \right) <t_{h}<\tilde{\tau}\left( \tilde{\sigma}\left( \bar{z}
\right) \right) $ and $\tilde{\sigma}\left( z_{h}\right) <s_{h}<\tilde{%
\sigma }\left( \bar{z}\right) $.
\end{lemma}

The posterior market belief $\hat{\mu}$ follows $\tilde{\mu}$ for $s\in %
\left[ \hat{\sigma}(z_{\ell }),\hat{\sigma}(z_{h})\right) $ but it is $\hat{%
\mu}(s)=G(z|z_{h}\leq z\leq \overline{z})$ for $s=s_{h}.$ Using the
(stronger) monotonicity, off-path $\hat{\mu}$ is determined as follows: $%
\hat{\mu}(s)=z_{h}$ if $s\in \lbrack \lim_{z\nearrow z_{h}}\hat{\sigma}%
(z),s_{h})$ and $\hat{\mu}(s)=\overline{z}$ if $s>s_{h}$. The equilibrium
matching $\hat{m}$ follows that (i) $\hat{m}(s)=n(\hat{\mu}(s))$ if $s\in %
\left[ \hat{\sigma}(z_{\ell }),\hat{\sigma}(z_{h})\right) $, (ii) $\hat{m}%
(s)=\left[ x_{h},\overline{x}\right] $ if $s=s_{h}.$

Theorem \ref{thm_unique_well_behaved_e} below shows that if $\underline{t}%
\leq t_{\ell }<t_{h}<\tilde{\tau}(\tilde{\sigma}\left( \overline{z}\right) )$
, then $\left\{ \hat{\sigma},\hat{\mu},\hat{\tau},\hat{m}\right\} $ is a
unique monotone CSE.

\begin{theorem}
\label{thm_unique_well_behaved_e}Fix a wage band to $T=[t_{\ell },t_{h}]$
with $\underline{t}\leq t_{\ell }<t_{h}<\tilde{\tau}(\tilde{\sigma}\left(
\overline{z}\right) )$. A unique monotone CSE is $\left\{ \hat{\sigma},\hat{%
\mu},\hat{\tau},\hat{m}\right\} $ and it exists
\end{theorem}

Theorem \ref{thm_unique_separating_eq_e} and Lemmas \ref%
{lemma_bottom_types_e} and \ref{lemma_well_behaved_e} establish the
existence of a unique well-behaved monotone CSE $\left\{ \hat{\sigma},\hat{
\mu},\hat{\tau},\hat{m}\right\} $. Because of Lemma \ref%
{theorem_all_eq_w/o_separating_e}, we can establish Theorem \ref%
{thm_unique_well_behaved_e} by showing that there is no monotone pooling CSE
if $\underline{t}\leq t_{\ell }<t_{h}<\tilde{\tau}(\tilde{\sigma}(\bar{z}))$
(See the proof of Theorem \ref{thm_unique_well_behaved_e} in Appendix \ref%
{App_thm_unique_well_behaved_e}).

\paragraph{Random Matching versus Cost Savings}
When the \emph{maximum wage} is less than $\tilde{\tau}(\tilde{\sigma}\left(
\overline{z}\right))$, it incurs the pooled education $s_{h}$ lower than $\tilde{
\sigma }\left( \bar{z}\right)$ as shown in Lemma \ref{lemma2_e}. If there are welfare gains from imposing a maximum wage, the education cost savings associated with $s_{h}$ must outweigh, the inefficiency induced by random matching in the pooling region.

Our policy analysis in Section \ref{secnumerical_analysis} demonstrates that the maximum
wage becomes a less effective policy instrument as the mean and variance of the firm productivity distribution increase
or the elasticity of marginal utility of consumption decreases. This is because such changes make the matching inefficiency
from high-type pooling outweighs the savings from reduced education costs.

\subsubsection{Pooling CSE}

As $t_{h}\rightarrow t_{\ell },$ the strictly well-behaved monotone CSE
converges to a pooling monotone CSE with no separating part. When $t_{\ell
}=t_{h}=t^{\ast },$ we have a pooling monotone CSE where every worker of
type above $z^{\ast }$ enters the market with the pooled signal $s^{\ast }$.
Further,
\begin{gather}
u(t^{\ast },s^{\ast },z^{\ast })\geq 0,  \label{pooling_sender_e} \\
\mathbb{E}\left[ g\left( t^{\ast },n\left( z^{\ast }\right) ,s^{\ast
},z^{\prime }\right) |z^{\prime }\geq z^{\ast }\right] \geq 0,\text{ }
\label{pooling_receiver_e}
\end{gather}%
where each condition holds with equality $z^{\ast }>\underline{z}$. The
existence of a unique monotone pooling CSE hinges on the existence of a
unique solution $\left( s^{\ast },z^{\ast }\right) $ that solves (\ref%
{pooling_sender_e}) and (\ref{pooling_receiver_e}).

\begin{lemma}
\label{lemma_pooling_e}Given $\underline{t}<t^{\ast }<\overline{t}^{\ast }$,
there exists a unique solution $\left( s^{\ast },z^{\ast }\right) \in
%TCIMACRO{\U{211d} }%
%BeginExpansion
\mathbb{R}
%EndExpansion
_{++}\times (\underline{z},\overline{z})$ that solves %
\eqref{pooling_sender_e} and \eqref{pooling_receiver_e} with equality, where
$\overline{t}^{\ast }$ is the wage that makes \eqref{pooling_sender_e} and %
\eqref{pooling_receiver_e} with equality given $z^{\ast }=\overline{z}.$
\end{lemma}

Now we establish the existence of a unique monotone CSE given any singleton
wage $t^{\ast }$ such that $\underline{t}\leq t^{\ast }<\overline{t}^{\ast
}. $

\begin{theorem}
\label{thm:pooling}Fix any singleton wage $t^{\ast }$ such that $\underline{t%
}\leq t^{\ast }<\overline{t}^{\ast }$, then there exists a unique monotone
CSE and it is pooling such that all workers of type no less than $z^{\ast }$
enter the market with a pooled signal $s^{\ast }$ and all firms of type no
less than $n(z^{\ast }).$
\end{theorem}

One implication of Theorem \ref{thm:pooling} is that \eqref{pooling_sender_e}
and \eqref{pooling_receiver_e} hold with equality in a unique monotone CSE
given any singleton wage $t^{\ast }$ such that $\underline{t}\leq t^{\ast }<%
\overline{t}^{\ast }$.

\subsection{Policy maker's optimization strategy}

\label{section_welfare_max}

Subsection \ref{section:unique_CSE} shows that there exists a unique
monotone CSE given any wage band $T\in \mathcal{T}$ when Assumptions A - F
are satisfied. Therefore, the optimal wage band problem satisfies%
\begin{equation}
\max_{T\in \mathcal{T}}\left[ \max_{c_{T}\in \mathcal{C}_{T}}W(c_{T})\right]
=\max_{T\in \mathcal{T}}W(c_{T}),  \label{max_uniqueCSE}
\end{equation}%
where $c_{T}$ on the right hand side is a unique monotone CSE given a wage
band $T\in \mathcal{T}$

Recall that $t^{\ast }(\underline{z})$ is the highest minimum wage that
induces every worker to participate in the market. Let $t_{h}^{\ast
}(t_{\ell })$ denote the lowest maximum wage that induces no pooled
education chosen at the top. Without loss of generality, the policymaker
can focus on the following set of wage bands $\mathcal{T}^{\ast }\subset
\mathcal{T}$:

\begin{equation*}
\mathcal{T}^{\ast }:=\left\{ [t_{\ell },t_{h}]:t^{\ast }(\underline{z})\leq
t_{\ell }\leq \hat{t},\text{ }t_{\ell }\leq t_{h}\leq t_{h}^{\ast }(t_{\ell
})\right\} \text{,}
\end{equation*}%
that is
\begin{equation}
\max_{T\in \mathcal{T}}W(c_{T})=\max_{T\in \mathcal{T}^{\ast }}W(c_{T}).
\label{max_uniqueCSE*}
\end{equation}

Lemma \ref{lem_unified_well_behaved_e} shows that the policymaker only
needs to consider a strictly well-behaved monotone CSE because it also
covers a monotone separating CSE and a monotone pooling CSE.

\begin{lemma}
\label{lem_unified_well_behaved_e}As $z_{h}\rightarrow \bar{z},$ a strictly
well-behaved monotone CSE $\left\{ \hat{\sigma},\hat{\mu},\hat{\tau},\hat{m}%
\right\} $ converges to the separating monotone CSE with the same lower
threshold worker type $z_{\ell }.$ As $z_{h}\rightarrow z_{\ell },$ $\left\{
\hat{\sigma},\hat{\mu},\hat{\tau},\hat{m}\right\} $ converges to the pooling
monotone CSE in which $t_{\ell }$ is a single feasible wage, $z_{\ell }$ is
the threshold worker type for market entry, and $s_{\ell }$ is the pooled
signal for workers in the market.
\end{lemma}

Combining with Lemma \ref{lem_unified_well_behaved_e}, the next two
propositions reduce the policymaker's optimal choice of a wage band to the
optimal choice of $z_{\ell }$ and $z_{h}$ subject to $z_{\ell }\in Z$ and $%
z_{h}\geq z_{\ell }$, given that a unique monotone CSE is well-behaved. Once
she identifies $z_{\ell }$ and $z_{h},$ she can retrieve $T=[t_{\ell
},t_{h}] $ that induces the two worker threshold types, $z_{\ell }$ and $%
z_{h}$.

\begin{proposition}
\label{prop_unbounded_design}For any given $z_{\ell }\in \lbrack \underline{z%
},\overline{z}),$ there exists a unique solution $(s_{\ell },t_{\ell })$ of (%
\ref{lem1e}) and (\ref{lem2e}). For any given $t_{\ell }\in \lbrack t^{\ast
}(\underline{z}),\hat{t})$, there exists a unique solution $(z_{\ell
},s_{\ell })$ of (\ref{lem1e}) and (\ref{lem2e}). Furthermore, given any $%
z_{\ell }\in \lbrack \underline{z},\overline{z}),$ $(s_{\ell },t_{\ell })\in
%TCIMACRO{\U{211d} }%
%BeginExpansion
\mathbb{R}
%EndExpansion
_{+}\times \lbrack t^{\ast }(\underline{z}),\hat{t})$ is a solution of (\ref%
{lem1e}) and (\ref{lem2e}) if and only if $(z_{\ell },s_{\ell })$ is a
solution of (\ref{lem1e}) and (\ref{lem2e}).
\end{proposition}

Note that (\ref{lem1e}) and (\ref{lem2e}) are the utility indifference
conditions that must satisfy at the bottom match. They also hold when $%
z_{\ell }=\underline{z}$ because we normalize $g(t^{\ast }(\underline{z}),n(
\underline{z}),s^{\ast }(\underline{z}),\underline{z})=0$ in (\ref%
{g_normalized}).

\begin{proposition}
\label{prop_bounded_design}For any given $z_{h}\in (z_{\ell },\overline{z}),$
there exists a unique $(s_{h},t_{h})$ of (\ref{jumping_seller_e}) and (\ref%
{jumping_buyer_e}). For any given $t_{h}\in \left( t_{\ell },t_{h}^{\ast
}(t_{\ell })\right) $, there exists a unique solution $(z_{h},s_{h})$ of (%
\ref{jumping_seller_e}) and (\ref{jumping_buyer_e}). Furthermore, given any $%
z_{h}\in (z_{\ell },\overline{z}),$ $(s_{h},t_{h})$ is a solution of (\ref%
{jumping_seller_e}) and (\ref{jumping_buyer_e}) if and only if $%
(z_{h},s_{h}) $ is a solution of (\ref{jumping_seller_e}) and (\ref%
{jumping_buyer_e}).
\end{proposition}

Note that (\ref{jumping_seller_e}) and (\ref{jumping_buyer_e}) are the
utility indifference conditions for the pooled education choice at the top.
Let $Z=[z_{\ell },z_{h}]$ denotes the separating region,
which means that all workers below $z_{\ell }$ stay out of the market
and all workers above $z_{h}$ choose a pooled education level. Define
\begin{equation*}
\mathcal{Z}^{\ast }:=\left\{ [z_{\ell },z_{h}]:\underline{z}\leq z_{\ell
}\leq z_{h}\leq \overline{z}\right\} .
\end{equation*}%
Let $\hat{c}_{Z}=\left\{ \hat{\sigma},\hat{\mu},\hat{\tau},\hat{m}\right\} $
denote a strictly well-behaved monotone CSE that induces a separating region
$Z=[z_{\ell },z_{h}]\in \mathcal{Z}^{\ast }$. Then, firm surplus and worker
surplus are expressed by%
\begin{eqnarray*}
R(\hat{c}_{Z}) &=&\int_{\hat{z}_{\ell }}^{\hat{z}_{h}}g(\hat{\tau}(\hat{%
\sigma}(z)),\hat{\sigma}(z),z,n(z))dG\left( z\right) +\int_{\hat{z}_{h}}^{%
\bar{z}}\mathbb{E}_{\hat{\mu}}\left[ g(t_{h},s_{h},z^{\prime
},n(z))|z^{\prime }\geq z_{h}\right] dG(z), \\
S(\hat{c}_{Z}) &=&\int_{\hat{z}_{\ell }}^{\hat{z}_{h}}u(\hat{\tau}(\hat{%
\sigma}(z)),\hat{\sigma}(z),z)dG\left( z\right) +\int_{\hat{z}_{h}}^{\bar{z}%
}u(t_{h},s_{h},z)dG(z).
\end{eqnarray*}%
where the first integration in $R(\hat{c}_{Z})$ denotes the firm surplus in
the separating region and the second denotes the firm surplus in the pooling
region. The social welfare function is expressed by $W(\hat{c}_{Z})=\omega R(%
\hat{c}_{Z})+(1-\omega )S(\hat{c}_{Z})$.

Now our key theorem is presented below.

\begin{theorem}[Optimal Threshold Problem]
\label{thm:optimal_band}The policymaker's optimal band problem is the \emph{%
optimal threshold problem}:%
\begin{equation}
\max_{T\in \mathcal{T}}\left[ \max_{c_{T}\in \mathcal{C}_{T}}W(c_{T})\right]
=\max_{Z\in \mathcal{Z}^{\ast }}W(\hat{c}_{Z}).  \label{optimal_band1}
\end{equation}
\end{theorem}

\begin{proof}
(\ref{max_uniqueCSE}) and (\ref{max_uniqueCSE*}) induce%
\begin{equation}
\max_{T\in \mathcal{T}}\left[ \max_{c_{T}\in \mathcal{C}_{T}}W(c_{T})\right]
=\max_{T\in \mathcal{T}^{\ast }}W(c_{T}).  \label{optimal_band2}
\end{equation}

Lemma \ref{lem_unified_well_behaved_e} implies that the policymaker can
focus on a strictly well-behaved monotone CSE. Furthermore, Propositions \ref%
{prop_bounded_design} and \ref{prop_unbounded_design} imply that there
exists a bijection $\varphi :\mathcal{T}^{\ast }\rightarrow \mathcal{Z}%
^{\ast }$ such that for any $T=[t_{\ell },t_{h}],$ $\varphi \left( T\right)
=[z_{\ell },z_{h}]$ is the separating region of worker types in the
corresponding strictly well-behaved monotone CSE. This implies
\begin{equation}
\max_{T\in \mathcal{T}^{\ast }}W(c_{T})=\max_{Z\in \mathcal{Z}^{\ast }}W(%
\hat{c}_{Z}).  \label{optimal_band3}
\end{equation}%
(\ref{optimal_band2}) and (\ref{optimal_band3}) lead to (\ref{optimal_band1}%
).
\end{proof}

\bigskip

Theorem \ref{thm:optimal_band} implies that the optimal band problem is
isomorphic to the \emph{optimal threshold problem} in a strictly
well-behaved monotone CSE where the policymaker finds $z_{\ell }$ and $z_{h}$
that maximize the weighted sum of firm surplus and worker surplus. Once the
optimal two ability thresholds $(z_{\ell },z_{h})$ are derived, we can
derive the corresponding optimal wage band $T=[t_{\ell },t_{h}]$ by solving (%
\ref{lem1e}) and (\ref{lem2e}) for $(s_{\ell },t_{\ell })$ given $t_{\ell }$
and solving (\ref{jumping_seller_e}) and (\ref{jumping_buyer_e}) for $%
(s_{h},t_{h})$ given $z_{h}$, respectively.

It is also interesting to examine the relation between the optimal threshold
problem and the optimal design of the posterior belief. Let $\Pi ^{\ast
}:=\{\mu _{T}:T\in \mathcal{T}^{\ast }\}$ be the set of all possible
equilibrium posterior beliefs. There exists a bijection $\gamma :\mathcal{Z}%
^{\ast }\rightarrow \Pi ^{\ast }$ because each posterior belief in $\Pi
^{\ast }$ is uniquely identified by $[z_{\ell },z_{h}]$ such that: (i) $%
z_{\ell }$ is the lowest worker type in the separating part with $\mu
(s)=\sigma ^{-1}(s)$ if $s\in \left[ \sigma (z_{\ell }),\sigma
(z_{h})\right) $, and (ii) $z_{h}$ is the lowest worker type in the pooling
region with $\mu (s_{h})=G(z|z_{h}\leq z\leq \overline{z}).$ Therefore, we
have that
\begin{equation*}
\mathcal{T}^{\ast }\text{ }\overset{\text{bijection }\varphi }{\rightarrow }%
\text{ }\mathcal{Z}^{\ast }\text{ }\overset{\text{bijection }\gamma }{%
\rightarrow }\text{ }\Pi ^{\ast }.
\end{equation*}%
Let $\hat{c}_{\mu }$ denote a strictly well-behaved monotone CSE that
induces a posterior belief $\mu \in \Pi ^{\ast }.$ We can derive the social
welfare function $W(\hat{c}_{\mu })$ analogous to $W(\hat{c}_{Z})$. Then, it
is straightforward to show the following corollary.

\begin{corollary}
The policymaker's optimal band problem is the \emph{optimal posterior
belief problem}:%
\begin{equation*}
\max_{T\in \mathcal{T}}\left[ \max_{c_{T}\in \mathcal{C}_{T}}W(c_{T})\right]
=\max_{\mu \in \Pi ^{\ast }}W(\hat{c}_{\mu }).
\end{equation*}
\end{corollary}

\subsubsection{An example}

In this subsection, we illustrate how the optimal wage band problem can be
implemented in practice. Although the optimal wage band problem, the optimal
threshold problem, and the optimal posterior belief problem are all
isomorphic, the threshold formulation is the most analytically tractable,
and it is the approach we adopt in what follows.
% In this example, we show how to put the optimal threshold problem into work.

A worker's ability $z$ follows a uniform distribution on $Z=[0,3]$. While
our theory consider heterogeneity on both sides of the market, for
simplicity, let us assume that firms are all homogeneous, $X=\{1\}$. We use
quasilinear profit and utility functions. The production function is $%
v(z)=2z+1$ and hence the firm's profit function becomes
\begin{equation*}
g(t,x,s,z)=2z+1-t.
\end{equation*}%
The worker's utility function is
\begin{equation*}
u(t,s,z)=t-1-c(s,z),
\end{equation*}%
where
\begin{equation*}
c(s,z)=\left\{
\begin{array}{cc}
0 & \text{if }s=0\text{ and }z=0 \\
\infty & \text{if }s>0\text{ and }z=0 \\
s^{2}/2z & \text{otherwise}%
\end{array}%
\right. .
\end{equation*}%
We normalize the lowest possible wage to $\underline{t}=1$ to ensure that
the utility of consumption $t-1\geq 0$ for all $t\geq \underline{t}$.

We begin with the group of workers and firms in a separating region. Given
the market wage schedule $\tau (s)$ and the posterior belief $\mu (s)$, the
first-order conditions, \eqref{worker_optimal_e} and %
\eqref{receiver_optimal_e} from the worker's and firm's optimization
problems are expressed by
\begin{equation*}
2\mu ^{\prime }(s)-\tau ^{\prime }\left( s\right) =0\text{ and }\tau
^{\prime }\left( s\right) -\frac{s}{\mu (s)}=0,
\end{equation*}%
respectively. These conditions yield the following first-order differential
equation for $\mu (s)$:
\begin{equation*}
\mu ^{\prime }(s)=\frac{s}{2\mu (s)},
\end{equation*}%
with the initial condition $s=s_{\ell }.$ The unique solution for the
differential equation is
\begin{equation*}
\mu (s)=\sqrt{\frac{s^{2}}{2}-\frac{s_{\ell }^{2}}{2}+z_{\ell }^{2}}.
\end{equation*}%
The worker's equilibrium choice of education is the inverse of $\mu (s)$,
given by $\sigma \left( z\right) =\sqrt{2z^{2}-2z_{\ell }^{2}+s_{\ell }^{2}}$%
. At the bottom match, the firm and the worker receive their reservation
profit and reservation utility respectively ((\ref{lem1e}) and (\ref{lem2e}%
)).\todo{\small Q. (8) and (9) are not in
Appendix B?} Given $z_{\ell }$, this leads to
\begin{equation*}
\left( t_{\ell },s_{\ell }\right) =\left( 2z_{\ell }+1,\sqrt{2z_{\ell
}^{2}+z_{\ell }}\right)
\end{equation*}%
\todo{Please check: $s_{\ell}=2z_{\ell}$?} Further, the market wage function
is $\tau \left( s\right) =2\mu \left( s\right) +1$.

We next consider the group in a pooling region on the top. This occurs when
the maximum wage $t_{h}$ is sufficiently low to induce workers above $z_{h}$
to pool their education levels at a common level $s_{h}.$ Given $z_{h}<3$, $%
t_{h}$ and $s_{h}$ are determined by the following indifference conditions
(i.e., \eqref{jumping_seller_e} and \eqref{jumping_buyer_e}):
\begin{align*}
t_{h}-1-\frac{s_{h}^{2}}{2z_{h}}& =\tau \left( \sigma \left( z_{h}\right)
\right) -1-\frac{\sigma \left( z_{h}\right) ^{2}}{2z_{h}}, \\
\mathbb{E}_{\mu (s)}[2z^{\prime }|z^{\prime }\geq z_{h}]+1-t_{h}&
=2z_{h}+1-\tau \left( \sigma \left( z_{h}\right) \right) .
\end{align*}%
In particular, the pooled education level for the pooling region is given by
$s_{h}=\sqrt{6z_{h}+z_{\ell }}$.

Recall that if firms are as well as workers, the utility for the worst
worker type $\underline{z}$ and the (expected) profit for the worst firm
type $\underline{x}$ are both equal to zero in the pooling equilibrium with
the single wage $t^{\ast }=\underline{t}$ where all workers and firms enter
the market, as shown in Theorem \ref{thm:pooling}. However, if firms are
homogeneous, that is not the case. Nonetheless, it is easy to show that we
can convert the optimal wage band problem to the optimal threshold problem.

Therefore, we focus on the policymaker's optimal threshold problem: The
policymaker maximizes the social welfare $W(\hat{c}_{Z})$ over $Z\in
\mathcal{Z}^{\ast }$. With $\omega =0.5$, the optimal threshold problem
becomes
\begin{equation}
\max_{Z\in \mathcal{Z}^{\ast }}W(\hat{c}_{Z}):=\frac{1}{2}\int_{z_{\ell
}}^{z_{h}}\left( 2z-\frac{\sigma \left( z\right) ^{2}}{2z}\right) \frac{1}{3}%
dz+\frac{1}{2}\int_{z_{h}}^{3}\left( \mathbb{E}[2z^{\prime }|z^{\prime }\geq
z_{h}]-\frac{s_{h}^{2}}{2z_{h}}\right) \frac{1}{3}dz.
\label{eq:obj function. simple example}
\end{equation}%
Note that the first term of \eqref{eq:obj function. simple example}
represents the total surplus in the separating region, and the second term
corresponds to the surplus in the pooling region on the top.

The optimal solution is $(z_{\ell},z_h)=(0,0)$, i.e.\ the pooled equilibrium.

\begin{comment}
The optimal solution is $(z_{\ell },z_{h})=(0,3)$. This outcome suggests
that the optimal policy in this example involves no binding wage band, i.e.\
neither a minimum nor a maximum wage is imposed. In contrast, the optimal
outcome changes significantly when firms, as well as workers, are
heterogenous as shown in the next section. The change arises because intense
competition among workers to match with a better firm amplifies
overinvestment in education. This creates a room for a binding wage band.
\end{comment}
% This solution is incorrect.

\begin{comment}
While there is a closed form solution for the market belief function, $\mu $
and the market wage function $\tau $, the optimal threshold problem does not
yield an analytical solution even in this simple example. Generally, we need
to conduct numerical analysis to derive a solution.
\end{comment}
% Repetition. We remark this at the beginning of the next section.

\section{Comparative Statics and Policy Analysis\label{secnumerical_analysis}%
}

In this section, we examine the effects of the optimal wage band policy and
its welfare implications. Recall the environment of the optimal wage band
problem described in Section \ref{section: preliminaries}. The worker's
utility function and the firm's profit function are represented in Equations
\eqref{eq: work's
utility}--\eqref{eq: firm_production}.

We normalize the lowest possible subsistence level of reaction to $%
\underline{t}=1$. Let $\underline{z}=0$. Recall that $n(z)$ denotes the firm
productivity at the same percentile as the worker with ability $z$, i.e., $H(n(z))=G(z)$. We assume that the following functional form
of $n(z)$:
\begin{align}
n(z)=kz^{q},
\end{align}%
where $k>0$ represents the scale parameter and $q\geq 0$ the relative
spacing parameter. The relative spacing parameter $q$ represents the
heterogeneity of firm productivity relative to worker ability.

Substituting into the first-order nonlinear ODE system in (\ref%
{system_diff_eq_e}), we obtain:
\begin{align}
\tau ^{\prime }(s)& =\frac{\beta b s^{b-1}\tau (s)^{\rho }}{\mu (s)}
\label{tau_prime} \\
\mu ^{\prime }(s)& =\frac{\beta b s^{b-1}\tau (s)^{\rho }-akAs^{a-1}\mu
(s)^{q+2}}{ kA\left( s^{a}+1\right) \mu (s)^{q+1}}  \label{mu_prime}
\end{align}%
with initial conditions $\tau (s_{\ell })=t_{\ell }$ and $\mu (s_{\ell
})=z_{\ell }.$ Analytical closed-form solutions for coupled nonlinear
systems are generally intractable, and our system is no exception. Instead,
we solve the system numerically over a wide range of parameter values.

\subsection{Baseline Model}

We first specify a baseline design and solve the optimal wage band problem in this
subsection. Comparative statics with respect to these parameter values are
presented in the following subsection.

The worker ability distribution $G$ follows a uniform distribution on
$[0,3]$.
%As shown in Figure \ref{figure: density beta(5,5)} in the appendix, the Beta(5,5) distribution is symmetric and bell-shaped.
Table \ref{table:parameters} summarizes the parameter values used in the baseline
model. In this baseline specification, education contributes positively to
production growth ($a=0.5$) and imposes quadratic disutility ($b=2$) on workers. Firms and
workers have the same degree of heterogeneity ($q=1$), and workers have
linear utility functions ($\rho=0$).

\begin{table}[bp]
\caption{Baseline Parameter Values}
\label{table:parameters}\centering
\begin{tabular}{ccc}
\toprule {Parameters} & \centering{Names} & {Values} \\
\midrule $a $ & \multicolumn{1}{l}{Education Contribution} & 0.5 \\
$b $ & \multicolumn{1}{l}{Education Cost} & 2 \\
$q $ & \multicolumn{1}{l}{Firm Heterogeneity (Relative Scale)} & 1 \\
$\rho $ & \multicolumn{1}{l}{Elasticity of Marginal Utility} & 0 \\
$\beta $ & \multicolumn{1}{l}{Cost Coefficient} & 0.5 \\
$A $ & \multicolumn{1}{l}{Technology} & 1 \\
$k $ & \multicolumn{1}{l}{Matching Scale} & 1 \\
\bottomrule &  &
\end{tabular}%
\end{table}

\begin{figure}[tbp]
    \centering
    \caption{Surplus Possibility and Frontier}
    \label{figure: surplus possiblity}
    \includegraphics[scale=0.4]{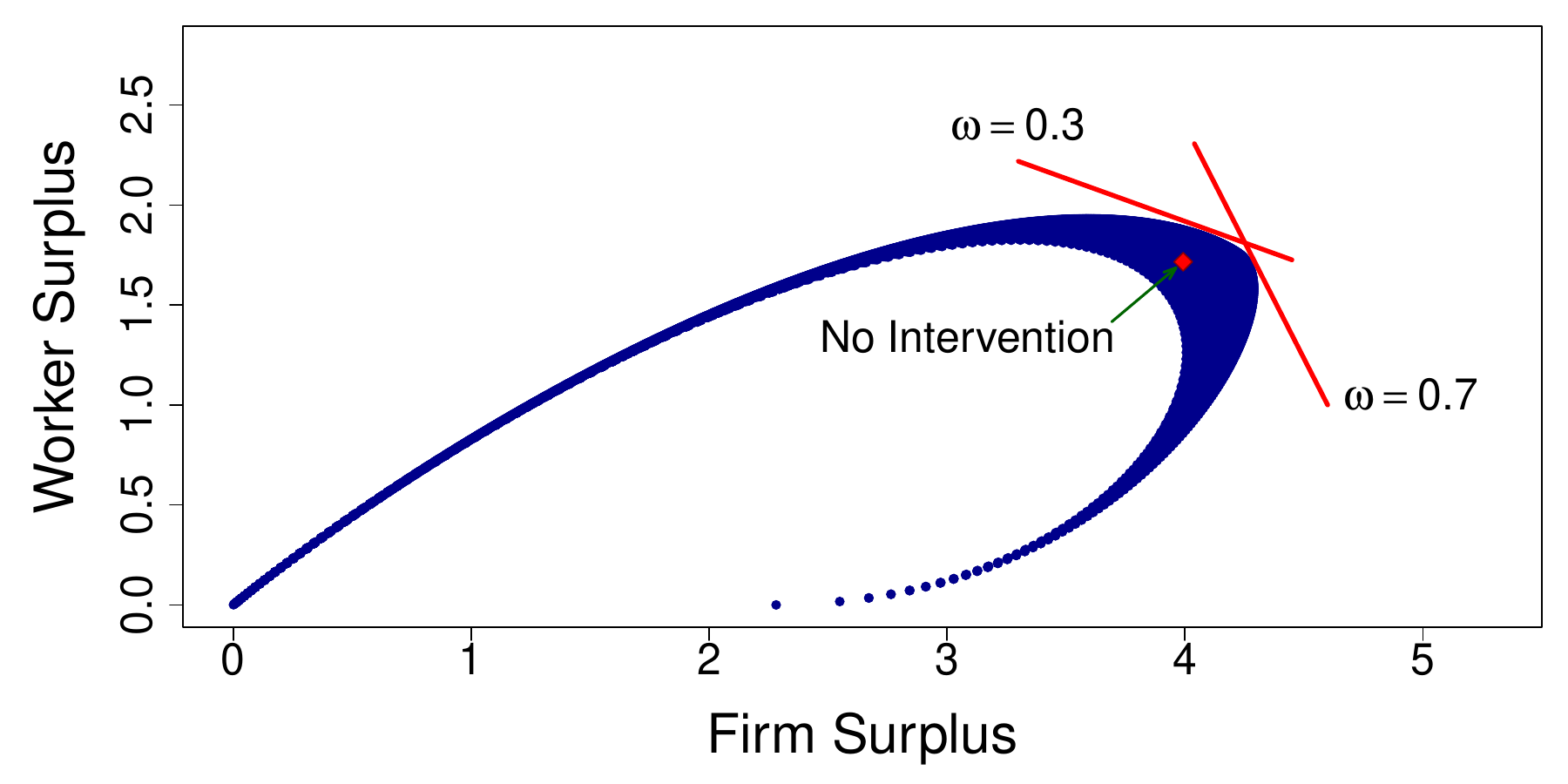}
    \captionsetup{font=small}
    \caption*{\textit{Notes}: Each blue point represents the equilibrium pair of
        worker surplus and firm surplus
    given different values of $(z_{\ell},z_h)$.}
\end{figure}

Figure \ref{figure: surplus possiblity} illustrates the surplus possibility
set and optimal outcomes given two different values of weight $\omega$. The
surplus possibility set represents equilibrium pairs of worker surplus and
firm surplus across different values of $(z_{\ell}, z_h)$. The efficient
frontier is located in the upper-right region, and we can find the optimal
policy outcome for any fixed weight $\omega$.\footnote{%
In Proposition~\ref{prop_bounded_design}, we showed that finding the optimal
wage band is equivalent to finding the optimal ability band.} In the figure,
we depict two optimal outcomes for $\omega=0.3$ and $\omega=0.7$. The surplus frontier in
Figure \ref{figure: surplus possiblity} is strictly convex. As a result, the
optimal solution moves smoothly as the weight value varies. Notice that the no-intervention outcome lies strictly
within the possibility set. Therefore, the policymaker can improve welfare by
choosing an optimal wage band.

\begin{figure}[thp]
    \centering
    \caption{Optimal Ability and Wage Bands}
    \label{figure: optimal bands}
    \begin{minipage}[b]{0.48\textwidth}
        \centering
        \includegraphics[width=\textwidth]{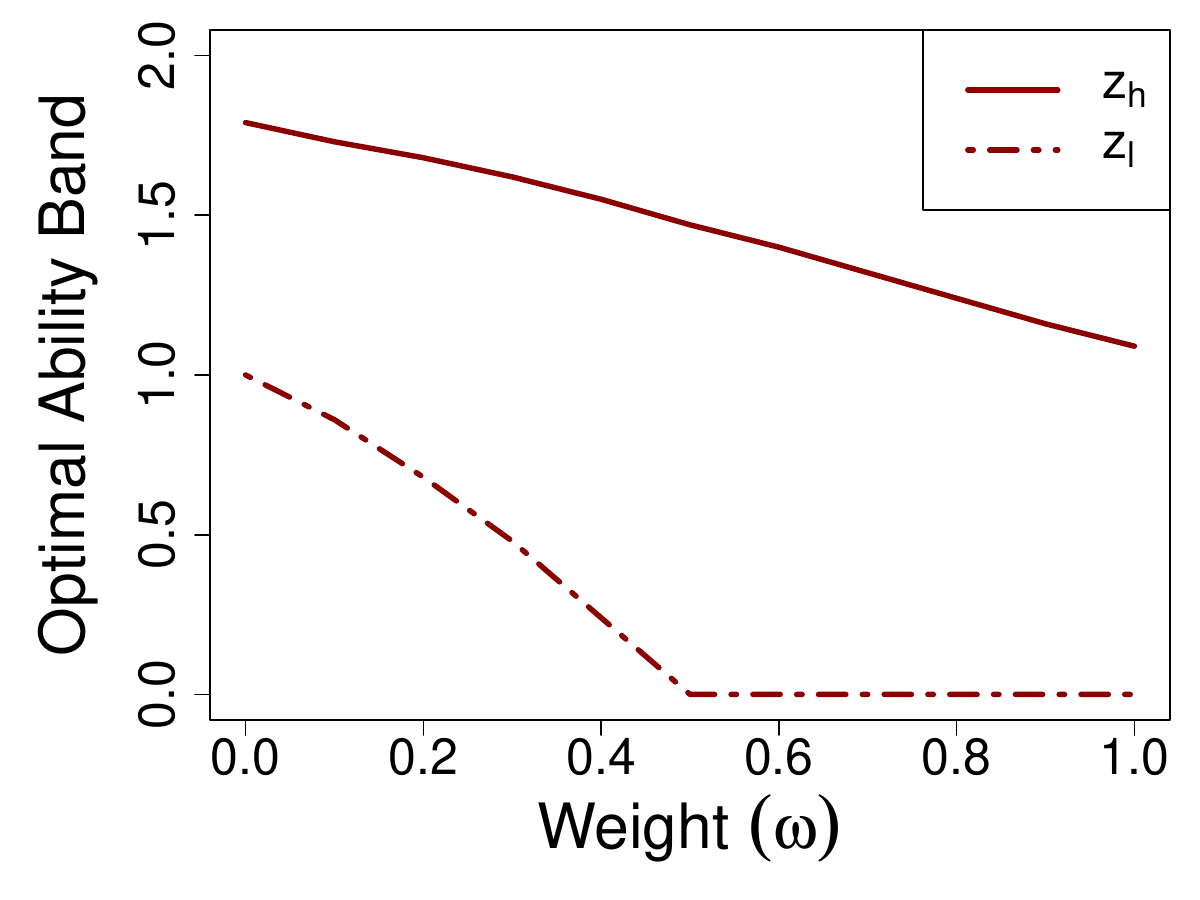}
    \end{minipage}
    \hfill
    \begin{minipage}[b]{0.48\textwidth}
        \centering
        \includegraphics[width=\textwidth]{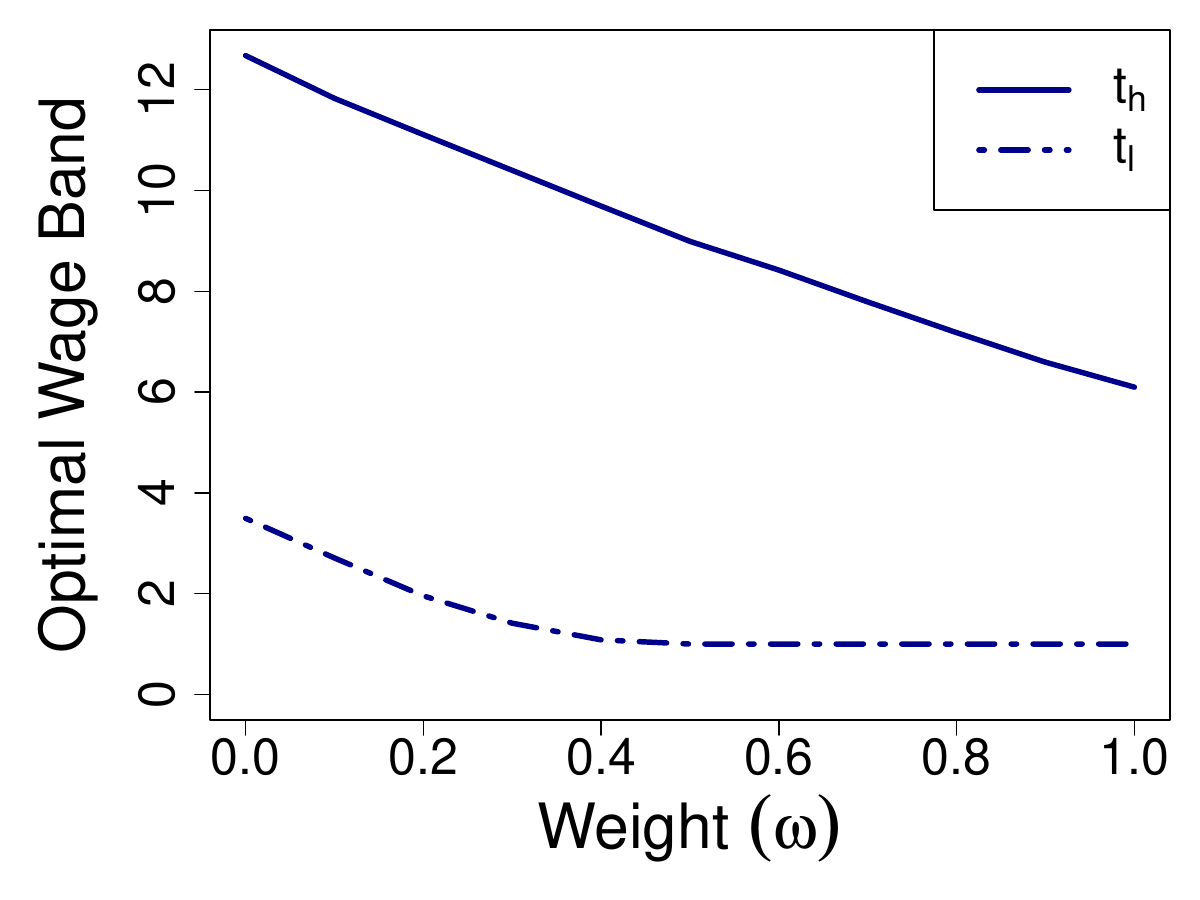}
    \end{minipage}
\end{figure}

Figure \ref{figure: optimal bands} presents the optimal ability bands and
the optimal wage bands across different values of the weight parameter $\omega$. Both the lower bound and the upper bound increase as the
policymaker places greater weight on workers (i.e., $\omega$ decreases). When the policy
maker places more weight on firms, the optimal minimum wage is at the
corner solution, i.e., the subsistence (no-intervention) level $\underline{t}=1$. However, as the
weight shifts toward workers and falls below $0.5$, the optimal minimum wage
becomes strictly higher than the subsistence  level.

\begin{figure}[tb]
    \centering
    \caption{Education Choice and Worker Utility Functions}
    \label{figure: mechanism}
    \begin{minipage}[b]{0.48\textwidth}
        \centering
        \includegraphics[width=\textwidth]{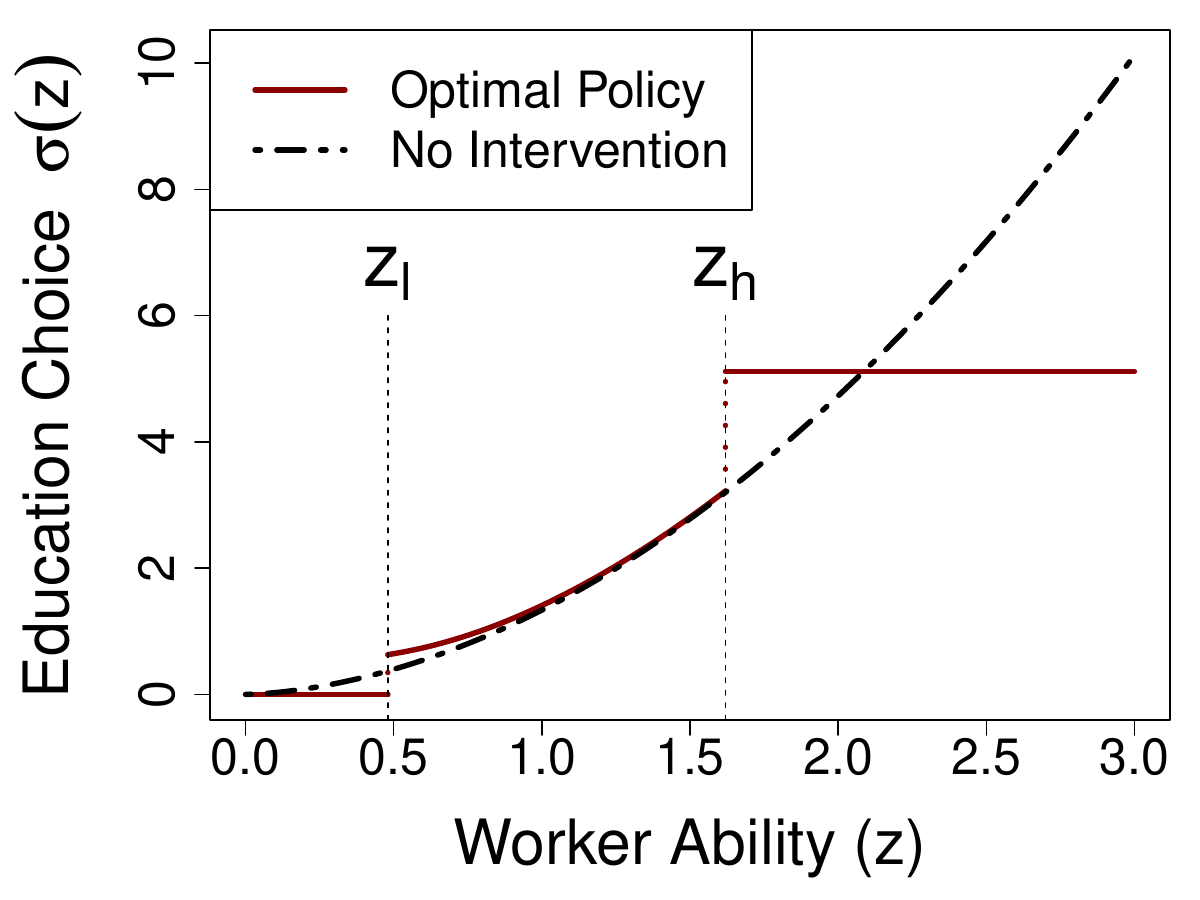}
    \end{minipage}
    \hfill
    \begin{minipage}[b]{0.48\textwidth}
        \centering
        \includegraphics[width=\textwidth]{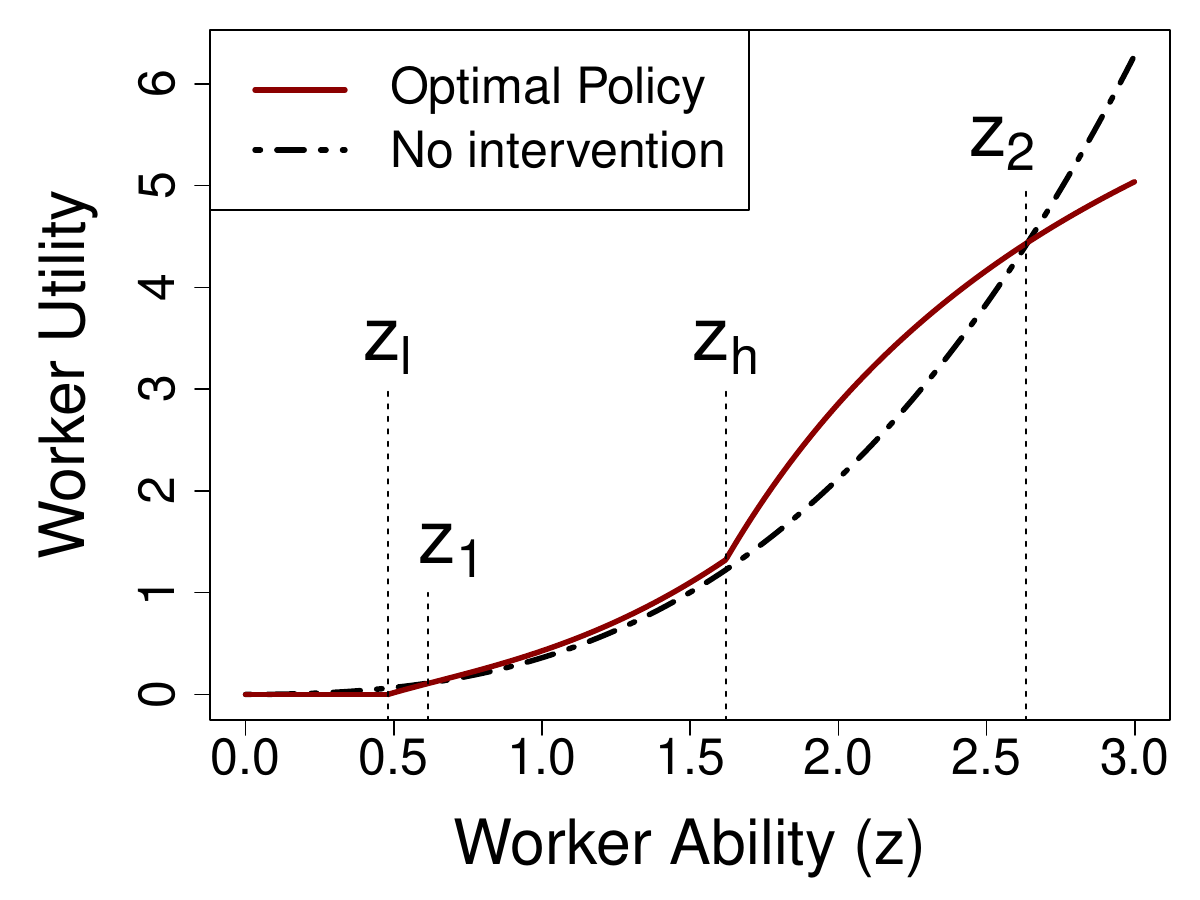}
    \end{minipage}
\end{figure}

The mechanism behind the welfare gains is an interesting feature of the model
and deserves careful investigation. Figure \ref{figure: mechanism}
presents education choice functions and worker utility levels under the optimal
policy and the no-intervention case. We set the relative weight at $\omega=0.3$
hereafter, which allows us to examine both minimum and maximum wages simultaneously.
The separating equilibrium in the no-intervention case yields smooth
increasing functions that reflect assortative matching. In contrast, the
well-behaved equilibrium with the optimal band generates two discontinuous
points for the education function and two kinked points for the utility
function. They are caused by the two pooling regions separated by the lower
and upper ability thresholds. Notice that low-skilled workers below the
lower threshold choose zero education and remain unemployed, while high-skilled
workers above the upper threshold also form another pooled group by choosing a
uniform level of education and receive the same wage. Therefore, the different
utility levels among high-skilled workers within the upper pooling region
($z \in [z_h,\overline{z}]$) in the right panel
reflect only the heterogeneity in their education costs.

Although both the minimum wage and maximum wage create pooling regions, the
mechanisms behind their welfare gains are quite different.  First, the upper threshold triggers
the pooling and reduce `over-investment' in education caused by
incomplete information. When we compare the worker utility curves under the
optimal policy and the no-intervention case in Figure
\ref{figure: mechanism}, random matching in the (upper) pooling region ($z \in
[z_h,\overline{z}]$) leads to net utility gains for workers whose abilities are relatively
lower ($z \in [z_h, z_2]$).
Meanwhile, top-skilled workers in $z \in [z_2, \overline{z}]$  experience a net utility loss relative to the
separating equilibrium in the no-intervention case.
Therefore, the optimal threshold level $z_h$ is determined by the relative utility gains
or losses of these two groups of workers.
Basically, a maximum wage creates a trade-off in terms of social welfare between
the inefficiency associated with random matching and cost savings associated
with the pooled education choice. When the latter is bigger, it is optimal to
impose a maximum wage.\footnote{For clarity of illustration, we focus only
    on worker utility here. As we show below, similar arguments apply to
    the firm profit side.}

Second, the lower threshold (i.e., the
minimum wage) achieves welfare gains by encouraging greater education
investments among mid-range workers, resulting in increased labor productivity.
A worker just above the
threshold $z_{\ell}$ takes a higher level of education than in the no-intervention
case as shown in the left panel of Figure \ref{figure: mechanism}. Furthermore, the separating property in the middle-ability
region generates a ripple effect that induces additional education increase for all the workers in
this range, which in turn increases wages for those workers.
This ripple effect extends to all workers up to the
threshold level $z_h$.
Unlike in the upper pooling region, a loss in equilibrium utility occurs for workers around the
threshold $z_{\ell}$, including unemployed workers and those whose
additional education investment exceeds the associated wage increase (i.e., $z
\in [\underline{z},z_1]$). Meanwhile, workers with higher ability levels in the separating region (i.e.,
$z \in [z_1, z_h]$) experience net utility gains.

The ripple effect is well-established
in the empirical literature \citep[e.g.,][]{phelan2019hedonic,
engbom2022earnings}, but it is limited to the direct impact of the minimum wage
on the wages of remaining workers. Our model reveals a novel mechanism through which the wage ripple effect can emerge via additional education investments in a competitive market with asymmetric
information. It also provides a detailed mechanism of the distributional
consequences in terms of utility gains and losses across heterogeneous workers.

\citet{lee2012optimal} also shows efficiency gains from minimum wages in a perfectly
competitive labor market without education investments. In their model, the
crucial mechanisms are: (i) efficient rationing, i.e., workers who lose a job
due to the minimum wage are
those with the least surplus from working; and (ii) a policymaker placing
greater weight on low-skilled workers. \citet{berger2025minimum} shows that in a general
equilibrium model of oligopsonistic markets, the redistributive effect of a
minimum wage--benefiting low-income workers--exceeds its efficiency gain from limiting firms’ market power.

In contrast to \citet{lee2012optimal} and \citet{berger2025minimum}, our analysis demonstrates that
the utility ripple effect of the minimum wage leads to
different implications when education signals worker ability: (employed) low-ability workers may be worse off, even though the total
social welfare increases.

\begin{figure}[tb]
    \centering
    \caption{Wage and Firm Profit Functions}
    \label{figure: mechanism firm side}
    \begin{minipage}[t]{0.48\textwidth}
        \centering
        \includegraphics[width=\linewidth]{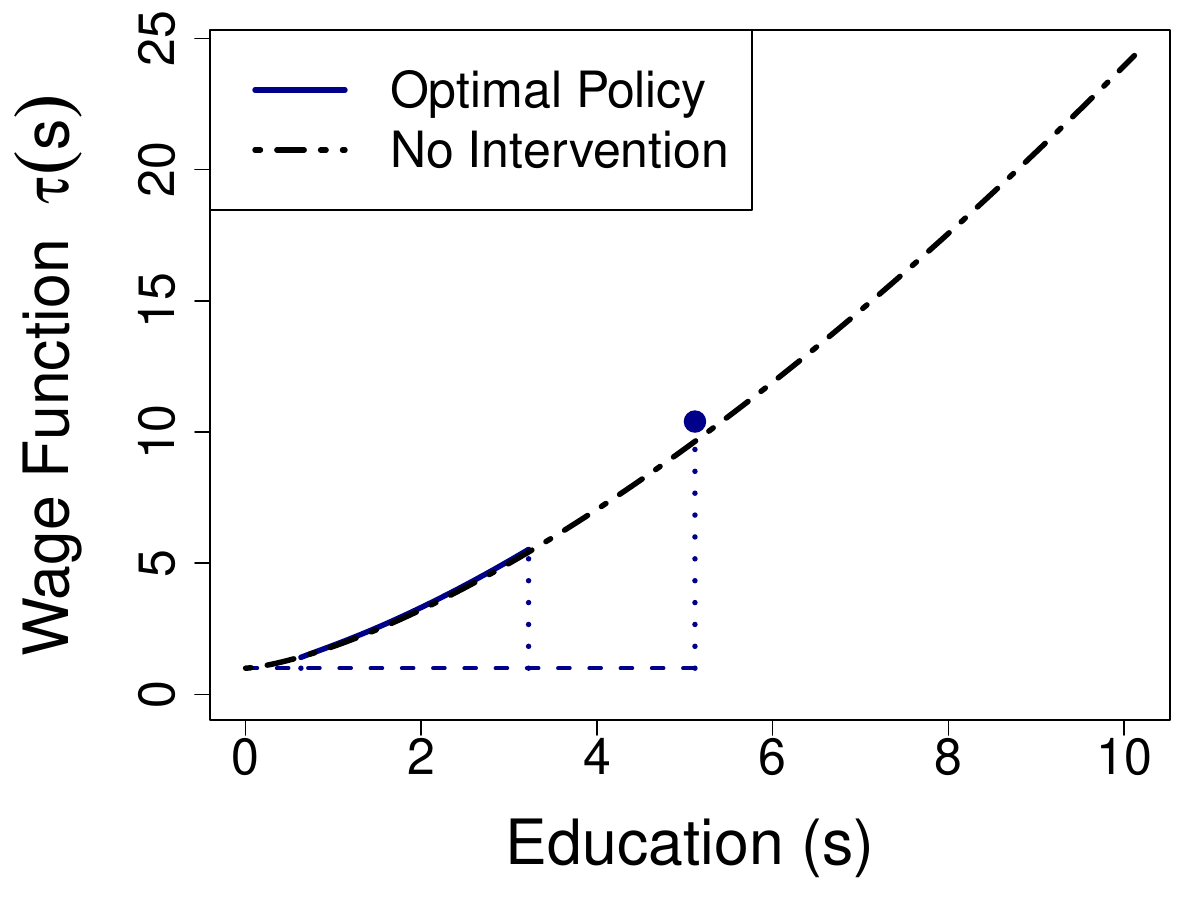}
    \end{minipage}
    \hfill
    \begin{minipage}[t]{0.48\textwidth}
        \centering
        \includegraphics[width=\linewidth]{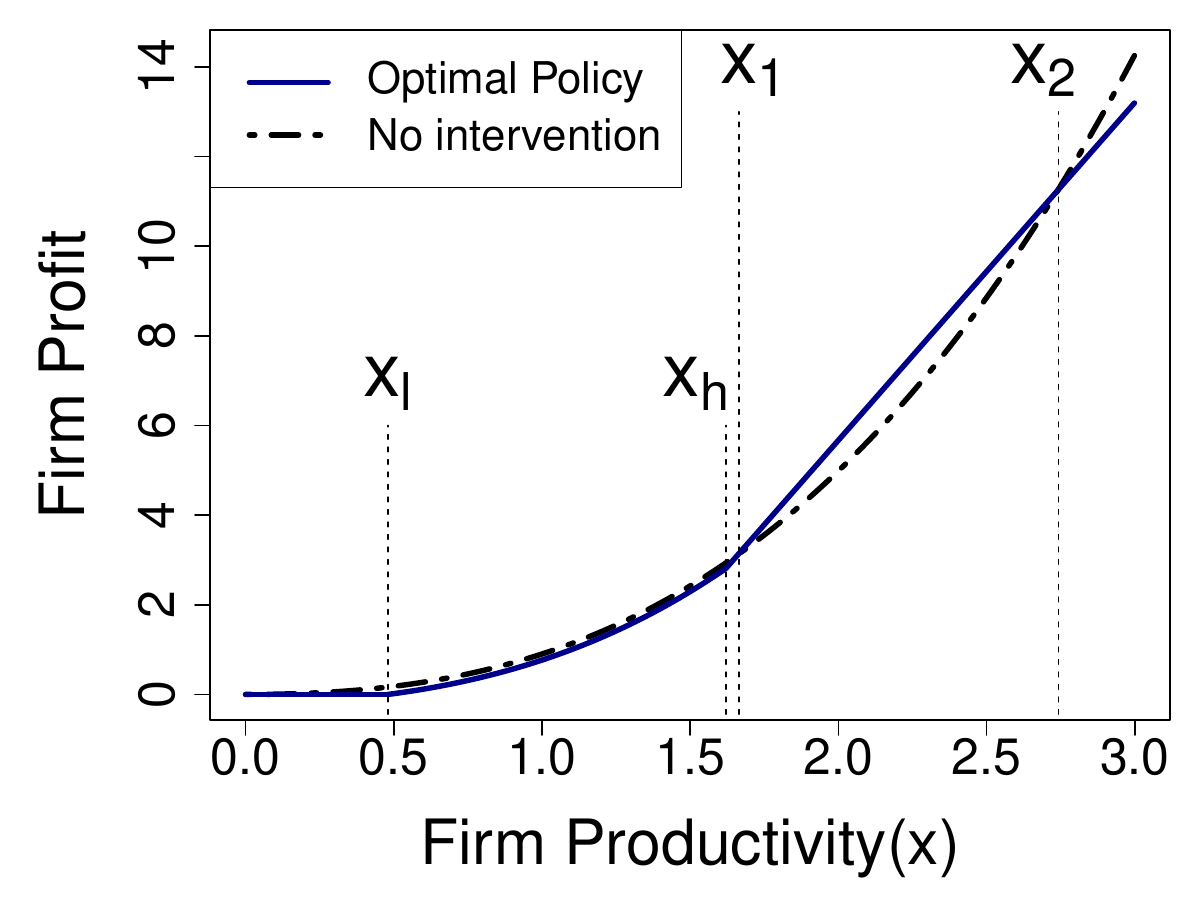}
    \end{minipage}
\end{figure}

Finally, we present the equilibrium wage function and the firm profit function in
Figure \ref{figure: mechanism firm side}. The isolated point on the wage
function represents the common wage level assigned to all workers in the pooling
equilibrium
region. Recall that the optimal wage band maximizes the weighted sum of
worker utility and firm profit. The distributional consequences of the
optimal wage band on the profit function resembles those on the utility function.
Low-productivity firms in $x\in[\underline{x},x_1]$ and high-productivity firms in
$x\in [x_2,\overline{x}]$ experience profit losses relative to the
no-intervention case, while mid-productivity firms in $x \in [x_1, x_2]$ realize
profit gains.

\subsection{Comparative Statics}
In this subsection, we conduct comparative statics with respect to the following parameters:
education contribution ($a$), firm heterogeneity ($q$), elasticity of marginal utility of consumption ($\rho$),
and education cost ($\beta$). All parameter values begin with those specified in
the baseline model. As before, we set the weight at $\omega=0.3$ to
examine the effect on the minimum and maximum wages simultaneously.

\begin{figure}[tbh]
    \centering
    \caption{Comparative Statics: $a$}
    \label{figure: comp statics a}

    \begin{minipage}[t]{0.48\textwidth}
        \centering
        \includegraphics[width=\linewidth]{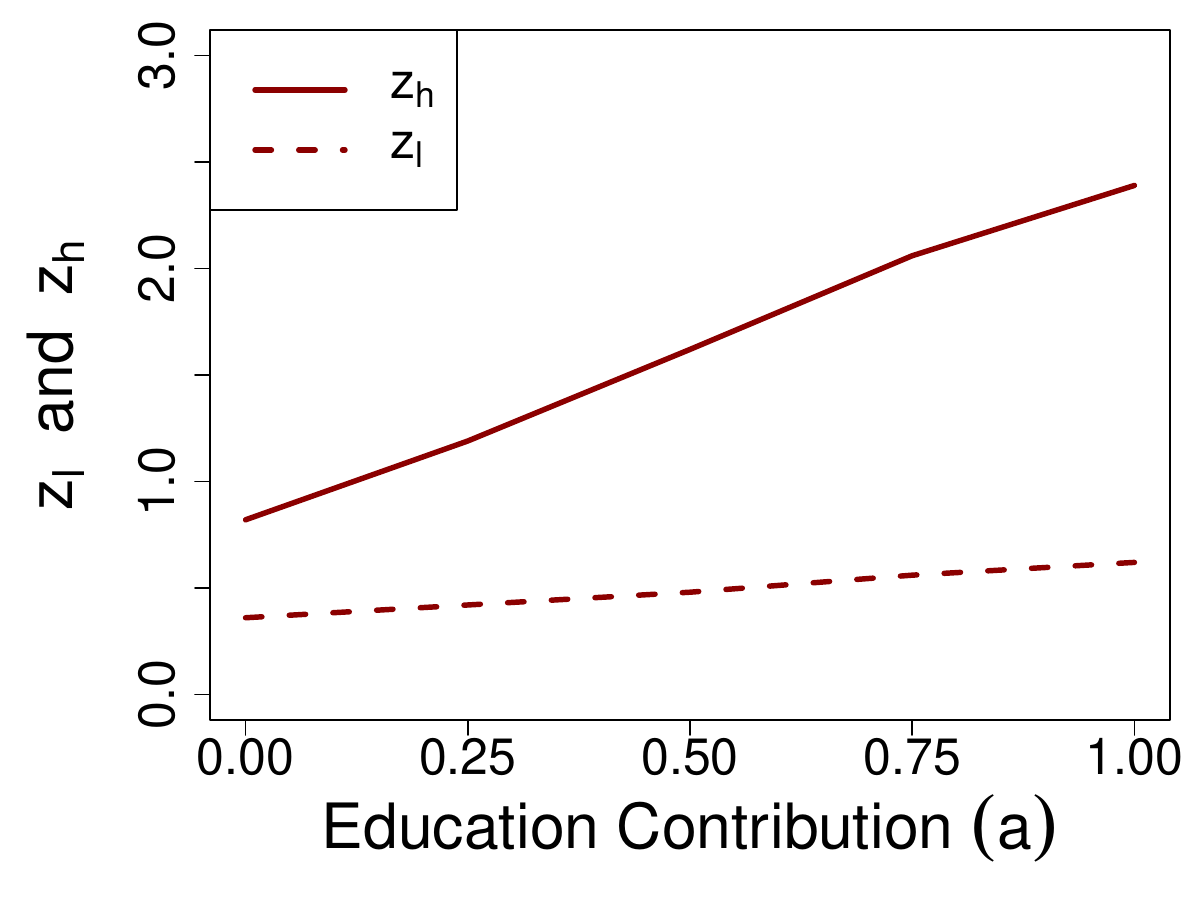}
    \end{minipage}
    \hfill
    \begin{minipage}[t]{0.48\textwidth}
        \centering
        \includegraphics[width=\linewidth]{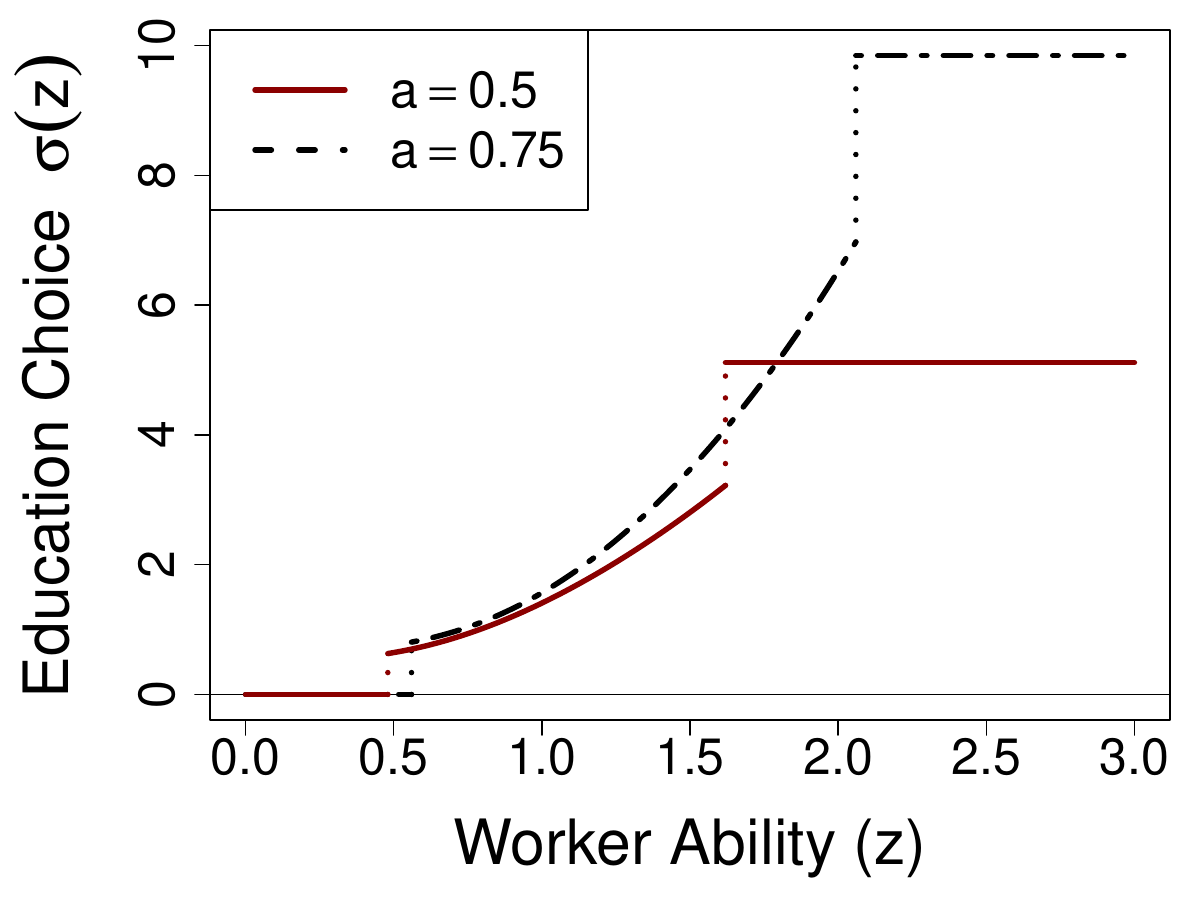}
    \end{minipage}

    \vspace{1em}

    \begin{minipage}[t]{0.48\textwidth}
        \centering
        \includegraphics[width=\linewidth]{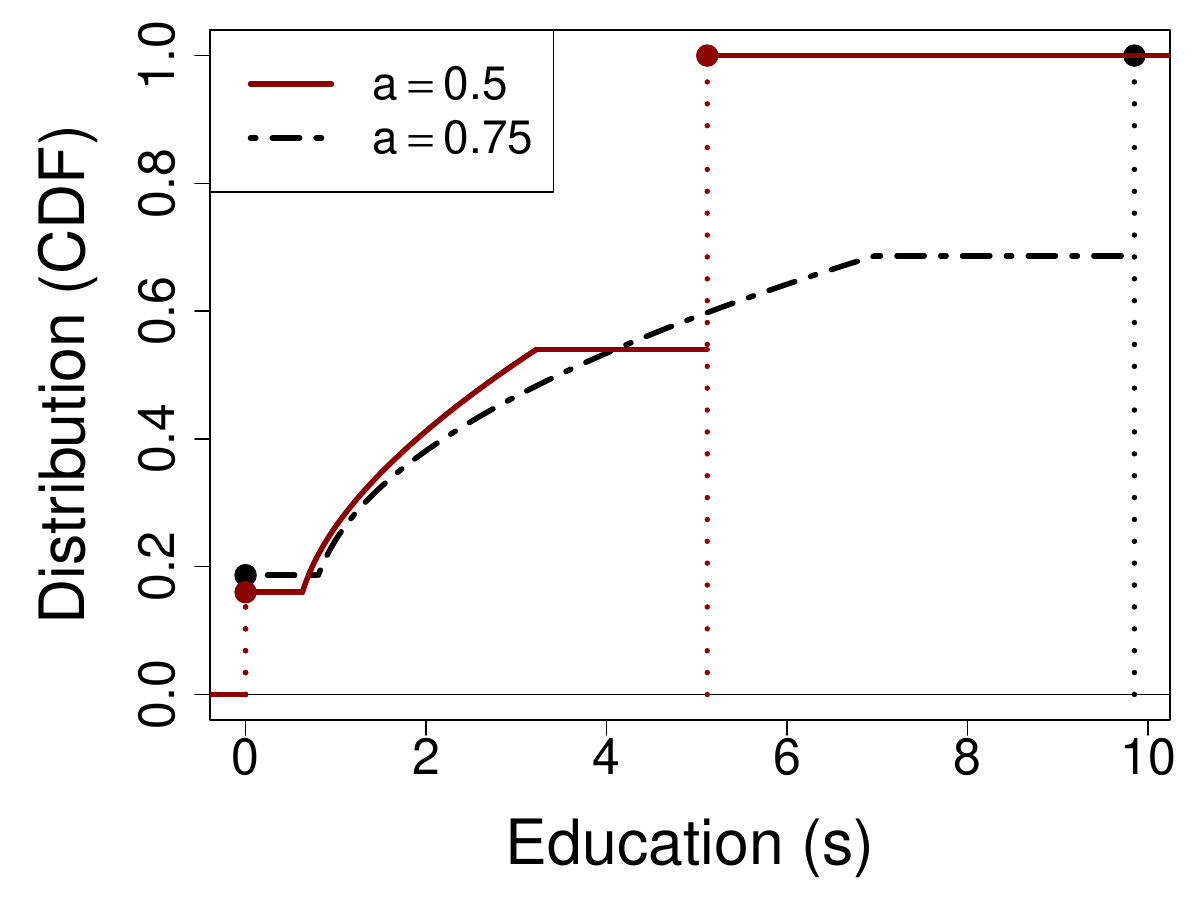}
    \end{minipage}
    \hfill
    \begin{minipage}[t]{0.48\textwidth}
        \centering
        \includegraphics[width=\linewidth]{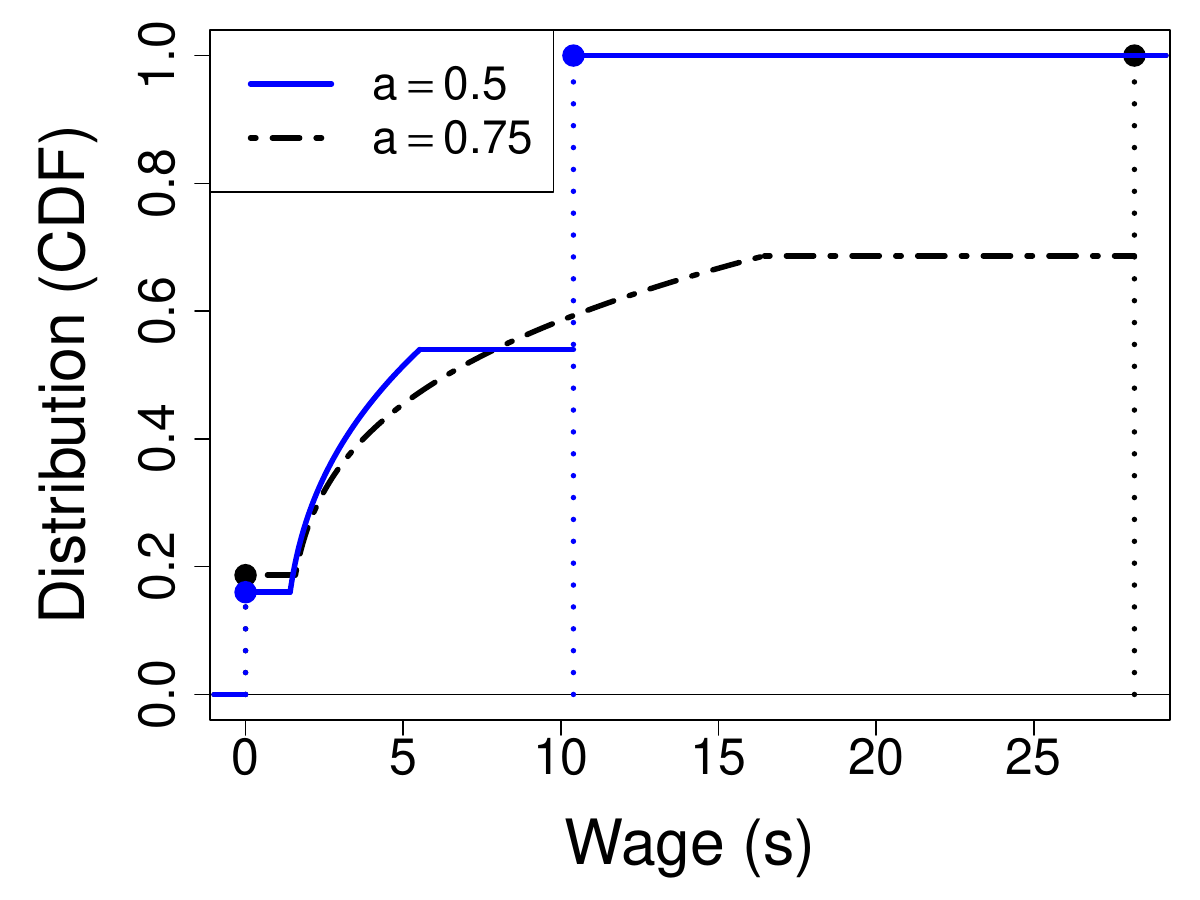}
    \end{minipage}
\end{figure}

\paragraph{Education Contribution ($a$).}
We begin the comparative statics by varying the education contribution parameter
$a$. Notice that a higher value of $a$ increases labor productivity for any
given ability level, resulting in higher output and wages. The upper-left panel
in Figure \ref{figure: comp statics a} illustrates the optimal lower and
upper ability thresholds as the parameter $a$ varies from 0 to 1. First, both thresholds
increase as $a$ becomes larger. Second, the separating region
(i.e., the length of the wage band) becomes narrower as $a$ decreases, reaching
its minimum when $a=0$, i.e., when education functions as a pure signaling device. This implies
that the cost savings from the pooled education investment outweigh the surplus
losses caused by random matching. The remaining three panels in Figure \ref{figure: comp statics a}
illustrate how education choice functions, education distributions, and wage
distributions shift as $a$ increases from 0.5 to 0.75. As $a$ increases, the
separating region of the education choice gets wider. The slope of the
education function becomes steeper as well. In the resulting equilibrium, both the mean and variance of education and wage distributions increase, reflecting broader and more dispersed investments in human capital.

\begin{figure}[tbh]
    \centering
    \caption{Comparative Statics: $q$}
    \label{figure: comp statics q}

    \begin{minipage}[t]{0.48\textwidth}
        \centering
        \includegraphics[width=\linewidth]{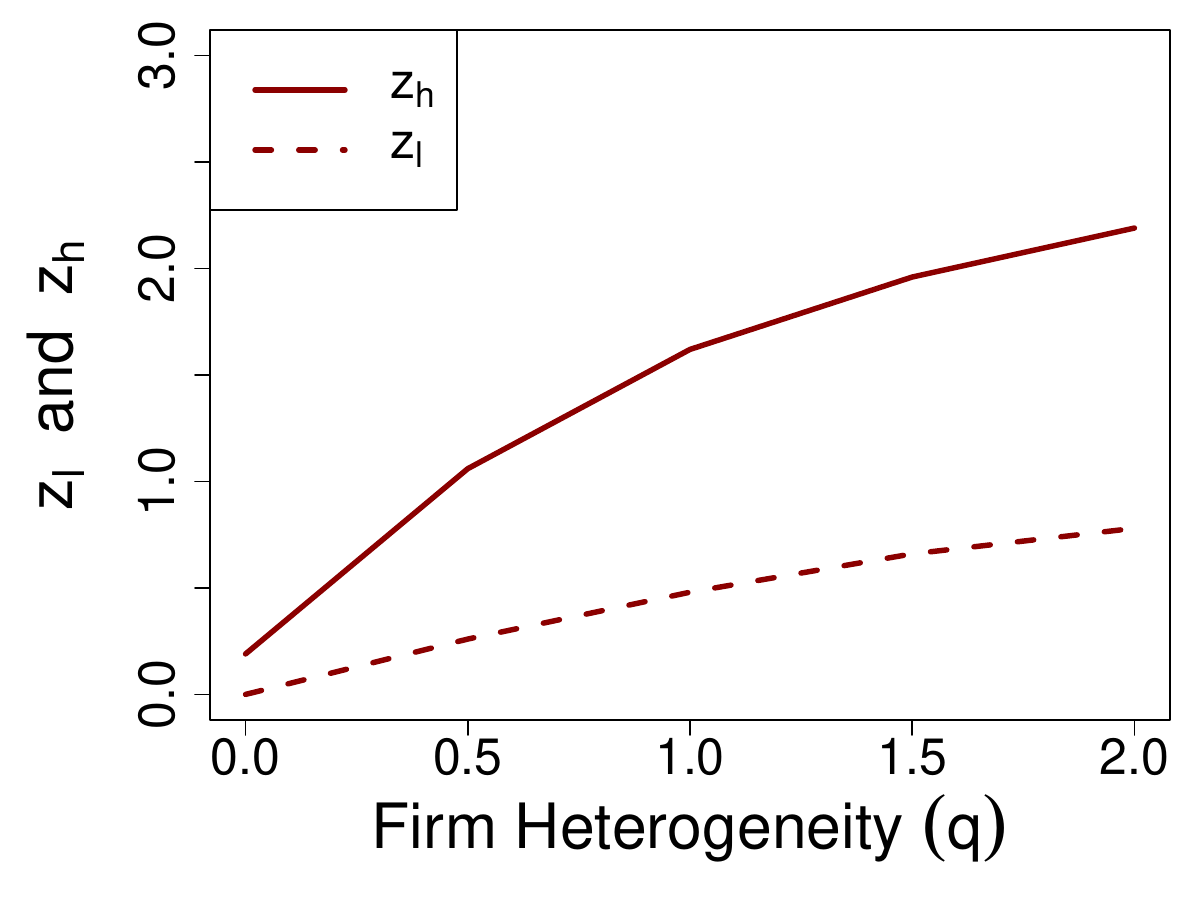}
    \end{minipage}
    \hfill
    \begin{minipage}[t]{0.48\textwidth}
        \centering
        \includegraphics[width=\linewidth]{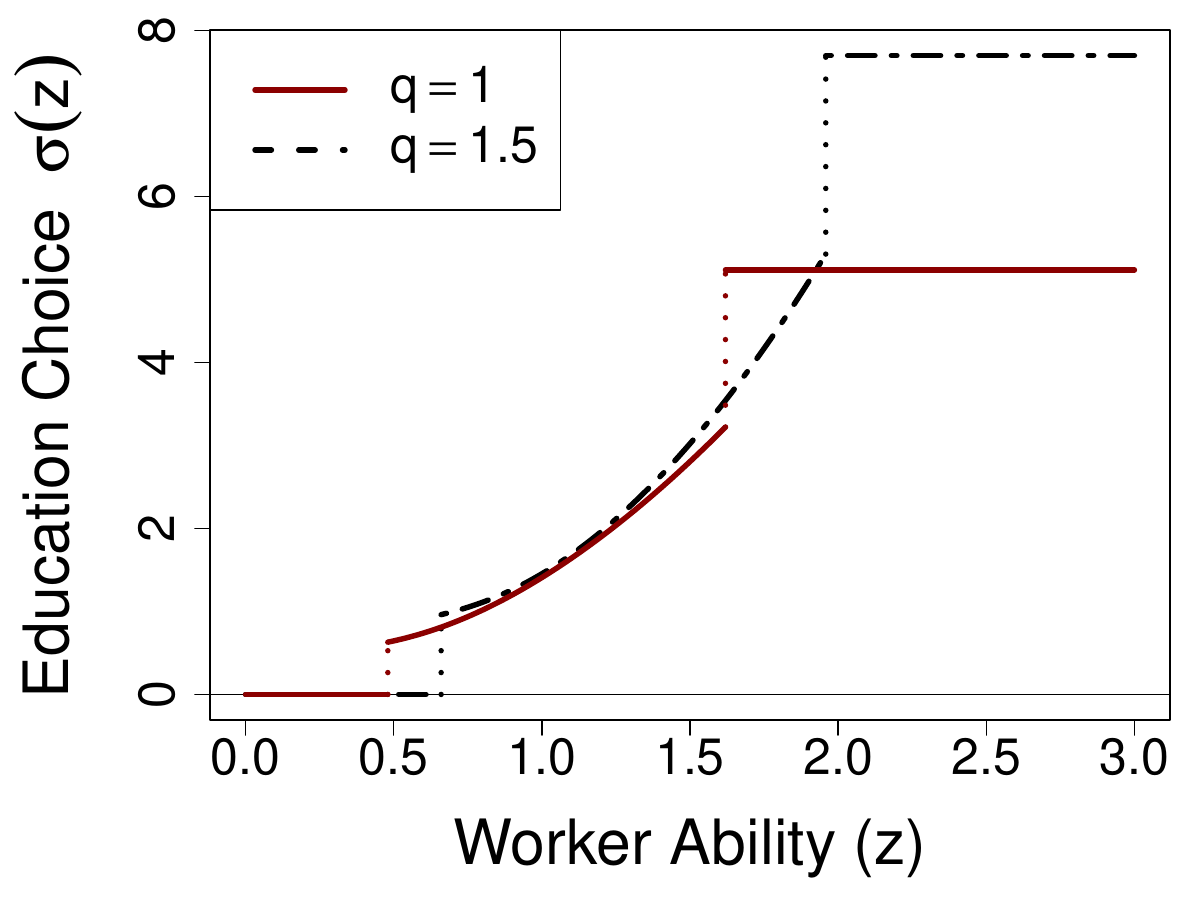}
    \end{minipage}

    \vspace{1em}

    \begin{minipage}[t]{0.48\textwidth}
        \centering
        \includegraphics[width=\linewidth]{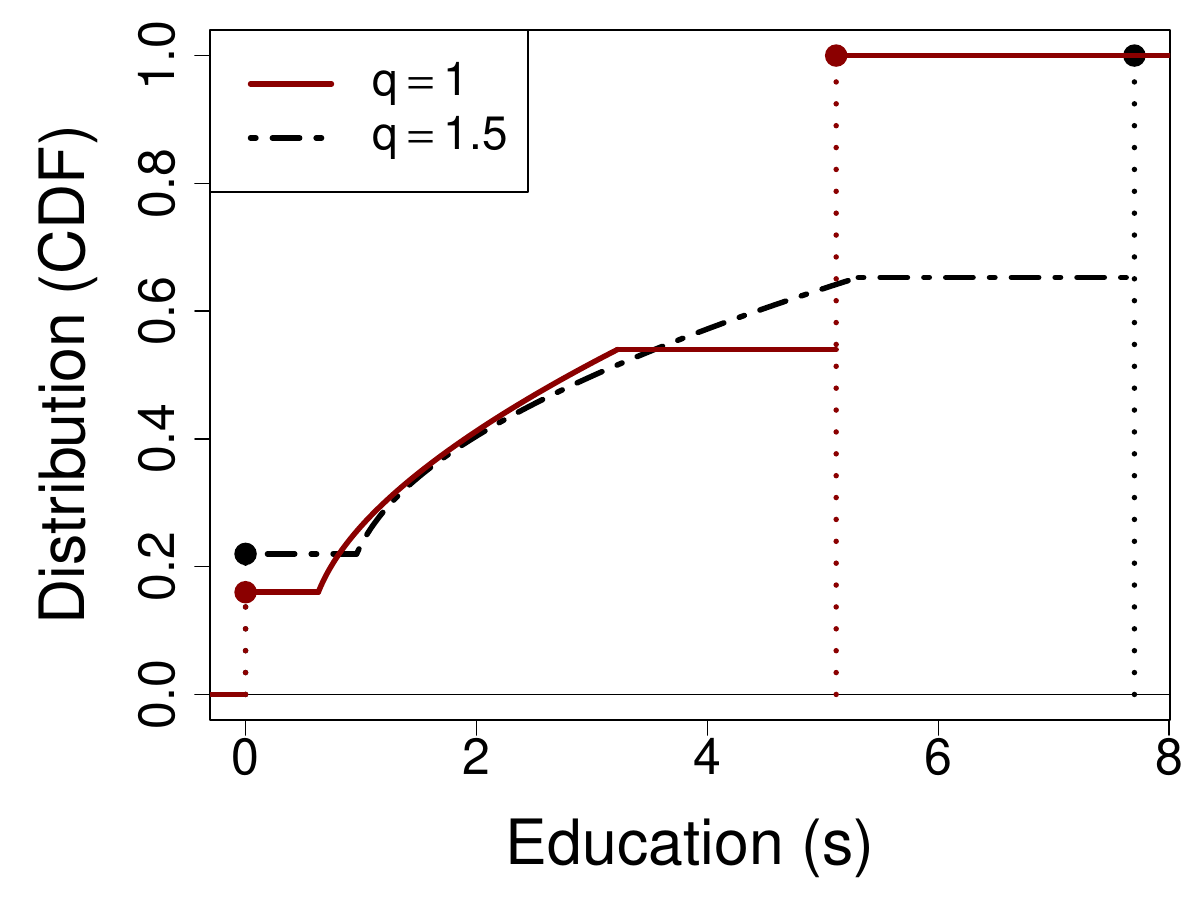}
    \end{minipage}
    \hfill
    \begin{minipage}[t]{0.48\textwidth}
        \centering
        \includegraphics[width=\linewidth]{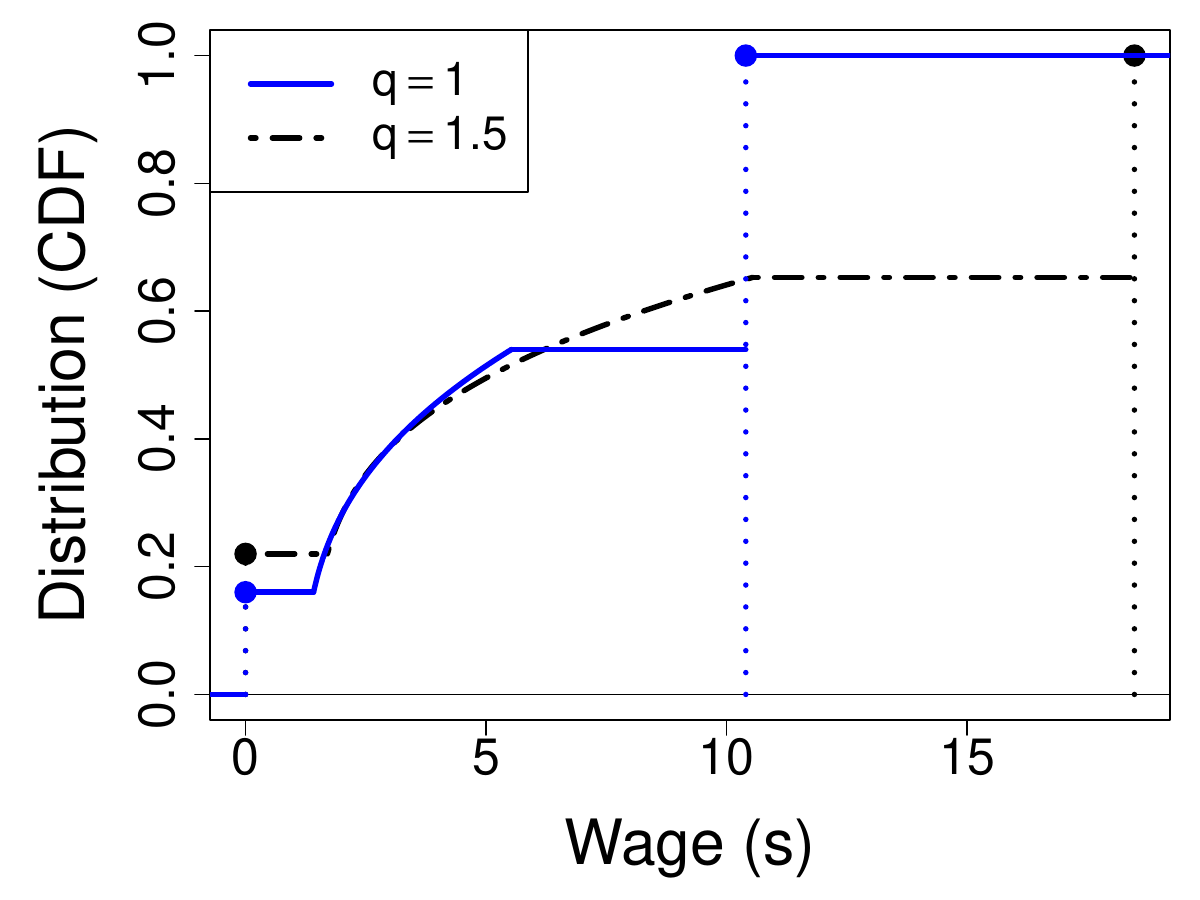}
    \end{minipage}
\end{figure}

\paragraph{Firm Heterogeneity ($q$).}
To understand how variation in the firm productivity distribution affects equilibrium
outcomes, we conduct comparative statics with respect to the firm heterogeneity
parameter $q$. Figure \ref{figure: comp statics q} summarizes equilibrium
outcomes as the parameter $q$ varies from 0 to 2. The lower and upper
ability thresholds exhibit a pattern similar to that observed for the parameter
$a$. Notice that firm productivity is homogeneous when $q=0$ and that
the matching surplus (i.e., production) for any given worker ability increases as $q$ increases.
Therefore, the effect of a larger $q$ is quite similar to that of the education
contribution parameter $a$. The remaining three panels confirm this intuition
showing similar shifts in education choices, education and wage distributions.

Similar to the education contribution parameter $a$, the firm heterogeneity
parameter $q$ has a positive relationship with matching surplus (i.e., firm
production). An increase in firm heterogeneity amplifies the ripple effect of
the minimum wage by raising the equilibrium net utilities of high-ability
workers and their contribution to worker surplus, resulting in a substantial gain in overall surplus.
Therefore, when the weight on worker surplus ($1-\omega$) is high,
such an increase pushes up the minimum wage ($t_{\ell }$)
and the minimum ability threshold ($z_{\ell }$), as shown in the upper-left and bottom-right
panels in Figure \ref{figure: comp statics q}.
We also observe that the upper ability threshold ($z_{h}$) and the maximum wage
($t_{h}$) rise as firm heterogeneity increases. Since
matching surplus grows by higher firm heterogeneity, the inefficiency from random matching
outweigh the cost savings from the pooled education choice.

It is interesting to relate this result with recent empirical findings on the firm
size distribution. \citet{poschke2018firm} shows that both the mean and variance of the firm size distribution are larger in rich countries and have increased over time for U.S. firms.
When the firm size is positively correlated with productivity, increases in the mean and
variance of the firm size distribution can be interpreted as an increase in firm
heterogeneity (i.e., $q$). In this case, the result above
provides a possible channel to explain the long-run decline in the top marginal income tax rate.
In 1942, Franklin D.\ Roosevelt proposed a maximum income of \$25,000 USD, with a 100\%
tax on all income above this level, though the proposal was never enacted. Instead,
the Revenue Act of 1942 introduced an 88\% marginal tax rate on the top income
bracket, along with a 5\% ``Victory Tax'' with post-war credits,
effectively resulting in a temporary top tax rate of 93\%. By the 1980s, the top rate
had fallen to 50\%, and by 2025 it stands at 37\%.

\begin{figure}[tbh]
    \centering
    \caption{Comparative Statics: $\rho$}
    \label{figure: comp statics rho}

    \begin{minipage}[t]{0.48\textwidth}
        \centering
        \includegraphics[width=\linewidth]{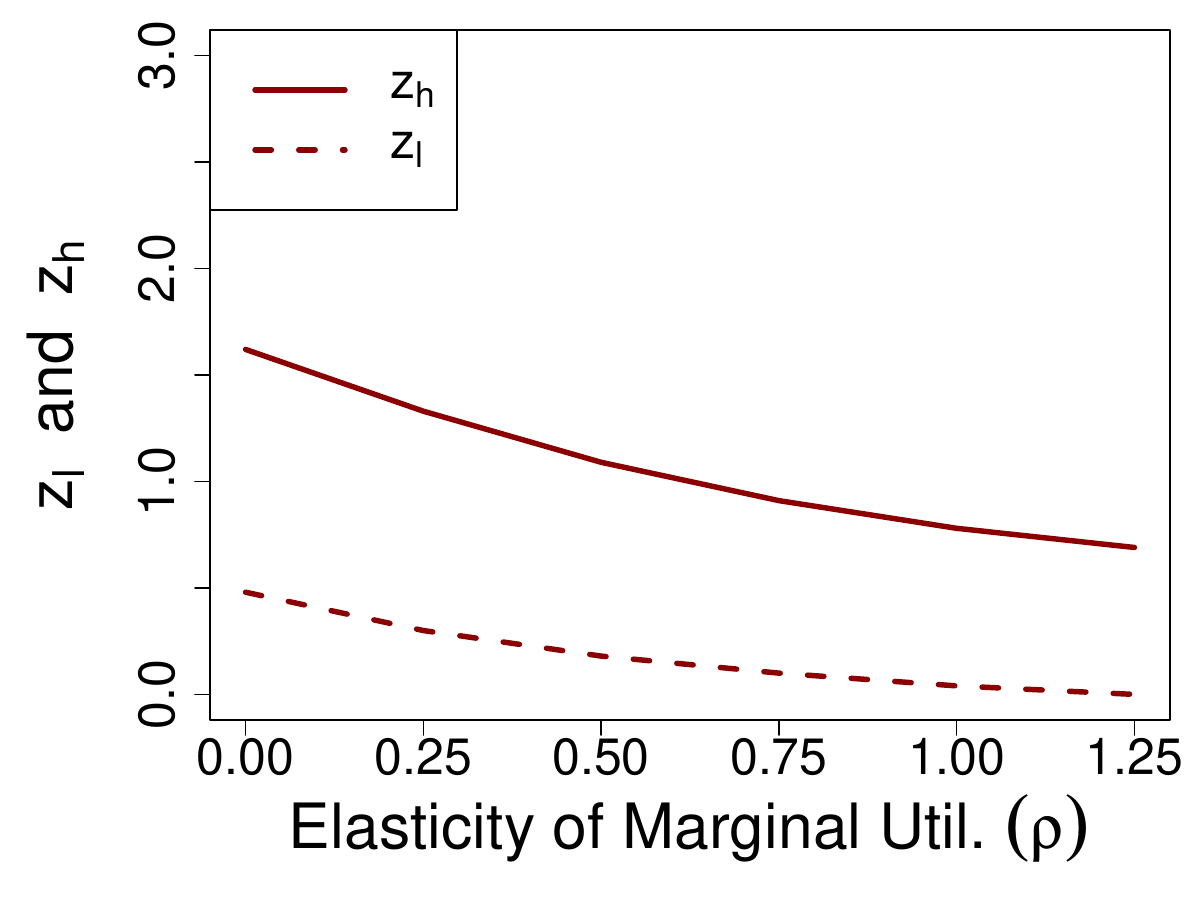}
    \end{minipage}
    \hfill
    \begin{minipage}[t]{0.48\textwidth}
        \centering
        \includegraphics[width=\linewidth]{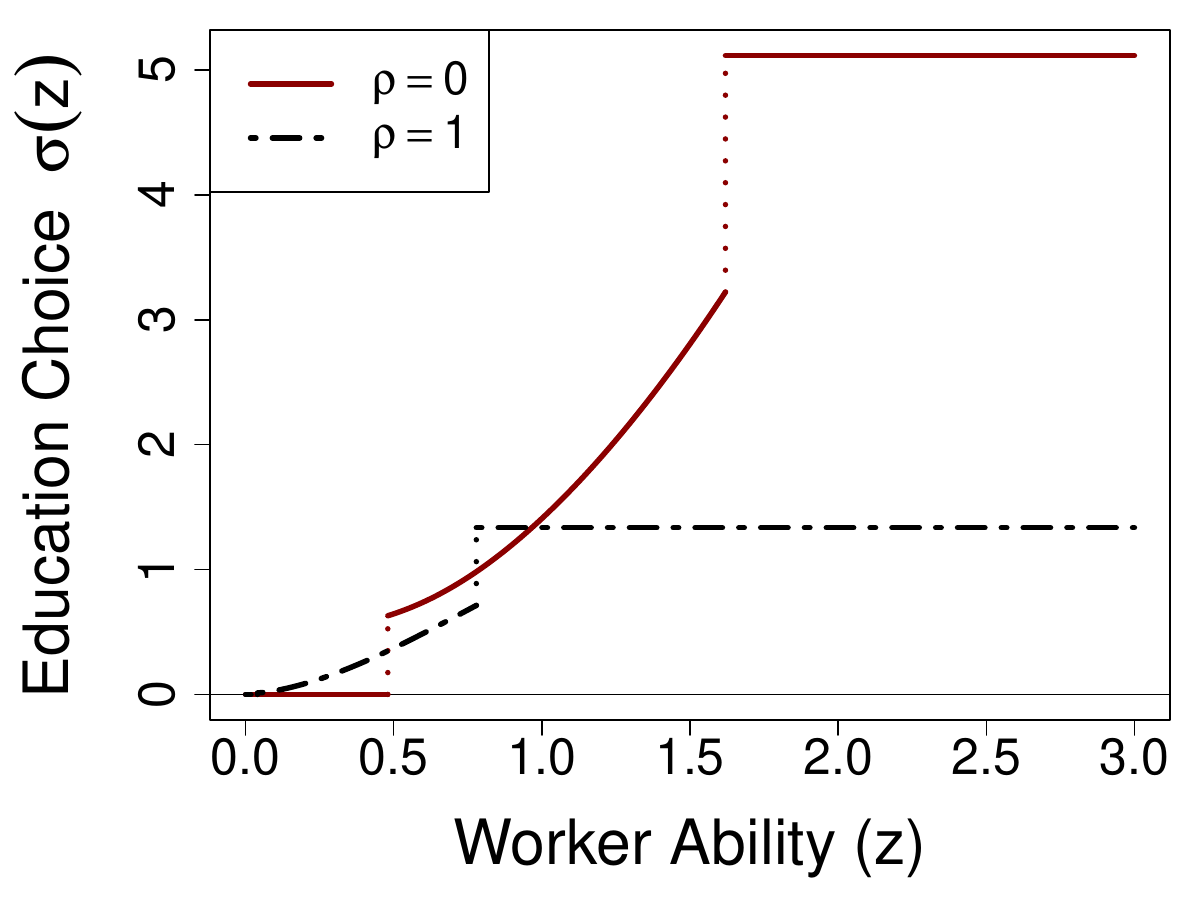}
    \end{minipage}

    \vspace{1em}

    \begin{minipage}[t]{0.48\textwidth}
        \centering
        \includegraphics[width=\linewidth]{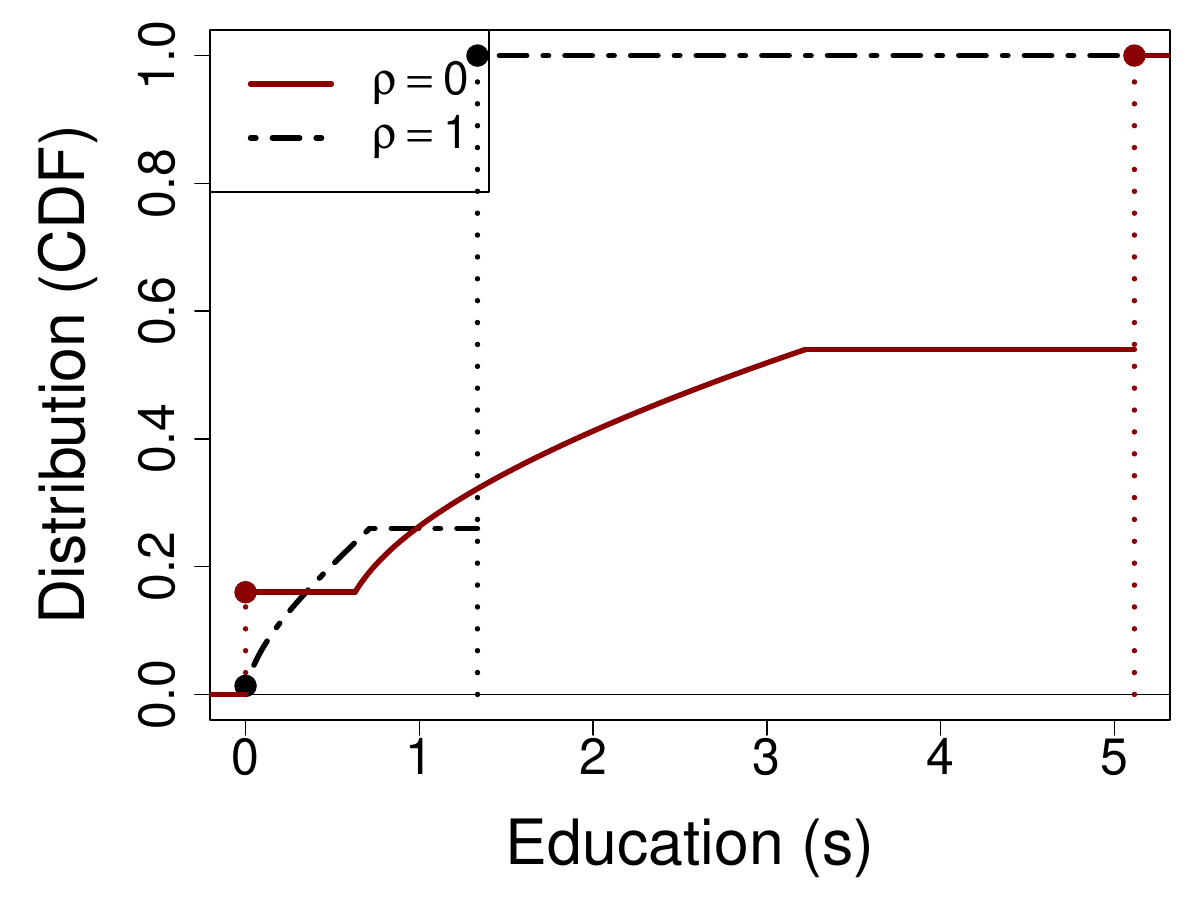}
    \end{minipage}
    \hfill
    \begin{minipage}[t]{0.48\textwidth}
        \centering
        \includegraphics[width=\linewidth]{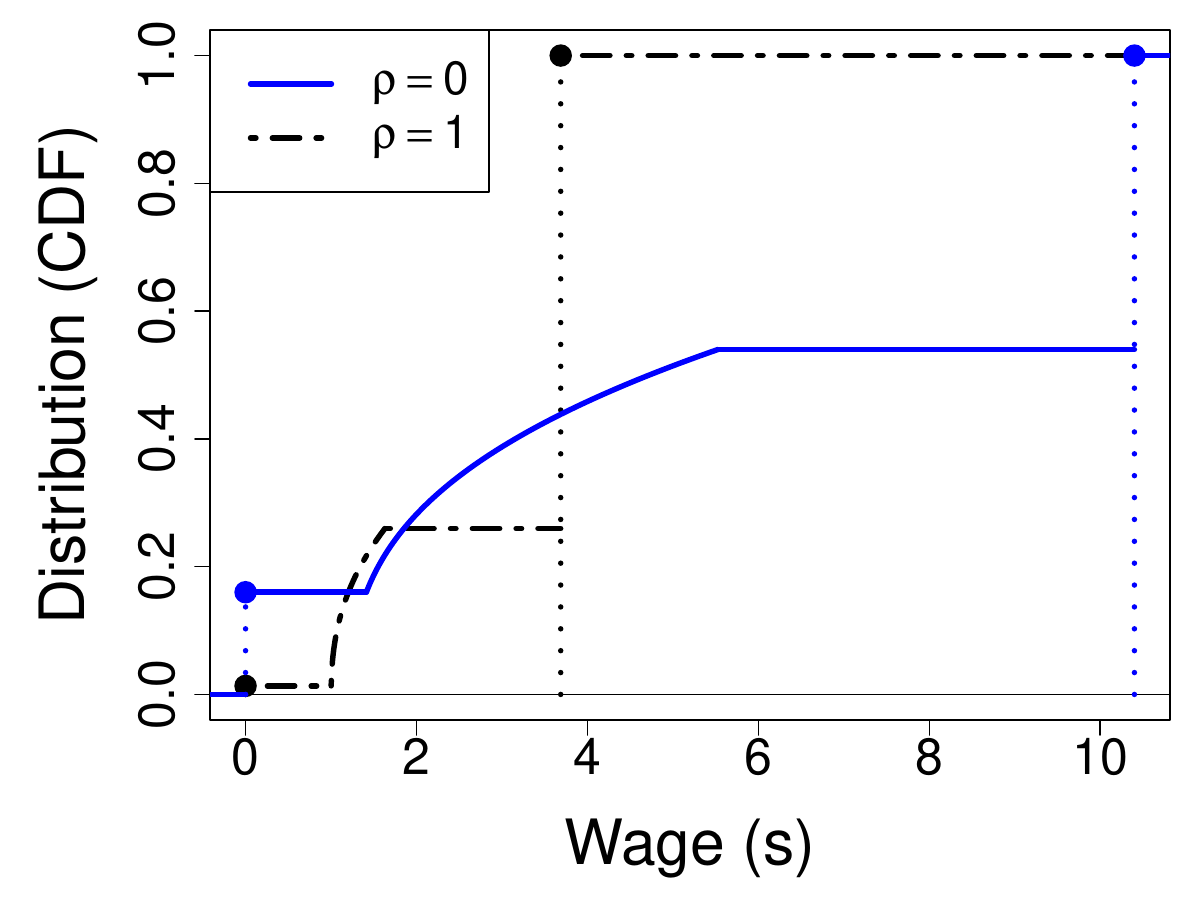}
    \end{minipage}
\end{figure}

\paragraph{Elasticity of Marginal Utility of Consumption($\rho$).}
We now turn to the role of the elasticity of marginal utility of consumption, $\rho$.
It governs the curvature of the iso-elastic utility function. Interestingly, it has a
similar role to the policymaker's implicit weighing of low-income workers.
For example, as $\rho$ increases, the utility function becomes more concave,
meaning that one dollar of wages earned by lower-income workers contributes more
to the social welfare function than the same amount earned by high-income
workers.
Figure \ref{figure: comp statics rho} presents the equilibrium outcomes as
$\rho$ varies from 0 to 1.25. Both the minimum and maximum ability thresholds
shift downward as $\rho$ increases.
Accordingly, the education choice function and two distribution functions shift
to the left when we compare the cases of $\rho=0$ and $\rho=1$ in the remaining
three panels.

Under the efficient rationing assumption, \citet{lee2012optimal} show that a
minimum wage may increases social welfare in a perfectly competitive market of
complete information with no education investment when
the policymaker places greater weights on low-skilled workers relative to
high-skilled workers, based on the \emph{local} analysis of the market wage changes mainly with a substitution effect.
Recently, \citet{berger2025minimum}  show the dominance of the redistributive effect of a
minimum wage--benefiting low-income workers--over its efficiency gain from limiting firms’ market power in oligopsonistic labor markets of complete information.

It is worth noting that our results suggest
that the optimal minimum wage may be negatively related to the
policymaker's weight on low-skilled workers in job matching markets with
incomplete information,
where education serves as a signaling mechanism prior to matching.

This is due to the asymmetric utility ripple effect.
The minimum wage makes low-skilled workers worse off but high-skilled workers
better off, as shown in the right panel of Figure \ref{figure: mechanism}. Therefore, when the policymaker places greater weights on low-skilled workers, they would reduce the minimum wage because the welfare loss from low-skilled workers outweighs the welfare gains from high-skilled workers.

The upper-right panel in Figure \ref%
{figure: comp statics rho} shows that as $\rho $ increases, the optimal two ability threshold $z_{\ell }$ and $z_{h}$ decrease, resulting in lower minimum and maximum wages, $t_{\ell }$ and $%
t_{h}$. Because the marginal utility of consumption is the effective marginal return on education investment,
a higher $\rho$ implies that the marginal return quickly decreases as the worker's consumption level is higher.
Therefore, when $\rho$ is high, it is optimal to impose a lower maximum wage $t_{h}$ to enlarge the pooling region on the
top.

\begin{figure}[bth]
    \centering
    \caption{Comparative Statics: $b$}
    \label{figure: comp statics beta}

    \begin{minipage}[t]{0.48\textwidth}
        \centering
        \includegraphics[width=\linewidth]{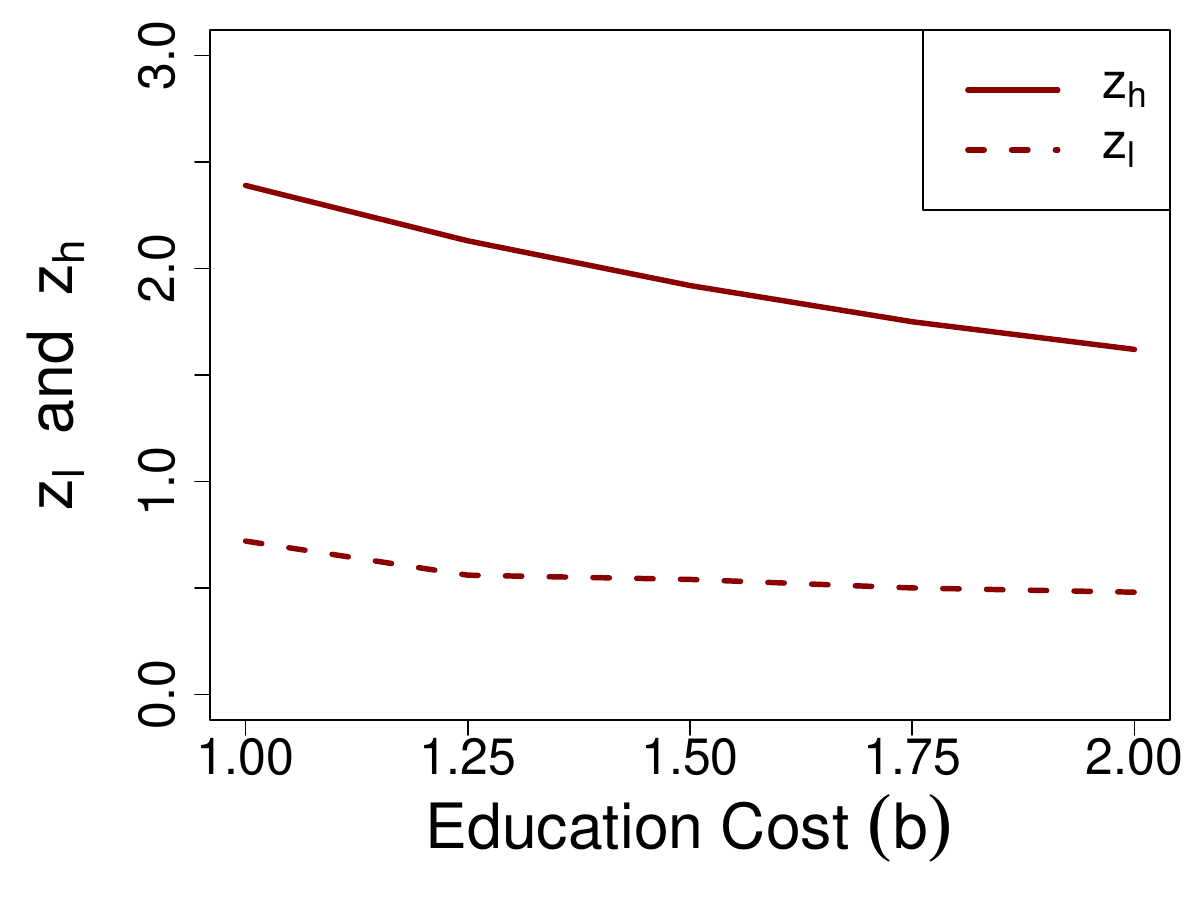}
    \end{minipage}
    \hfill
    \begin{minipage}[t]{0.48\textwidth}
        \centering
        \includegraphics[width=\linewidth]{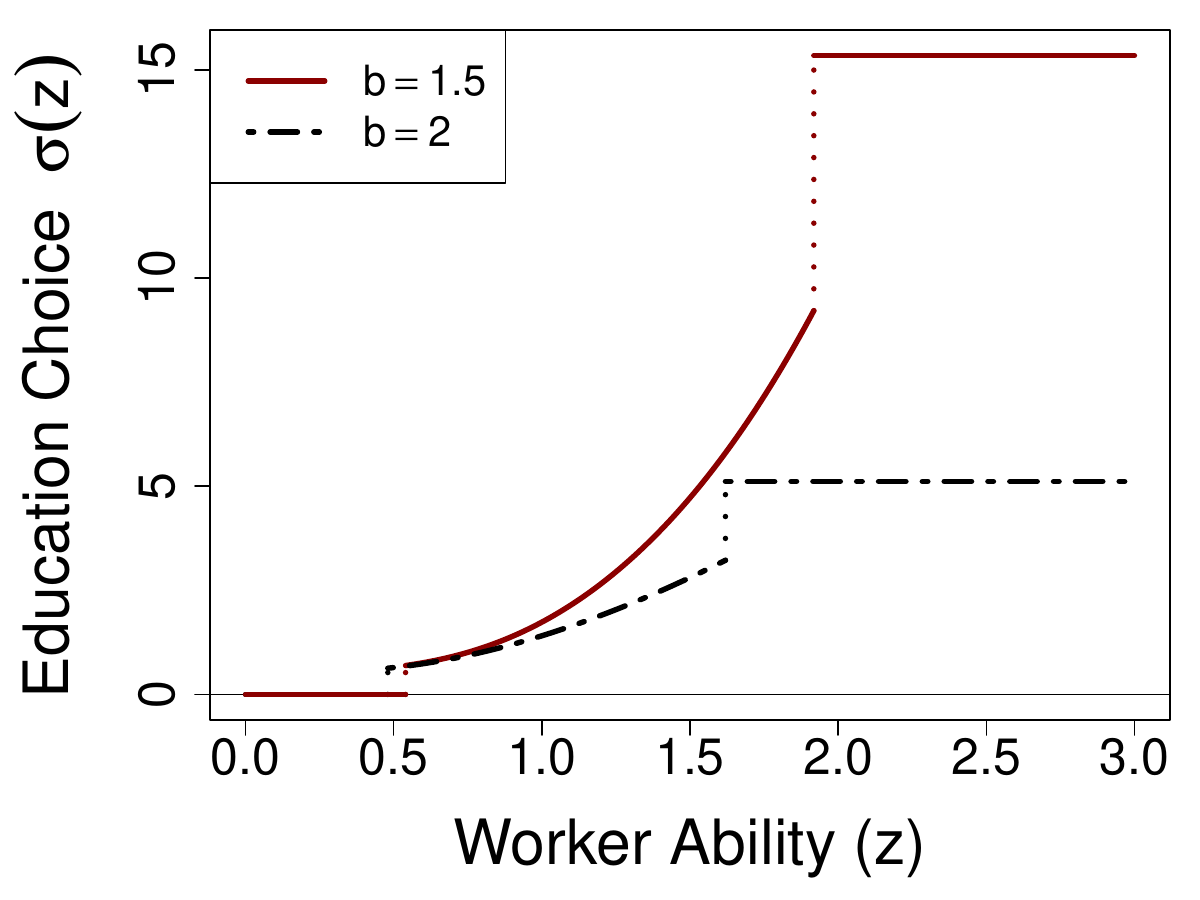}
    \end{minipage}

    \vspace{1em}

    \begin{minipage}[t]{0.48\textwidth}
        \centering
        \includegraphics[width=\linewidth]{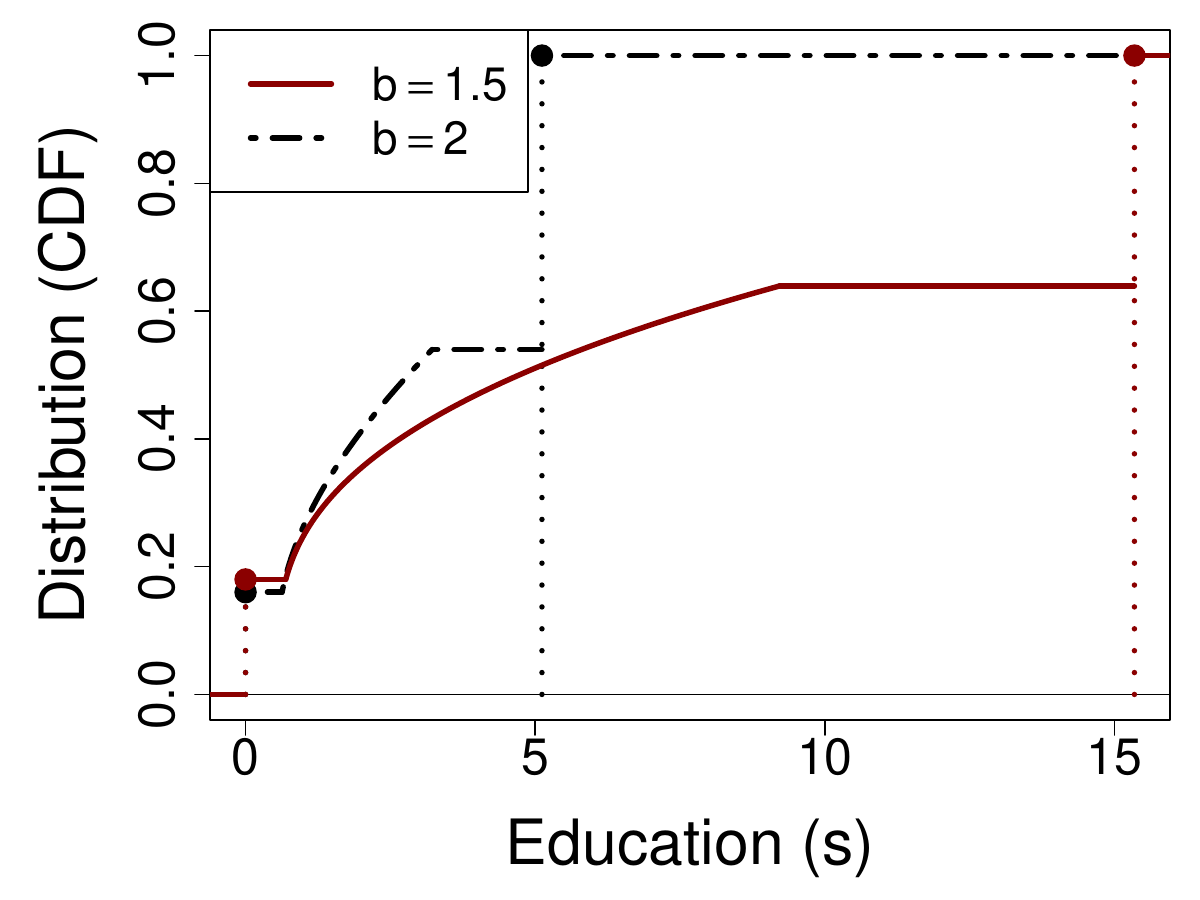}
    \end{minipage}
    \hfill
    \begin{minipage}[t]{0.48\textwidth}
        \centering
        \includegraphics[width=\linewidth]{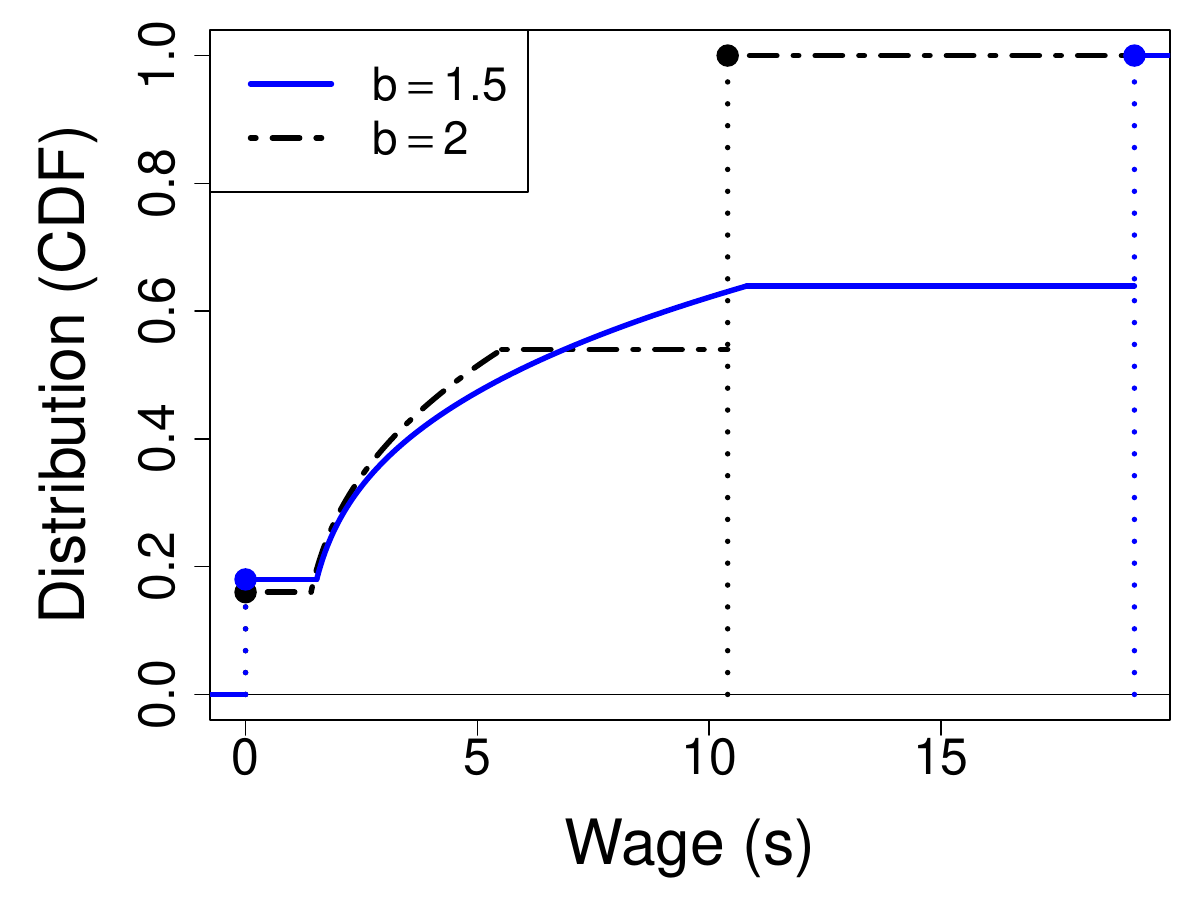}
    \end{minipage}
\end{figure}

\paragraph{Education Cost ($b$).}
Finally, Figure \ref{figure: comp statics beta} illustrates the effect of the
education cost parameter $b\in[1,2]$ on the equilibrium outcome. These results
align with our intuition: as education costs increase, surplus
benefits decline given any ability level, leading to lower optimal threshold
values. As a result, the education choice function and related distributions
shift toward lower values.

\subsection{Welfare Analysis}
In this subsection, we compute how much the welfare can be improved by the
optimal wage band policy relative to the no-intervention case. In addition to
the optimal policy that sets both minimum and maximum wages, we also
examine a sub-optimal policy that sets only a minimum wage, a social policy commonly
observed in many developed economies.

\begin{figure}[tbh]
    \centering
    \caption{Welfare Analysis}\label{figure: welfare analysis}

    \begin{minipage}[t]{0.48\textwidth}
        \centering
        \includegraphics[width=\linewidth]{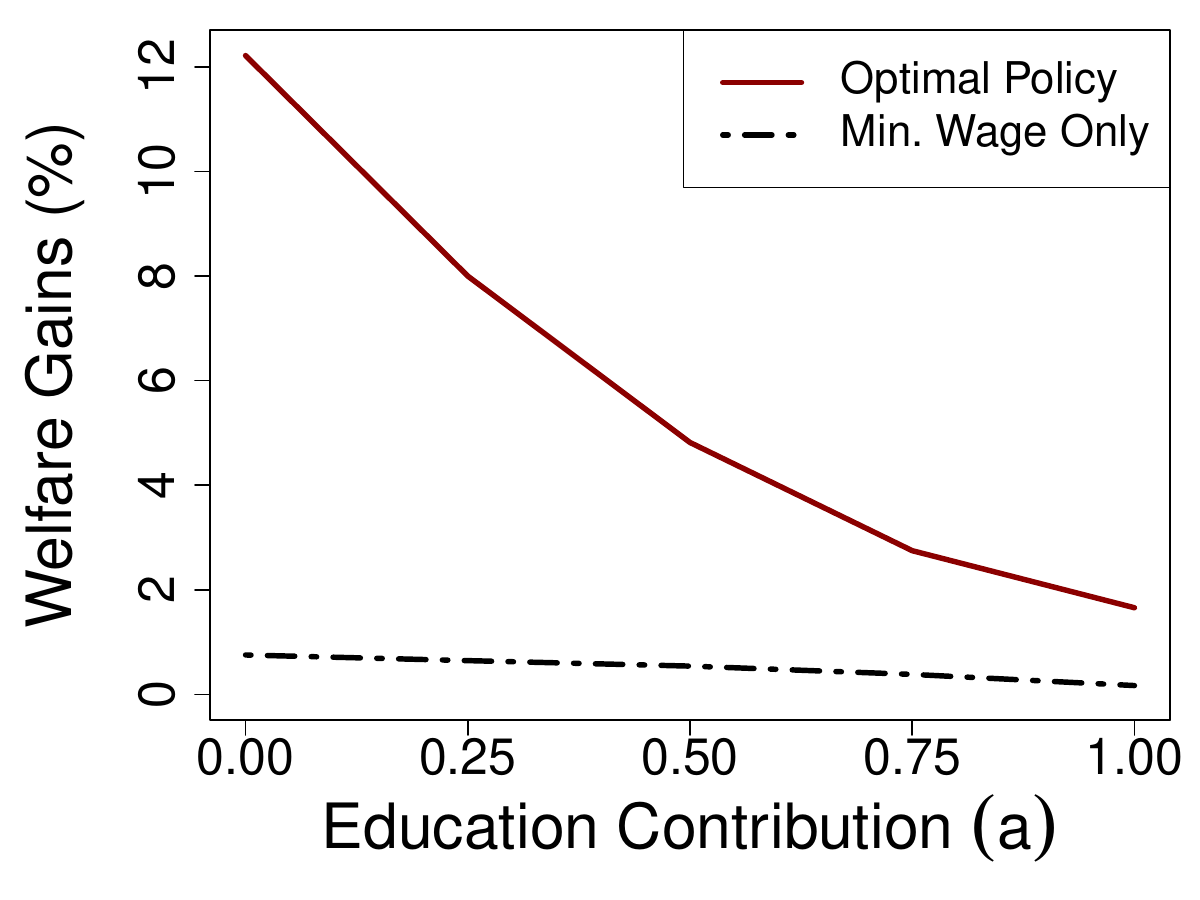}
    \end{minipage}
    \hfill
    \begin{minipage}[t]{0.48\textwidth}
        \centering
        \includegraphics[width=\linewidth]{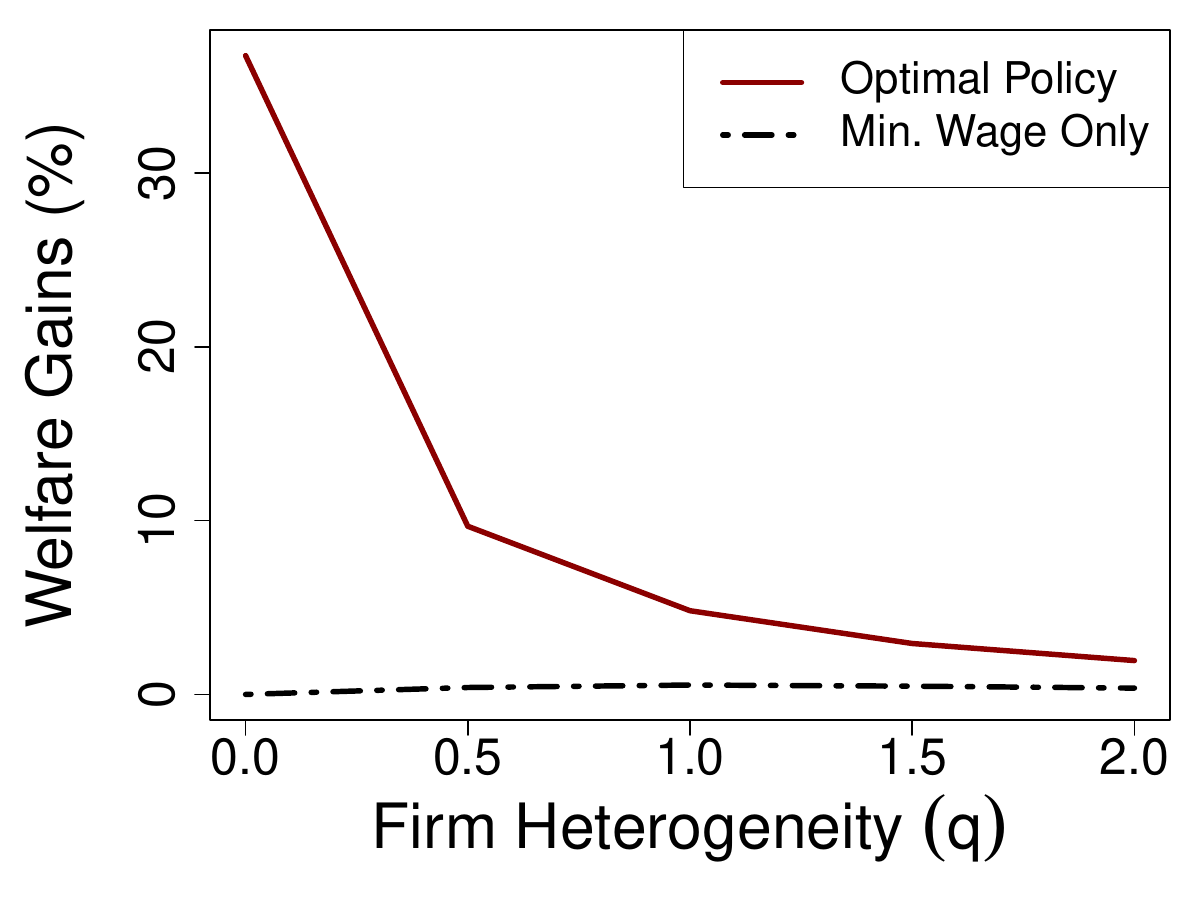}
    \end{minipage}

    \vspace{1em}

    \begin{minipage}[t]{0.48\textwidth}
        \centering
        \includegraphics[width=\linewidth]{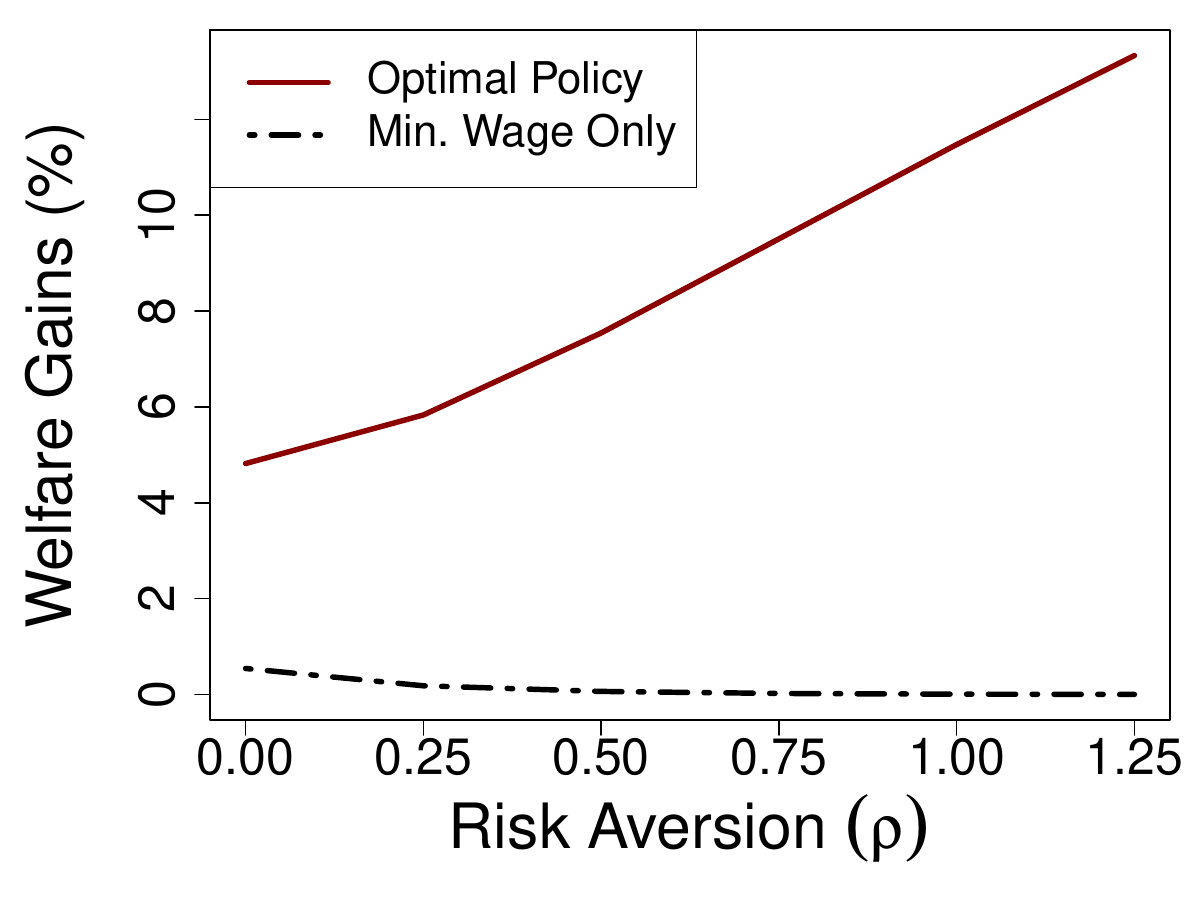}
    \end{minipage}
    \hfill
    \begin{minipage}[t]{0.48\textwidth}
        \centering
        \includegraphics[width=\linewidth]{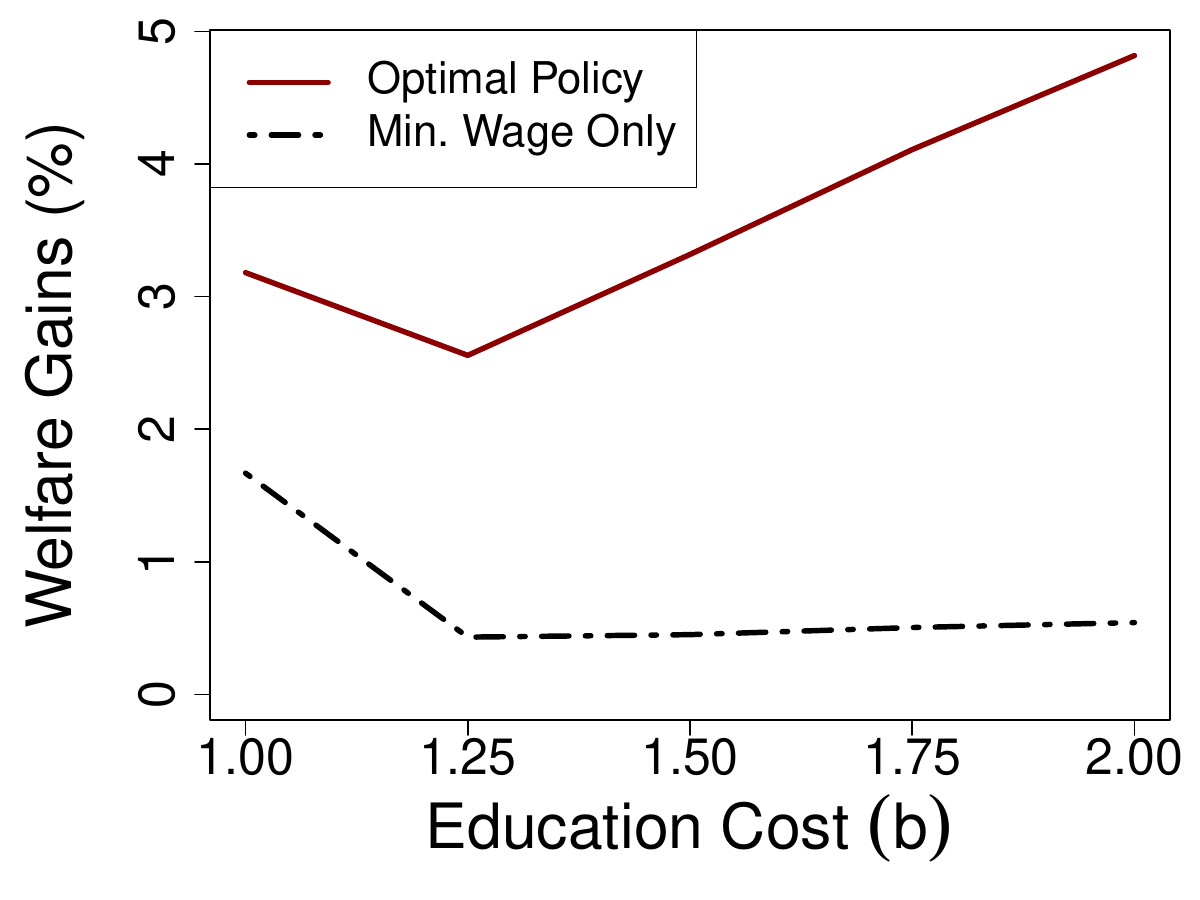}
    \end{minipage}

\end{figure}

Figure \ref{figure: welfare analysis} illustrates the welfare gains, expressed in
percentage points, from different policy interventions, evaluated across a range
of model parameter values. These gains are directly related to the
optimal threshold bands ($z_{\ell}$ and $z_h$) discussed in the previous
subsection. As the wage band widens, the optimal policy approaches the no-intervention
case, resulting in smaller welfare improvements. This pattern is confirmed when you
compare the two first panels in Figures \ref{figure: welfare analysis} and
\ref{figure: comp statics a}. As the education parameter $a$ increases, the
optimal band becomes wider, the the welfare gain declines to $1.7\%$ relative to
the no-intervention case when $a=1$. In contrast, the welfare gain reaches as
high as $12.2\%$ when $a=0$, i.e., when education functions purely as a signaling
device with no contribution to production. The welfare improvement from the minimum-wage-only policy
is not substantial, but there are still modest gains from $0.2\%$ to $0.8\%$ for
different $a$ values.

In case of the firm heterogeneity parameter,  the welfare improvement from the
optimal policy is higher when $q$ is lower, i.e.\ more homogeneous firms. The
minimum-wage-only policy achieves the highest improvement of $0.5\%$ when
$q=1$ is in the middle.

We now turn to the effect of the elasticity of the marginal utility of consumption $\rho$, which has the similar role to the policymaker's implicit relative weight placed on
low-skilled workers. As $\rho$ increases, the relative welfare gain from the minimum-wage-only policy decreases, while the
gain from the fully optimal policy increases. The welfare improvements range from
$0\%$ to $0.5\%$ under the minimum-wage-only policy and from $4.8\%$ to $13.3\%$ for the
optimal policy.

The final panel illustrates the results for different values of the education
cost parameter $b$. The welfare gains from both policies exhibit a convex shape,
with minimum values around $b=1.25$. The minimum-wage-only policy yields improvements of from $0.4\%$ to $1.7\%$, while the fully optimal policy yields gains from $2.6\%$ to $4.8\%$.

In sum, we observe modest welfare gains from the minimum-wage-only policy across
a \emph{wide} range of parameter values. Substantially larger gains arise from the optimal wage
band policy that includes a maximum wage. The variation in welfare gains is
also much greater when the maximum wage is implemented.

\section{Concluding remarks}

We study the optimal wage band problem in which a policymaker chooses a
wage band prior to the actions of workers and firms in job marching markets,
where workers make costly education investments. We show that the monotone
\editYS{competitive} signaling equilibrium is unique and
that this problem is isomorphic to an optimal threshold problem, which is more
tractable even in highly nonlinear settings.

In contrast to the wage ripple effect of the minimum wage \citep[e.g.,][]{engbom2022earnings}, we find that the utility ripple effect is asymmetric:
it makes low-skilled workers worse off while benefiting high-skilled workers.
This leads to a new implication for the optimal minimum wage, differing from models that do not account for education investment \citep[e.g.,][]{lee2012optimal}.

Increases in the mean and variance of the firm productivity distribution, or decreases in the elasticity of marginal utility of consumption,
%lead to increases
\editYS{raise}
not only the minimum wage but also the maximum wage. This is because the maximum wage becomes less effective when the matching inefficiency from high-type pooling outweighs the savings from reduced education costs.

Furthermore, the proposed optimal threshold problem is tractable, making the methodology applicable to various matching markets with signaling, such as college admissions, medical residency matching, and venture capital–startup matching. Interestingly, a wage band can be interpreted as two tax brackets with a minimum wage, where the lower bracket has a zero income tax rate and the upper bracket a 100\% rate. It would be worthwhile to explore how our model and methodology could be extended to derive optimal income tax schedules with multiple tax brackets. These extensions, however, are beyond the scope of this paper and are left for future research.

\bibliographystyle{chicago}
\bibliography{reference}

\section*{Appendix}

\addcontentsline{toc}{section}{Appendices} \renewcommand{\thesubsection}{		%
\Alph{subsection}}

\subsection{Definition of CSE\label{App_def_CSE}}

We present the formal definition of a competitive signaling equilibrium
given a wage band $T=[t_{\ell },t_{h}]$, along with several related concepts
that we used for modeling labor market matching under incomplete
information, following \citet*{HAN_et_al}.

\begin{definition}
\label{def_profitable_sender_deviation}Given $\{\sigma _{T},\mu _{T},\tau
_{T},m_{T}\}$\emph{, a profitable worker deviation to an off-path education}
exists if there is a worker $z^{\prime }$ such that there exist an education
level $s^{\prime }\notin \sigma _{T}(\mathcal{Z})$\thinspace , a wage $%
t^{\prime }\in T$, and a firm $x^{\prime }\in \mathcal{X}^{\ast }$
satisfying:
\begin{align}
& \text{(a) }\mathbb{E}_{\mu (s^{\prime })}\left[ g(t^{\prime },s^{\prime
},z,x^{\prime })\right] >\mathbb{E}_{\mu (m^{-1}(x^{\prime }))}\left[
g\left( \tau _{T}\left( m^{-1}(x^{\prime })\right) ,m_{T}^{-1}(x^{\prime
}),z,x^{\prime }\right) \right]  \label{receiver_matching_utility} \\
& \text{(b) }%
\begin{cases}
u(t^{\prime },s^{\prime },z^{\prime })>u(\tau _{T}(\sigma _{T}(z^{\prime
})),\sigma _{T}(z^{\prime }),z^{\prime }) & \text{ if }\sigma _{T}(z^{\prime
})\in S^{\ast } \\
u(t^{\prime },s^{\prime },z^{\prime })>0 & \text{ if }\sigma _{T}(z^{\prime
})\notin S^{\ast }.%
\end{cases}
\label{sender_matching_partner}
\end{align}
\end{definition}

Recall that $\mathbb{E}_{\mu _{T}}$ denotes integration over $z$ with
respect to the measure $\mu $. Conditions $(a)$ and $(b)$ imply that,
instead of following $\{\sigma _{T},\tau _{T},m_{T}\}$, worker $z^{\prime }$
and firm $x^{\prime }$ choose $s^{\prime }$ and $t^{\prime }$, respectively,
and enjoy strictly higher utility and expected profit. If either $(a)$ or $%
(b)$ is violated, we say that there is no profitable worker deviation to an
off-path education.

Next, we define the equilibrium matching outcome.

\begin{definition}
\label{def_stable_matching}Given $\{\sigma _{T},\mu _{T}\}$, a pair $\{\tau
_{T},m_{T}\}$\ is \emph{an equilibrium matching outcome} if the following
conditions hold:

\begin{enumerate}
\item[(a)] $m_{T}$\ is stable: there is no worker-firm pair $(z^{\prime
},x^{\prime })$ with $\sigma _{T}(z^{\prime })=s^{\prime }\in S^{\ast }$ and
$x^{\prime }\notin m\left( s^{\prime }\right) $ such that there exist $%
t^{\prime }\in T$ satisfying:
\begin{align}
& \text{(i) }\mathbb{E}_{\mu _{T}(s)}\left[ g(t^{\prime },s^{\prime
},z,x^{\prime })\right] >\mathbb{E}_{\mu _{T}(m^{-1}(x^{\prime }))}\left[
g\left( \tau _{T}\left( m_{T}^{-1}(x^{\prime })\right) ,m_{T}^{-1}(x^{\prime
}),z,x^{\prime }\right) \right]  \label{stable_matching1} \\
& \text{(ii) }u(t^{\prime },s^{\prime },z^{\prime })>u(\tau _{T}(s^{\prime
}),s^{\prime },z^{\prime })  \label{stable_matching2}
\end{align}

\item[(b)] $\tau _{T}$ clears the markets: for all $S^{\prime }\in \mathcal{B%
}(S^{\ast })$ we have
\begin{equation*}
H\left( \left\{ x|x\in m_{T}(\xi _{T}(x))\text{, }\xi _{T}(x)\in S^{\prime
}\right\} \right) =G\left( \left\{ z|\sigma _{T}\left( z\right) \in
S^{\prime }\right\} \right) \mathbf{.}
\end{equation*}
\end{enumerate}
\end{definition}

Condition $(a)$ implies that the induced matching function $m_{T}$\
characterizes equilibrium matches in which no worker-firm pair can be
strictly better off by deviating to form a new match. Condition $(b)$
implies that the market-clearing wage function $\tau _{T}$\ induces a
measure preserving matching function $m_{T}$. Recall that a firm $x\in
\mathcal{X}^{\ast }$\ hires a worker with signal $\xi _{T}(x)\in S^{\ast }$,
and the market wage becomes $\tau _{T}(\xi _{T}(x))$\ under the equilibrium
matching outcome $\{\tau _{T},m_{T}\}$ given $\{\sigma _{T},\mu _{T}\}$.

Finally, we are ready to define a competitive signaling equilibrium.
% Moved the citation of Ramey (1996) to the end of Definition 3 for smoother flow.

\begin{definition}
\label{definition1} Given a wage band $T=[t_{\ell },t_{h}]$, the tuple $%
\{\sigma _{T},\mu _{T},\tau _{T},m_{T}\}$\ constitutes \emph{a competitive
signaling equilibrium (CSE)} if the following conditions hold:
% Any signaling game involves incomplete information. "with incompelte information" seems to be repetitive. Drop it for a succinct expression.

\begin{enumerate}
\item For all worker $z\in \mathcal{Z}$, the education choice $\sigma
_{T}(z) $\ is optimal.

\item The posterior belief $\mu _{T}$ is Bayesian consistent:

\begin{enumerate}
\item if $s\in \sigma _{T}(\mathcal{Z})$\ satisfies $G(\{z|\sigma
_{T}(z)=s\})>0,$\ then $\mu _{T}(s)$\ is determined from $G$ \ and $\sigma
_{T}$\ using Bayes' rule.

\item if $s\in \sigma _{T}(\mathcal{Z})$\ but $G(\{z|\sigma _{T}(z)=s\})=0$,
then $\mu _{T}(s)$\ can be any probability distribution with
\begin{equation*}
\supp\mu _{T}(s)=\cl\left\{ z|\sigma _{T}(z)=s\right\} ,
\end{equation*}%
where $\supp$ and $\cl$ denote support and closure, respectively.

\item if $s\notin \sigma _{T}(\mathcal{Z})$, then $\mu _{T}(s)$ is
unrestricted.
\end{enumerate}

\item Given $\{\sigma _{T},\mu _{T}\}$, a pair $\{\tau _{T},m_{T}\}$ is an
equilibrium matching outcome.
\end{enumerate}
\end{definition}

%Note that we adopt the definition of consistency for $\mu$ as given in \citet*{RAMEY1996508}.
The notion of CSE is appropriate for our model, as we consider a large
matching market with heterogeneous workers and firms, where firms observe
only the education level of workers and have \emph{incomplete information}
about their ability. The market is \emph{competitive}, so the equilibrium
wage schedule function $\tau$ is taken as given by both workers and firms.
Finally, the consistency of the \editYS{posterior} belief $\mu$ follows \cite%
{RAMEY1996508}.

%%%%%%%%%%%%%%%%%%%%%%%%%%%%%%%%%%%%%%%%%%%%%%%%%%%%%%%%%%%%%%%%%%%%%%%%%%%%%%%%%%%%%%%%%%%%%
% Removed the following paragraph, which is a copy from the JET paper. Can add a citation of Han et al. (2024) if we want to refer further discussions and relationship with the existing literature. However, it does not seem to be necessary in the current context. We already mentioned that the concept comes directly from Han et al. (2024).
\begin{comment}Since a single worker or firm has no significant market power in a large matching market, workers and firms take the market wage function $\tau $\ as given when they make their decision and that $\tau $\ is fully reinforced by their optimal decisions through the market clearing condition. Therefore, $\tau $\ is endogenous similar to how prices are determined endogenously in
the general equilibrium framework. The notion of CSE indeed follows the
tradition of how to formulate a competitive equilibrium in a large matching
market in the literature (e.g., Mailath, Postlewaite, Samuelson (2013,
2017), Cole, Mailath, Postlewaite (2001) among many). The notion of CSE
contributes to the literature in that it proposes the new notion of
competitive \textquotedblleft signaling\textquotedblright\ equilibrium,
combining signaling and matching in a large market with two-sided
heterogeneity.
\end{comment}
%%%%%%%%%%%%%%%%%%%%%%%%%%%%%%%%%%%%%%%%%%%%%%%%%%%%%%%%%%%%%%%%%%%%%%%%%%%%%%%%%%%%%%%%%%%%%

\subsection{Assumptions\label{App_separating_CSE}}

To facilitate the analysis of monotone CSE, we introduce a few assumptions.

\begin{description}
\item[Assumption A] \label{Assumption1}$u(t,s,z)$ is (i) decreasing in $s$,
strictly increasing in $t$ and $z$, satisfies (ii) the single crossing
property in $(t,z)$, the strict single crossing property in $(s,z)$, (iii)
the (weakly) decreasing difference property in $(t,s)$, (iv) is continuous
in all its arguments, and (v) twice differentiable everywhere.\textbf{\ }

\item[Assumption B] \label{Assumption2} $g(t,s,z,x)$ (i) is strictly
increasing in $x$ and $z,$ non-decreasing in $s$ and strictly decreasing in $%
t$ (ii) is supermodular in $(t,s,z)$ and satisfies the single crossing
property in $(\left( t,s,z\right) ;x)$ and the strict single crossing
property in $(z;x)$ at each $(s,t)$, (iii) satisfies the (weakly) decreasing
difference property in $(t,s)$, (iv) is continuous in all its arguments, and
(v) twice differentiable everywhere.
\end{description}

Assumptions A.(i) -(iii) and B.(i) - (iii) are monotone-supermodular
assumptions on worker and \editYS{firm's} utility and profit functions. Using these
assumptions, we can establish that a CSE is monotone in the stronger set
order if and only if it passes Criterion D1 (Han, Sam, and Shin (2024)). The
equivalence between the stronger monotonicity of a belief and Criterion D1
is crucial in deriving a unique (stronger) monotone CSE given any wage band
even if the wage band induces no separating CSE.

We also assume the concavity of the utility functions (Assumption C) and
technical properties (Assumptions D and E).

\begin{description}
\item[Assumption C] \label{ass5_e}$g(\underline{t},x,s,z)\geq 0$ for all $%
(x,s,z).$ $g$ is concave in $t$ and $u$ is concave in $(s,t)$ with $%
\lim_{s\rightarrow 0}u_{s}(t,s,z)=0$ and $\lim_{s\rightarrow \infty }$ $%
u_{s}(t,s,z)=-\infty $. Either $g(t, x, s, z)$ is strictly concave in $t$,
or, if not, it is strictly concave in $s$. Also, $\lim_{s \to 0} g_s(t, x,
s, z) = \infty$ and $\lim_{s \to \infty} g_s(t, x, s, z) = 0$.

%		
%		
%		
%		
%		If $g(t,x,s,z)$ is strictly increasing in $s$, it is
%		concave in ($s,t)$ and either $g$ or $u$ is strictly concave in $s$ with $%
%		\lim_{s\rightarrow 0}g_{s}(t,x,s,z)=\infty $ and $\lim_{s\rightarrow \infty
%		}g_{s}(t,x,s,z)=0.$ \textbf{If }$g(t,x,s,z)$\textbf{\ is not strictly
%			increasing in }$s,$\textbf{\ }$g$\textbf{\ is strictly concave in }$t.$

\label{Assumption4}

\item[Assumption D] $0<G^{\prime }(z)<\infty $ for all $z\in \left[
\underline{z},\overline{z}\right] $ and $0<H^{\prime }(x)<\infty $ for all $%
x\in \lbrack \underline{x},\overline{x}]$.

\label{Assumption5}

\item[Assumption E] \label{ass7 copy(1)_e}$\lim_{z\rightarrow \underline{z}%
}u(t,s,z)=-\infty $ for all $s>0$, $u(\underline{t},0,\underline{z})=0,$ $g(%
\underline{t},\underline{x},0,z)=0$ for all $z\geq \underline{z}$ and $g(%
\underline{t},x,0,z)+u(\underline{t},0,z)>0$ for all $x>\underline{x},$ all $%
z>\underline{z}$.
\end{description}

Define $f(t,z):=g(t,n(z),\kappa (t,z),z),$ where $\kappa (t,z)$ is the
education level that satisfies
\begin{equation}
u(t,\kappa (t,z),z)=0.  \label{constraint}
\end{equation}

\noindent \textbf{Assumption F} For any $z>\underline{z},$ $%
\lim_{t\rightarrow \underline{t}}f(t,z)>0$ and $\lim_{t\rightarrow \infty
}f(t,z)<0.$

\subsection{Lemma \protect\ref{lemma_bottom_types_e}\label%
{App_lem_bottom_match}}

Define $z^{\circ }(t_{\ell })$\ as the worker type such that $t_{\ell }$\ is
the bilaterally efficient wage in a match between the worker type $z^{\circ
}(t_{\ell })$\ and the firm type $n(z^{\circ }(t_{\ell }))$, i.e., $t^{\ast
}(z^{\circ }(t_{\ell }))=t_{\ell }$. Lemma \ref{lemma_bottom_types_e} below
shows the existence of the lowest worker type who enters the market and her
education given the minimum wage $t_{\ell }.$

\begin{lemma}
\label{lemma_bottom_types_e}

\begin{enumerate}
\item If $\underline{t}\leq t_{\ell }\leq t^{\ast }(\underline{z})$,
everyone enters the market: $z_{\ell }=\underline{z}$ and worker type $%
z_{\ell }$'s education is $s_{\ell }=s^{\ast }(\underline{z})$ and the
equilibrium wage $\tau (s_{\ell })$ chosen by firm type $n(z_{\ell })=%
\underline{x}$ is $\tau (s_{\ell })=t^{\ast }(\underline{z}).$

\item If $t_{\ell }\in (t^{\ast }(\underline{z}),\hat{t})$ and Assumption F
is satisfied, there exists a unique solution $(s_{\ell },z_{\ell })\in
%TCIMACRO{\U{211d} }%
%BeginExpansion
\mathbb{R}
%EndExpansion
_{++}\times (\underline{z},z^{\circ }(t_{\ell }))$ that solves (\ref{lem1e})
and (\ref{lem2e}), and the equilibrium wage $\tau (s_{\ell })$ chosen by $%
n(z_{\ell })$ is $t_{\ell }$.
\end{enumerate}
\end{lemma}

We first present some results that help us prove Lemma \ref%
{lemma_bottom_types_e}. Note that the maximization problem in (\ref%
{maximization}) is equivalent to the unconstrained maximization problem: $%
\max_{t\in
%TCIMACRO{\U{211d} }%
%BeginExpansion
\mathbb{R}
%EndExpansion
_{+}}f(t,z).$

\begin{lemma}
\label{lemma_concavity}$f$ is strictly concave in $t$ and strictly
supermodular in $(t,z)$.
\end{lemma}

\begin{proof}
Note that
\begin{equation*}
\kappa _{tt}=\frac{-u_{s}^{2}u_{tt}+2u_{t}u_{s}u_{ts}-u_{t}^{2}u_{ss}}{
u_{s}^{3}}\leq 0
\end{equation*}
because $-u_{s}^{2}u_{tt}+2u_{t}u_{s}u_{ts}-u_{t}^{2}u_{ss}\geq 0$ due to $%
u_{s}<0$ and the quasiconcavity of $u$ in $(s,t)$ (the concavity of $u$
(Assumption C) implies the quasiconcavity). Then, it implies that $\kappa $
is concave in $t$. Because of Assumption C, either (i) $g$ is strictly
concave in $t$, or (ii) $g$ is strictly concave in $s$. Therefore, because $%
g $ is non-decreasing in $s$, strictly decreasing in $t$ (Assumption B), and
$\kappa $ is concave in $t$, we conclude that $f$ is strictly concave in $t.$
Furthermore, $n(z)$ is increasing in $z$ and $\kappa (t,z)$ are increasing
in $z$ and $t$, respectively. Therefore, given Assumptions A.(ii), A.(iii),
B.(ii), and B.(iii), $f$ is strictly supermodular in $(t,z)$.
\end{proof}

\begin{lemma}
\label{lemma_increasing_t_s}Given Assumption F, $(s^{\ast }(z),t^{\ast }(z))$
is unique and strictly increasing in $z$.
\end{lemma}

\begin{proof}
Because $u$ is increasing in $z,$ $D(z)\geq _{s}D(z^{\prime })$ for $%
z>z^{\prime }.$ o show the uniqueness of $(s^{\ast }(z),t^{\ast }(z))$, note
that the constraint holds with equality at an optimal solution because $u$
is monotone increasing in $t$ and $g$ is monotone increasing in $t.$ The
maximization problem in (\ref{maximization}) is equivalent to $\max_{t\in
\in
%TCIMACRO{\U{211d} }%
%BeginExpansion
\mathbb{R}
%EndExpansion
_{+}}f(t,z).$ The first-order condition is
\begin{equation}
f_{t}=g_{t}+g_{s}\cdot \kappa _{t}=0,  \label{FOC}
\end{equation}
where $\kappa _{t}=-\frac{u_{t}}{u_{s}}.$ Because $f$ is strictly concave
(Lemma \ref{lemma_concavity}), $(s^{\ast }(z),t^{\ast }(z))$ is uniquely
determined by (\ref{constraint}) and (\ref{FOC}) given Assumption F.

Finally, we need to show that $(s^{\ast }(z),t^{\ast }(z))$ is strictly
increasing. Because of Lemma \ref{lemma_concavity}, $f$ is strictly
supermodular in $(t,z).$ Therefore, $f_{t}(t^{\ast }(z),z^{\prime })>0$ for
all $z^{\prime }>z,$ which implies that $t^{\ast }(z)<t^{\ast }(z^{\prime })$
for all $z^{\prime }>z$ given the strict concavity of $f$ in $t$. This also
implies that $s^{\ast }(z)<s^{\ast }(z^{\prime })$ for all $z<z^{\prime }$.
\end{proof}

Since $f_{t}$ is continuous and differentiable, with $f_{tt}<0$ (Lemma \ref%
{lemma_concavity}), by the implicit function theorem, $t^{\ast }(z)$ is
continuously differentiable. We normalize
\begin{equation}
g(t^{\ast }(\underline{z}),n(\underline{z}),s^{\ast }(\underline{z}),%
\underline{z})=0.  \label{g_normalized}
\end{equation}%
If $t_{\ell }\leq t^{\ast }(\underline{z}),$ then the lowest worker and firm
types who enter the market are $z_{\ell }=\underline{z}$ and $x_{\ell }=%
\underline{x}=n(\underline{z})$ and they choose $t^{\ast }(\underline{z})$
and $s^{\ast }(\underline{z})$ respectively as their wage and education
choices. Furthermore, $(t^{\ast }(\underline{z}),s^{\ast }(\underline{z}))$
is bilaterally efficient because it is a solution for the problem (\ref%
{maximization}). Because it is bilaterally efficient, there is no
informational rent at the bottom match.

Suppose that $t_{\ell }>t^{\ast }(\underline{z})$. Then, we need to consider
the new maximization problem in (\ref{maximization}) with an additional
constraint $t\geq t_{\ell }$. Because $t^{\ast }(z)$ is increasing and
continuous, there exists $z^{\circ }(t_{\ell })$ such that
\begin{equation*}
t^{\ast }(z^{\circ }(t_{\ell }))=t_{\ell }.
\end{equation*}
Because $t^{\ast }(z)$ is increasing in $z$ (Lemma \ref{lemma_increasing_t_s}
), we have that $t^{\ast }(z)<t_{\ell }$ for all $z<z^{\circ }(t_{\ell }).$
Because $f(t,z)$ is strictly concave in $t$, the additional constraint is
binding at the solution for the new maximization problem for all $z\leq
z^{\circ }(t_{\ell })$ and it slacks for all $z>z^{\circ }(t_{\ell })$.

Given $z\leq z^{\circ }(t_{\ell }),$ the maximum utility for the firm of
type $n(z)$ is $g(t_{\ell },n(z),\kappa (t_{\ell },z),\underline{z}).$
Recall that $\kappa (t_{\ell },z)$ satisfies (\ref{lemmaAe_1}). The left
hand side of (\ref{lemmaAe_1}) is nonnegative at $s=0$. It is also strictly
negative as $s\rightarrow \infty $ because of Assumption C. Therefore, by
the intermediate value theorem there exists $\kappa (t_{\ell },z)\in \mathbb{%
\ R}_{++}$ that satisfies (\ref{lemmaAe_1}). Furthermore, $\kappa (t_{\ell
},z)$ is continuous in $z$.\footnote{%
To show that $\kappa (t_{\ell },z)$ is continuous, suppose towards a
contradiction that $\kappa (t_{\ell },z)$ is not continuous. Then there is
some $\tilde{z}\in (\underline{z},\overline{z})$ such that $\kappa (t_{\ell
},\tilde{z})\neq \lim_{z\rightarrow \tilde{z}}\kappa (t_{\ell },z):=\tilde{s}%
.$ This leads to a contradiction since $0=\lim_{z\rightarrow \tilde{z}}\left[
u(t_{\ell },\kappa (t_{\ell },z),z)\right] =u(t_{\ell },\tilde{s},\tilde{z}%
))\neq 0.$ The first equality holds because $\kappa (t_{\ell },z)$ is a
solution to (\ref{lemmaAe_1}). The second equality is holds because $%
u(t_{\ell },\kappa (t_{\ell },z),z)$ is continuous by Assumption A.(iv).}

\paragraph{Proof of Lemma \protect\ref{lemma_bottom_types_e}}

Item 1 is clear. Let us prove item 2 where $t_{\ell }\in (t^{\ast }(
\underline{z}),\hat{t})$. Because $g(t^{\ast }(\underline{z}),n(\underline{z}
),s(\underline{z}),\underline{z})=0$ by the normalization and $t_{\ell
}>t^{\ast }(\underline{z})$, we have $g(t_{\ell },n(\underline{z}),\kappa
(t_{\ell },\underline{z}),\underline{z})<0$.

At $z=z^{\circ }(t_{\ell }),$ we have that
\begin{equation*}
g(t_{\ell },n(z^{\circ }(t_{\ell })),\kappa (t_{\ell },z^{\circ }(t_{\ell
})),z^{\circ }(t_{\ell }))=g(t^{\ast }(z^{\circ }(t_{\ell })),n(z^{\circ
}(t_{\ell })),s^{\ast }(z^{\circ }(t_{\ell })),z^{\circ }(t_{\ell }))>0,
\end{equation*}
where the equality holds because $t^{\ast }(z^{\circ }(t_{\ell }))=t_{\ell }$
and $s^{\ast }(z^{\circ }(t_{\ell }))=\kappa (t_{\ell },z^{\circ }(t_{\ell
}))$. Because $g(t^{\ast }(\underline{z}),n(\underline{z}),s^{\ast }(
\underline{z}),\underline{z})=0,$ the second inequality holds since $%
g(t^{\ast }(z),n(z),s^{\ast }(z),z)$ is strictly increasing in $z$ (by the
envelope theorem, Assumption B.(i), and Assumption D). Then, by the
intermediate value theorem, there exists a unique solution $z_{\ell }\in (%
\underline{z},\overline{z})$ that solves $g(t_{\ell },n(z),\kappa (t_{\ell
},z),z)=0.$ Because $u(t_{\ell },\kappa (t_{\ell },z),z)=0$ is embedded into
$\kappa (t_{\ell },z)$ as shown in (\ref{lemmaAe_1}), we have that $s_{\ell
}=\kappa (t_{\ell },z)$. Consequently, $\tau (s_{\ell })=t_{\ell }.$
Finally, we do not need to consider any type above $z\geq z^{\circ }(t_{\ell
})$ for the solution $z_{\ell }$ because $g(t^{\ast }(z)),n(z),s^{\ast
}(z),z)>0$ for all $z\geq z^{\circ }(t_{\ell }). $

\subsection{Proof of Lemma \protect\ref{lemma_well_behaved_e}}

First, we show that if there is a solution $(s_{h},z_{h})$ of (\ref%
{jumping_seller_e}) and (\ref{jumping_buyer_e}), it is unique. Consider the
following set for workers:
\begin{equation}
\left\{ (s,z)\in
%TCIMACRO{\U{211d} }%
%BeginExpansion
\mathbb{R}
%EndExpansion
_{++}\times Z:u(t_{h},s,z)=u(\tilde{\tau}\left( \tilde{\sigma}\left(
z\right) \right) ,\tilde{\sigma}\left( z\right) ,z)\right\} .  \label{set1_e}
\end{equation}%
Because $(s_{h},z_{h})$ must satisfy (\ref{jumping_seller_e}) and $s_{h}>%
\tilde{\sigma}\left( z_{h}\right) $ (Lemma \ref{lemma2_e}), $(s_{h},z_{h})$
must belong to the set in (\ref{set1_e}). Applying the envelope theorem for $%
\tilde{\sigma}$, to the total differential of $u(t_{h},s,z)=u(\tilde{\tau}%
\left( \tilde{\sigma}\left( z\right) \right) ,\tilde{\sigma}\left( z\right)
,z)$ yields the slope of the equation as
\begin{equation}
\frac{dz}{ds}=\frac{u_{s}(t_{h},s,z)}{u_{z}(t_{h},s,z)-u_{z}\left( \tilde{%
\tau}\left( \tilde{\sigma}\left( z\right) \right) ,\tilde{\sigma}\left(
z\right) ,z\right) }>0\text{ for all }s>\tilde{\sigma}\left( z\right) ,
\label{slope_sellers_e}
\end{equation}%
where the sign holds because $u_{s}<0$ and $u_{z}(t_{h},s,z)-u_{z}\left(
\tilde{\tau}\left( \tilde{\sigma}\left( z\right) \right) ,\tilde{\sigma}%
\left( z\right) ,z\right) <0$ due to Assumption A.(ii).

Consider the following set for firms:
\begin{equation}
\left\{ (s,z)\in
%TCIMACRO{\U{211d} }%
%BeginExpansion
\mathbb{R}
%EndExpansion
_{++}\times Z:%
\begin{array}{c}
\mathbb{E}[g(t_{h},s,z^{\prime },n\left( z\right) )|z^{\prime }\geq z]=g(%
\tilde{\tau}\left( \tilde{\sigma}\left( z\right) \right) ,\tilde{\sigma}%
\left( z\right) ,z,n\left( z\right) ), \\
s>\tilde{\sigma}\left( z\right)%
\end{array}%
\right\} \text{.}  \label{set2_e}
\end{equation}%
Because $(s_{h},z_{h})$ must satisfy (\ref{jumping_buyer_e}) and $s_{h}>%
\tilde{\sigma}\left( z_{h}\right) $ (Lemma \ref{lemma2_e}), $(s_{h},z_{h})$
must belong to the set in (\ref{set2_e}). Applying the envelope theorem for $%
\tilde{\sigma},$ and $\tilde{\tau}$ to the total differential of $\mathbb{E}%
[g(t_{h},s,z^{\prime },n\left( z\right) )|z^{\prime }\geq z]=g(\tilde{\tau}%
\left( \tilde{\sigma}\left( z\right) \right) ,\tilde{\sigma}\left( z\right)
,z,n\left( z\right) )$, we can express the slope of the equation as
\begin{equation}
\frac{dz}{ds}=-\frac{\mathbb{E}g_{s}}{\mathbb{E}g_{x}n^{\prime }+\frac{%
\partial \mathbb{E}\left[ g|z^{\prime }>z\right] }{\partial z}-\left(
g_{x}n^{\prime }+g_{z}\right) }\leq 0\text{ for all }s>\tilde{\sigma}\left(
z\right) ,  \label{slope_buyers_e}
\end{equation}%
where the sign holds because $-\mathbb{E}g_{s}\leq 0$ ($g_{s}\geq 0$
according to Assumption B.(i)) and the denominator is positive due to
Assumptions B and D.

(\ref{slope_sellers_e}) and (\ref{slope_buyers_e}) imply that the two sets
in (\ref{set1_e}) and (\ref{set2_e}) have at most one element in common.
This implies that if there is a solution $(s_{h},z_{h})$ that solves (\ref%
{slope_sellers_e}) and (\ref{slope_buyers_e}), it must be unique.

Now, we demonstrate the existence of solution. (\ref{jumping_seller_e}) and
( \ref{jumping_buyer_e}) together become
\begin{equation}
F(z,s)=0,  \label{lemmaA-4_e}
\end{equation}
where $F\left( z,s\right) :=\mathbb{E}[g(t_{h},n\left( z\right) ,s,z^{\prime
})|z^{\prime }\geq z]-g\left( \tilde{\tau}\left( \tilde{\sigma}\left(
z\right) \right) ,n\left( z\right) ,\tilde{\sigma}\left( z\right) ,z\right)
+u\left( t_{h},s,z\right) -u(\tilde{\tau}\left( \tilde{\sigma}\left(
z\right) \right) ,\tilde{\sigma}\left( z\right) ,z).$ Note that $F\left(
z,s\right) $ is strictly positive at $s=\tilde{\sigma}\left( z\right) $
because
\begin{equation*}
\mathbb{E}[g(t_{h},n\left( z\right) ,\tilde{\sigma}\left( z\right)
,z^{\prime })|z^{\prime }\geq z]>g\left( \tilde{\tau}\left( \tilde{\sigma}
\left( z\right) \right) ,n\left( z\right) ,\tilde{\sigma}\left( z\right)
,z\right)
\end{equation*}
and is negative as $s\rightarrow \infty $ by Assumption C. Because $F\left(
z,s\right) $ is continuously differentiable on $\left( z_{\ell },\overline{z}
\right) \times (s_{\ell },\tilde{\sigma}(\bar{z}))$ given Assumptions A.
(iv) and B.(iv), this implies that there exists $s(z)\in (\tilde{\sigma}
\left( z\right) ,\tilde{\sigma}(\bar{z}))$ for each $\left( z_{\ell },
\overline{z}\right) $ that satisfies (\ref{lemmaA-4_e}) by the intermediate
value theorem.

To show $s(z)$ is continuously differentiable, for an arbitrary small $%
\varepsilon >0,$ pick a point $\left( \overline{z}-\varepsilon ,\tilde{%
\sigma }\left( \overline{z}-\varepsilon \right) \right) $ in the rectangle $%
\left( z_{\ell },\overline{z}\right) \times (s_{\ell },\tilde{\sigma}(\bar{z}%
)).$ Then, as $\varepsilon $ gets closer to zero, we have
\begin{multline*}
F\left( \overline{z}-\varepsilon ,\tilde{\sigma}\left( \overline{z}
-\varepsilon \right) \right) =\mathbb{E}[g(t_{h},n\left( \overline{z}
-\varepsilon \right) ,\tilde{\sigma}\left( \overline{z}-\varepsilon \right)
,z^{\prime })|z^{\prime }\geq \overline{z}-\varepsilon ] \\
-g\left( \tilde{\tau}\left( \tilde{\sigma}\left( \overline{z}-\varepsilon
\right) \right) ,n\left( \overline{z}-\varepsilon \right) ,\tilde{\sigma}
\left( \overline{z}-\varepsilon \right) ,\overline{z}-\varepsilon \right)
+u\left( t_{h},\tilde{\sigma}\left( \overline{z}-\varepsilon \right) ,
\overline{z}-\varepsilon \right) \\
-u\left( \tilde{\tau}\left( \tilde{\sigma}\left( \overline{z}-\varepsilon
\right) \right) ,\tilde{\sigma}\left( \overline{z}-\varepsilon \right) ,
\overline{z}-\varepsilon \right) =0.
\end{multline*}
By continuity of $g_{s}$, $n$, $\tilde{\sigma}$, and $u_{s}$, we also have
\begin{multline*}
F_{s}\left( \overline{z}-\varepsilon ,\tilde{\sigma}\left( \overline{z}
-\varepsilon \right) \right) =\mathbb{E}[g_{s}(t_{h},n\left( \overline{z}
-\varepsilon \right) ,\tilde{\sigma}\left( \overline{z}-\varepsilon \right)
,z^{\prime })|z^{\prime }\geq \overline{z}-\varepsilon ] \\
+u_{s}\left( t_{h},\tilde{\sigma}\left( \overline{z}-\varepsilon \right) ,
\overline{z}\right) <0
\end{multline*}
according to Assumption C. Therefore, according to the implicit function
theorem there exists an open rectangle $I\times J\subset \left( z_{\ell },
\overline{z}\right) \times (s_{\ell },\tilde{\sigma}(\bar{z}))$ that
contains the point $\left( \overline{z}-\varepsilon ,\tilde{\sigma}\left(
\overline{z}-\varepsilon \right) \right) $, and a unique continuously
differentiable function $s(z)\in J$ defined on $I$ that satisfies (\ref%
{lemmaA-4_e}). Because $u$, $\tilde{\tau}$, $\tilde{\sigma}$, and $s(z)$ are
continuous, $\Lambda \left( z,s(z)\right) :=u(t_{h},s,z)-u(\tilde{\tau}
\left( \tilde{\sigma}\left( z\right) \right) ,\tilde{\sigma}\left( z\right)
,z)$ is also continuous. Because $u\left( t_{h},s(\overline{z}),\overline{z}
\right) <u\left( \tilde{\tau}\left( \tilde{\sigma}\left( \overline{z}\right)
\right) ,\tilde{\sigma}\left( \overline{z}\right) ,\overline{z}\right) $,
and $u\left( t_{h},s(z_{\ell }),z_{\ell }\right) >u\left( \tilde{\tau}\left(
\tilde{\sigma}\left( z_{\ell }\right) \right) \tilde{\sigma}\left( z_{\ell
}\right) ,z_{\ell }\right) ,$ there exists $z_{h}\in (z_{\ell },\overline{z}
) $ such that $\Lambda \left( z_{h},s(z_{h})\right) =0$ by the intermediate
value theorem. Because $F(z_{h},s(z_{h}))=0,$ $\Lambda \left(
z_{h},s(z_{h})\right) =0$ implies that $\mathbb{E}[g(t_{h},n\left(
z_{h}\right) ,s(z_{h}),z^{\prime })|z^{\prime }\geq z_{h}]=g\left( \tilde{
\tau}\left( \tilde{\sigma}\left( z_{h}\right) \right) ,n\left( z_{h}\right)
, \tilde{\sigma}\left( z_{h}\right) ,z_{h}\right) .$

\subsection{Proof of Theorem \protect\ref{thm_unique_well_behaved_e}\label%
{App_thm_unique_well_behaved_e}}

Theorem \ref{thm_unique_separating_eq_e} and Lemma \ref{lemma_bottom_types_e}
and \ref{lemma_well_behaved_e} establish the existence of a unique
(strictly) well-behaved stronger monotone CSE.

To show that there is no other stronger monotone CSE, note that there is no
separating CSE with $T=[t_{\ell },t_{h}]$ with $\underline{t}\leq t_{\ell
}<t_{h}<\tilde{\tau}\left( \tilde{\sigma}\left( \overline{z}\right) \right)
. $ Therefore, it is sufficient to show that there is no pooling CSE because
of Lemma \ref{theorem_all_eq_w/o_separating_e}. On contrary, suppose that
there exists a pooling CSE. Because of Lemma \ref%
{lemma_binding_upper_bound_e}, $t_{h}$ is the equilibrium wage for workers
with pooled education $s^{\ast }.$ Therefore, $x^{\ast }=n(z^{\ast })$ and
the following system of equations is satisfied in a pooling CSE:
\begin{gather}
u(t_{h},s^{\ast },z^{\ast })\geq 0,  \label{bottom_seller1_0} \\
\mathbb{E}\left[ g\left( t_{h},n\left( z^{\ast }\right) ,s^{\ast },z^{\prime
}\right) |z^{\prime }\geq z^{\ast }\right] \geq 0,\text{ }
\label{bottom_buyer1_0}
\end{gather}%
where both inequalities hold with equality if $z^{\ast }>\underline{z}$.

Suppose that $z^{\ast }>\underline{z}$. Then, (\ref{bottom_seller1_0}) and ( %
\ref{bottom_buyer1_0}) hold with equality. Further, because both $t_{h}$ and
$z^{\ast }$ are positive, $s^{\ast }$ must be positive from (\ref%
{bottom_seller1_0}) with equality. On the other hand, there should be no
profitable downward deviation for workers. Therefore,
\begin{eqnarray*}
g(t^{\prime },n(z^{\ast }),s,z^{\ast })+u(t^{\prime },s,z^{\ast }) &\leq &%
\mathbb{E}\left[ g(t_{h},n\left( z^{\ast }\right) ,s^{\ast },z^{\prime
})|z^{\prime }\geq z^{\ast }\right] +u(t_{h},s^{\ast },z^{\ast })\text{ } \\
\text{for all }t^{\prime } &\in &T\text{ and all }s<s^{\ast }.
\end{eqnarray*}%
Because (\ref{bottom_seller1_0}) and (\ref{bottom_buyer1_0}) hold with
equality, this becomes
\begin{equation}
g(t^{\prime },n(z^{\ast }),s,z^{\ast })+u(t^{\prime },s,z^{\ast })\leq 0%
\text{ for all }t^{\prime }\in T\text{ and each }s<s^{\ast }.
\label{no_pooling}
\end{equation}

Given Assumption E, there exists $t^{\prime }\in T$ and $s<s^{\ast }$ that
violates (\ref{no_pooling}) because $g$ and $u$ are continuous. Therefore,
if there is a stronger monotone pooling CSE, it must be the case where $%
z^{\ast }=\underline{z}.$ In this case, $s^{\ast }=0$. Otherwise, the worker
type $z$ arbitrarily close to $0$ will get negative utility because $%
t_{h}<\infty $ and $\lim_{z\rightarrow \underline{z}}u(t,s,z)=-\infty $ for
all $s>0$ (Assumption E).

Given $z^{\ast }=\underline{z}$ and $s^{\ast }=0,$ every worker will get
positive utility upon being matched and will enter the market. Given
Assumption E, $s^{\ast }=0$\ makes $\mathbb{E}\left[ g\left( t_{h},n\left(
z^{\ast }\right) ,s^{\ast },z^{\prime }\right) |z^{\prime }\geq z^{\ast }%
\right] -t_{h}<0$\ because $\mathbb{E}\left[ g\left( n\left( z^{\ast
}\right) ,s^{\ast },z^{\prime }\right) |z^{\prime }\geq z^{\ast }\right] =0$
\ given $z^{\ast }=\underline{z}$, $n\left( z^{\ast }\right) =\underline{x}$
, and $s^{\ast }=0.$\ Because $\mathbb{E}\left[ g\left( n\left( z^{\ast
}\right) ,s^{\ast },z^{\prime }\right) |z^{\prime }\geq z^{\ast }\right] $\
is continuous in $x,$\ this implies that there is a positive measure of firm
types close to $\underline{x}$\ whose expected utility is negative.
Therefore the market clearing condition is violated and hence there is no
pooling CSE.

\newpage

\pagenumbering{arabic} \setcounter{page}{1}

\setcounter{section}{0}
\renewcommand\thesection{\arabic{section}}
\renewcommand\thesubsection{\thesection\arabic{subsection}}\renewcommand{\theequation}{S\arabic{equation}}%
\setcounter{equation}{0}

\begin{center}
\textbf{\LARGE Supplemental Material}

\bigskip

\textbf{{\LARGE for \textquotedblleft \textbf{Optimal Wage Band for Job Matching}}}

\bigskip

\textbf{{\LARGE \textbf{with Signaling\textquotedblright\ }}}

\bigskip
\end{center}

This supplemental appendix provides omitted proofs in \textquotedblleft Optimal Wage Band for Job Matching with Signaling\textquotedblright\ by Seungjin Han, Alex Sam, and Youngki Shin.

\section{Proof of Lemma \protect\ref{lemma_no_bottom_bunching_e}}

Let $Z(s)$ be the set of the types of workers who choose the same education $%
s$ and it has a positive measure. We start with the case where there exists $%
\max Z(s)$. Let $z^{\circ }:=\max Z(s).$ We first show that $z^{\circ }=%
\overline{z}$. Let $x^{\circ }:=\max X(s)$, where $X(s)$ be the set of types
of firms who are matched with a worker with $s$ in equilibrium. We prove by
contradiction. Suppose that bunching does not happen on the top, i.e., $%
z^{\circ }<\overline{z}$. Then we have that
\begin{equation}
s=\sigma (z^{\circ })\leq \lim_{z\searrow z^{\circ }}\sigma (z)
\label{no_bottom_bunching1_e}
\end{equation}%
This is due to the monotonicity of $\sigma $ in Lemma 1.(i) in Han, Sam, and
Shin (2024). We like to show that (\ref{no_bottom_bunching1_e}) holds with
strict inequality, i.e., $s<\lim_{z\searrow z^{\circ }}\sigma (z)$. In
equilibrium, we have that for any $z>z^{\circ },$
\begin{gather}
u(\tau (s),s,z^{\circ })\geq \lim_{z\searrow z^{\circ }}\left[ u(\tau
(\sigma (z)),\sigma (z),z^{\circ })\right]  \label{no_bottom_bunching1_0_e}
\\
\mathbb{E[}g(\tau (s),s,z^{\prime },x^{\circ })|z^{\prime }\in Z(s)]\geq
\lim_{z\searrow z^{\circ }}\mathbb{E[}g(\tau (\sigma (z)),\sigma
(z),z^{\prime \prime },x^{\circ })|z^{\prime \prime }\in Z(\sigma (z))].
\label{no_bottom_bunching1_1'_e}
\end{gather}%
For any $\sigma (z)\geq s$, we have that $z^{\prime \prime }\geq z^{\circ
}=\max Z(s)$ for any $z^{\prime \prime }\in Z(\sigma (z))$ because of the
monotonicity of $\sigma $ (Lemma 1.(i) in Han, Sam, and Shin (2024)).
Further $Z(s)$ has a positive measure. Therefore, the monotonicity of $g$ in
Assumption B.(i) implies that, for any $z>z^{\circ }$
\begin{equation}
\mathbb{E[}g(\tau (s),s,z^{\prime },x^{\circ })|z^{\prime }\in Z(s)]<\mathbb{%
\ E[}g(\tau (s),\sigma (z),z^{\prime \prime },x^{\circ })|z^{\prime \prime
}\in Z(\sigma (z))].  \label{no_bottom_bunching1_2_e}
\end{equation}%
(\ref{no_bottom_bunching1_1'_e}) and (\ref{no_bottom_bunching1_2_e}) imply
that
\begin{equation}
\tau (s)<\lim_{z\searrow z^{\circ }}\tau (\sigma (z)).
\label{no_bottom_bunching1_3_e}
\end{equation}%
Because $u$ is decreasing in $s$ (Assumption A.(i)), (\ref%
{no_bottom_bunching1_0_e}) and (\ref{no_bottom_bunching1_3_e}) induce that
\begin{equation}
s=\sigma (z^{\circ })<\lim_{z\searrow z^{\circ }}\sigma (z).
\label{no_bottom_bunching1_4}
\end{equation}%
Therefore, any $s^{\prime }\in (s,\lim_{z\searrow z^{\circ }}\sigma (z))$ is
not chosen in equilibrium given that the monotonicity of $\sigma $.

The support of $\mu (s)$ is $Z(s)$. On the other hand, we have that $%
\lim_{z\searrow z^{\circ }}\inf $ supp$(\mu (\sigma (z)))=z^{\circ }$. This
implies that there is the unique stronger monotone posterior belief on the
worker's type conditional on any $s^{\prime }\in (s,\lim_{z\searrow z^{\circ
}}\sigma (z))$ and it is equal to $\mu (s^{\prime })=z^{\circ }$.

Suppose that the worker of type $z^{\circ }$ deviates to education $%
s+\epsilon \in (s,\lim_{z\searrow z^{\circ }}\sigma (z))$. A firm of type $x$
who is currently matched with a worker with $s$ receives the matching
utility of $\mathbb{E[}g(\tau (s),x,s,z)|z\in Z(s)]$. Note that $\tau
(s)<t_{h}$ given ( \ref{no_bottom_bunching1_3_e}).

Because $g$ and $u$ are continuous and $Z(s)$ has a positive measure, we
have that
\begin{eqnarray}
\lim_{\epsilon ,\varepsilon \searrow 0}g(\tau (s)+\varepsilon ,s+\epsilon
,z^{\circ },x) &>&\mathbb{E[}g(\tau (s),s,z,x)|z\in Z(s)]
\label{no_bottom_bunching2_e} \\
\lim_{\epsilon ,\varepsilon \searrow 0}u(\tau (s)+\varepsilon ,s+\epsilon
,z^{\circ }) &=&u(\tau (s),s,z^{\circ }).  \label{no_bottom_bunching3_e}
\end{eqnarray}%
Because $g$ and $u$ are continuous, (\ref{no_bottom_bunching2_e}) and (\ref%
{no_bottom_bunching3_e}) imply that there is a profitable worker deviation
for the worker of type $z^{\circ }$ for some values of $\varepsilon \in
(0,t_{h}-\tau (s))$ and $\epsilon >0$. This contradicts that $s$ is an
equilibrium education chosen by all workers whose types are in $Z(s)$.

We can analogously prove that there exists a profitable worker deviation if $%
z^{\circ }<\overline{z}$ when $z^{\circ }$ is defined as $\sup Z(s)$ rather
than $\max Z(s)$. Assumption D implies that there is no atom in the worker
type distribution. Therefore, $Z(s)$ is an interval with $\max Z(s)=%
\overline{z}$ due to the monotonicity of $\sigma $ (Lemma 1.(i) in Han, Sam,
and Shin (2024)).

\section{Proof of Lemma \protect\ref{lemma_binding_upper_bound_e}}

Let $z^{\ast }$ be the minimum of $Z(s)$ (We can analogously prove the lemma
for the case where $z^{\ast }$ is infimum of $Z(s)$). If $Z(s)=[z^{\ast },%
\overline{z}]$ has a positive measure, we have that $z^{\ast }<\overline{z}$
given Assumption D on $G.$ Let $t^{\ast }$ be the wage to education $s$
chosen by the positive measure of workers. We prove by contradiction.

On the contrary, suppose that $\tau (s)<t_{h}$ in a stronger monotone CSE.
Because type $\overline{z}$ is one of workers who choose $s$ and $\overline{z%
}$ is the maximum of worker types, the stronger monotonicity of $\mu $
implies that $\mu (s^{\prime })=\overline{z}$ for any $s^{\prime }>s.$

Because $g$ and $u$ are continuous in the worker's education, $g$ is
increasing in $z$, and $Z(s)$ has a positive measure, we have that
\begin{eqnarray}
\lim_{\epsilon ,\varepsilon \searrow 0}g(\tau (s)+\varepsilon ,s+\epsilon ,%
\overline{z},x) &>&\mathbb{E[}g(\tau (s),s,z,x|z^{\ast }\leq z\leq \overline{%
z}]  \label{PSD_simple0_1_e} \\
\lim_{\epsilon ,\varepsilon \searrow 0}u(\tau (s)+\varepsilon ,s+\epsilon ,%
\overline{z}) &=&u(\tau (s),s,\overline{z}).  \label{PSD_simple0_2_e}
\end{eqnarray}

Because $g$ and $u$ are continuous, (\ref{PSD_simple0_1_e}) and (\ref%
{PSD_simple0_2_e}) imply that there is a profitable worker deviation for the
worker of type $\overline{z}$ for some values of $\varepsilon \in (0,\tau
(s)-t_{h})$ and $\epsilon >0$. This contradicts that $s$ is an equilibrium
education chosen by all workers whose types are in $Z(s)$. Therefore, the
only way to prevent such an upward deviation by the worker is $\tau
(s)=t_{h} $.

\section{Proof of Lemma \protect\ref{lemma2_e}\label{Appendix_lemma2_e}}

When all workers of type $z_{h}$ or above choose the same education $s_{h}$
in equilibrium, we have that for all $z>z_{h}$
\begin{equation}
u(t_{h},s_{h},z)\geq u(\tilde{\tau}\left( \tilde{\sigma}\left( z_{h}\right)
\right) ,\tilde{\sigma}\left( z_{h}\right) ,z).  \label{jumping_sellers20}
\end{equation}%
Since (\ref{jumping_seller_e}) holds at $(s_{h},z_{h})$, (\ref%
{jumping_seller_e}) and (\ref{jumping_sellers20}) imply that for all $%
z>z_{h},$
\begin{equation*}
u(t_{h},s_{h},z)-u(\tilde{\tau}\left( \tilde{\sigma}\left( z_{h}\right)
\right) ,\tilde{\sigma}\left( z_{h}\right) ,z)\geq u(t_{h},s_{h},z_{h})-u(%
\tilde{\tau}\left( \tilde{\sigma}\left( z_{h}\right) \right) ,\tilde{\sigma}%
\left( z_{h}\right) ,z_{h}),
\end{equation*}%
which implies that $s_{h}\geq \tilde{\sigma}\left( z_{h}\right) $ by the
strict supermodularity of $u$ (Assumption A.(ii)). Because $s_{h}\geq \tilde{%
\sigma}\left( z_{h}\right) ,$ both (\ref{jumping_seller_e}) and (\ref%
{jumping_sellers20}) imply that $t_{h}\geq \tilde{\tau}\left( \tilde{\sigma}%
\left( z_{h}\right) \right) .$

Because $s_{h}\geq \tilde{\sigma}\left( z_{h}\right) ,$ we have that
\begin{equation}
\mathbb{E}[g(t_{h},s_{h},n\left( z_{h}\right) ,z^{\prime })|z^{\prime }\geq
z_{h}]>g(\tilde{\tau}\left( \tilde{\sigma}\left( z_{h}\right) \right) ,
\tilde{\sigma}\left( z_{h}\right) ,n\left( z_{h}\right) ,z_{h}).
\label{eq_strict}
\end{equation}
Note that (\ref{jumping_buyer_e}) at $(s,z)=(s_{h},z_{h})$ is written as
\begin{equation*}
\mathbb{E}[g(t_{h},s_{h},z^{\prime },n\left( z_{h}\right) )|z^{\prime }\geq
z_{h}]=g(\tilde{\tau}\left( \tilde{\sigma}\left( z_{h}\right) \right) ,
\tilde{\sigma}\left( z_{h}\right) ,z_{h},n\left( z_{h}\right) ),
\end{equation*}
which implies $t_{h}>\tilde{\tau}\left( \tilde{\sigma}\left( z_{h}\right)
\right) $ given (\ref{eq_strict}). If $t_{h}>\tilde{\tau}\left( \tilde{%
\sigma }\left( z_{h}\right) \right) ,$ (\ref{jumping_seller_e}) at $%
(s_{h},z_{h})$ implies that $s_{h}>\tilde{\sigma}\left( z_{h}\right) .$ For
all $z,$ let $s_{h}^{s}(z)$ be the value of $s$ that satisfies $u\left(
t_{h},s_{h}^{s}(z),z\right) =u\left( \tilde{\tau}\left( \tilde{\sigma}\left(
z\right) \right) ,\tilde{\sigma}\left( z\right) ,z\right) .$ Because $t_{h}<
\tilde{\tau}\left( \tilde{\sigma}\left( \bar{z}\right) \right) ,$ $%
s_{h}^{s}( \bar{z})<\tilde{\sigma}\left( \bar{z}\right) .$ Moreover, $%
s_{h}=s_{h}^{s}(z_{h})<s_{h}^{s}(\bar{z})$ because $z_{h}< \bar{z}$ and $%
s_{h}^{s}$ is increasing in $z.$ Therefore, we have that $s_{h}<\tilde{\sigma%
}\left( \bar{z}\right) .$

\section{Proof of Theorem \protect\ref{thm:pooling}}
Lemma \ref{lemma_pooling_e} implies that if the policymaker chooses a
singleton wage band, $t^{\ast }$ such that $\underline{t}<t^{\ast }<%
\overline{t}^{\ast }$, then a unique monotone CSE is a pooling equilibrium
where (\ref{pooling_sender_e} ) and (\ref{pooling_receiver_e}) hold with
equality.

What if the singleton wage band is $t^{\ast }=\underline{t}$? Because $%
g\left( \underline{t},\underline{x},0,z\right) =0$ for all $z\geq \underline{%
z}$ (part of Assumption E) and $g\left( \underline{t},\underline{x}%
,s,z\right) $ is non-decreasing in $s$ (Assumption B.(i)), we have that $%
\mathbb{E}\left[ g\left( \underline{t},n\left( z^{\ast }\right) ,s^{\ast
},z^{\prime }\right) |z^{\prime }\geq z^{\ast }\right] \geq 0$ for all $%
s^{\ast }\geq 0$ and $z^{\ast }\geq \underline{z}$. Since the firm's
reservation profit is zero, this implies that all firms enter the market.
Therefore, all workers should enter the market as well for market clearing
in equilibrium, i.e., $z^{\ast }=\underline{z}$. Then, we must have $s^{\ast
}=0$. Otherwise, market \editYS{clearing} condition is not satisfied because $%
\lim_{z\rightarrow \underline{z}}u(t,s,z)=-\infty $ for all $s>0$ (part of
Assumption E). Finally, given $t^{\ast }=\underline{t},$ $z^{\ast }=%
\underline{z}$ and $s^{\ast }=0$, we have that
\begin{equation*}
u(t^{\ast },s^{\ast },z^{\ast })=0
\end{equation*}%
because $u(\underline{t},0,\underline{z})=0$ (part of Assumption E) and
\begin{equation*}
\mathbb{E}\left[ g\left( \underline{t},n\left( z^{\ast }\right) ,s^{\ast
},z^{\prime }\right) |z^{\prime }\geq z^{\ast }\right] =0
\end{equation*}%
because $n\left( z^{\ast }\right) =\underline{x}$ and $g\left( \underline{t},%
\underline{x},0,z\right) =0$ for all $z\geq \underline{z}$ (part of
Assumption E).

\section{Proof of Lemma \protect\ref{lemma_pooling_e}}

For $s^{\ast }>0$ and $t^{\ast }>0,$ $(s^{\ast },z^{\ast })$ must satisfy
\begin{align}
u(t^{\ast },s,z)& =0, & &  \label{lem10} \\
\mathbb{E}\left[ g\left( t^{\ast },n\left( z\right) ,s,z^{\prime }\right)
|z^{\prime }\geq z\right] & =0. & & \text{ }  \label{lem20}
\end{align}%
First, we show that if there is a solution $(s^{\ast },z^{\ast })$ that
solves (\ref{lem10}) and (\ref{lem20}), it is unique. Consider the firm's
indifference curve
\begin{equation*}
\left\{ (s,z)\in
%TCIMACRO{\U{211d} }%
%BeginExpansion
\mathbb{R}
%EndExpansion
_{++}\times \left[ \underline{z},\overline{z}\right] :\mathbb{E}\left[
g\left( t^{\ast },n\left( z\right) ,s,z^{\prime }\right) |z^{\prime }\geq z%
\right] =0\right\} .
\end{equation*}%
The slope of this indifference curve is
\begin{equation}
\frac{dz}{ds}=-\frac{g_{s}}{\mathbb{E}g_{x}n^{\prime }+\frac{\partial
\mathbb{E}\left[ g|z^{\prime }>z\right] }{\partial z}}\leq 0.
\label{MRS_buyer_e}
\end{equation}%
Given $g_{s}\geq 0$, $g_{x}>0,$ $g_{z}>0$ (Assumptions B(i)), the sign above
holds because of $n^{\prime }>0$. Note that $n^{\prime }>0$ holds because of
Assumption D.

On the other hand, $\left\{ (s,z)\in
%TCIMACRO{\U{211d} }%
%BeginExpansion
\mathbb{R}
%EndExpansion
_{+}\times \left[ \underline{z},\overline{z}\right] :u(t^{\ast
},s,z)=0\right\} $ is the worker's indifference curve based on (\ref{lem20})
and its slope is
\begin{equation}
\frac{dz}{ds}=-\frac{u_{s}}{u_{z}}>0  \label{MRS_seller_e}
\end{equation}%
where the sign holds due to Assumption A(i). (\ref{MRS_buyer_e}) and (\ref%
{MRS_seller_e}) imply that the two indifference curves intersect at most
once and the intersection becomes a unique solution for the system of
equations, (\ref{lem10}) and (\ref{lem20}).

Now, we show the existence of a solution satisfying (\ref{lem10}) and (\ref%
{lem20}). Putting (\ref{lem10}) and (\ref{lem20}) together gives
\begin{equation}
\mathbb{E}\left[ g\left( t^{\ast },n\left( z\right) ,s,z^{\prime }\right)
|z^{\prime }\geq z\right] +u(t^{\ast },s,z)=0  \label{lemmaA-30}
\end{equation}
for all $\left( z,s\right) \in (\underline{z},\overline{z})\times
%TCIMACRO{\U{211d} }%
%BeginExpansion
\mathbb{R}
%EndExpansion
_{++}.$ Assumption A(v) and B(v) imply that the left hand side of (\ref%
{lemmaA-30}) is continuously differentiable in the rectangle $(\underline{z}%
, \overline{z})\times
%TCIMACRO{\U{211d} }%
%BeginExpansion
\mathbb{R}
%EndExpansion
_{++}.$ The left hand side of (\ref{lemmaA-30}) is nonnegative at $s=0$. It
is also strictly negative as $s\rightarrow \infty $ because of Assumption C.
Furthermore, we have that
\begin{multline*}
\max_{s}\left[ \mathbb{E}\left[ g\left( t^{\ast },n\left( z\right)
,s,z^{\prime }\right) |z^{\prime }\geq z\right] +u(t^{\ast },s,z)\right] > \\
\max_{s}\left[ \mathbb{E}\left[ g\left( t^{\ast },n\left( \underline{z}
\right) ,s,z^{\prime }\right) |z^{\prime }\geq \underline{z}\right]
+u(t^{\ast },s,\underline{z})\right] =0.
\end{multline*}
Therefore, by the intermediate value theorem there exists $s(z)\in \mathbb{R}
_{++}$ that satisfies (\ref{lemmaA-30}). Furthermore, $s(z)$ is continuous.
To show that $s(z)$ is continuous, suppose towards a contradiction that $%
s(z) $ is not continuous. Then there is some $\tilde{z}\in (\underline{z},
\overline{z})$ such that $s(\tilde{z})\neq \lim_{z\rightarrow \tilde{z}
}s(z):=\tilde{s}.$ This leads to a contradiction since $0=\lim_{z\rightarrow
\tilde{z}}\left[ \mathbb{E}\left[ g\left( t^{\ast },n\left( z\right)
,s(z),z^{\prime }\right) |z^{\prime }\geq z\right] +u(t^{\ast },s(z),z) %
\right] =\mathbb{E}\left[ g\left( t^{\ast },n\left( \tilde{z}\right) ,\tilde{%
		s},z^{\prime }\right) |z^{\prime }\geq \tilde{z}\right] +u(t^{\ast },%
\tilde{s	},\tilde{z}))\neq 0.$ The first equality holds because $s(z)$ is a
solution to (\ref{lemmaA-30}). The second equality is holds because
\begin{equation*}
\mathbb{E}\left[ g\left( t^{\ast },n\left( z\right) ,s,z^{\prime }\right)
|z^{\prime }\geq z\right] +u(t^{\ast },s(z),z)
\end{equation*}
is continuous by Assumption A.(iv) and B.(iv). Let $\Lambda (z):=\mathbb{E} %
\left[ g\left( t^{\ast },n\left( z\right) ,s(z),z^{\prime }\right)
|z^{\prime }\geq z\right] $. Since $g$ is continuous and $n(z)$ and $s(z)$
are continuous, $\Lambda (z)$ is continuous. Because $\Lambda (\underline{z}
)<0$ and $\Lambda (\overline{z})>0,$ there exists $z^{\ast }\in (\underline{%
z },\overline{z})$ such that $\Lambda (z^{\ast })=0$ by the intermediate
value theorem. Because of (\ref{lemmaA-30}), $\Lambda (z^{\ast })=0$ also
implies that $u(t^{\ast },s(z^{\ast }),z^{\ast })=0,$ so that both (\ref%
{lem10}) and (\ref{lem20}) holds with equality at $s^{\ast }=s(z^{\ast })$
and $z^{\ast }. $

\section{Proof of Lemma \protect\ref{lem_unified_well_behaved_e}}

It is clear that if $z_{h}\rightarrow \bar{z},$ $\left\{ \hat{\sigma},\hat{%
\mu},\hat{\tau},\hat{m}\right\} $ converges to the stronger monotone
separating equilibrium with the same lower threshold worker type $z_{\ell }$
as the pooling region vanishes. Note that as $z_{h}\rightarrow z_{\ell }$, $%
\lim_{z_{h}\rightarrow z_{\ell }}t_{h}=t_{\ell }$ and $\lim_{z_{h}%
\rightarrow z_{\ell }}s_{h}=\tilde{\sigma}(z_{\ell })=s_{\ell }$ because ( %
\ref{jumping_seller_e}) and (\ref{jumping_buyer_e}) are satisfied only when $%
(t_{h},s_{h})=(t_{\ell },\tilde{\sigma}(z_{\ell }))$. Therefore, combining ( %
\ref{lem1e}) and (\ref{lem2e}) with (\ref{jumping_seller_e}) and (\ref%
{jumping_buyer_e}) yields that
\begin{gather*}
\lim_{z_{h}\rightarrow z_{\ell }}u(t_{h},s_{h},z_{h})=u(t_{\ell },s_{\ell
},z_{\ell })=0, \\
\lim_{z_{h}\rightarrow z_{\ell }}\mathbb{E}[g(t_{h},n\left( z_{h}\right)
,s_{h},z^{\prime })|z^{\prime }\geq z_{h}]=\mathbb{E}[g(t_{\ell },n\left(
z_{\ell }\right) ,s_{\ell },z^{\prime })|z^{\prime }\geq z_{\ell }]=0,
\end{gather*}%
where the inequality holds with equality if $z_{\ell }>\underline{z}.$ The
second equality of the first line and the inequality of the second line are
a consequence of Theorem 6 in Han, Sam, and Shin (2024). These imply that $%
\left\{ \hat{\sigma},\hat{\mu},\hat{\tau},\hat{m}\right\} $ converges the
stronger monotone pooling equilibrium where $t_{\ell }$ is the single
feasible wage, $z_{\ell }$ is the threshold worker type for market entry and
$s_{\ell }$ is the pooled education for workers in the market. Note that
when $z_{\ell }=\underline{z},$ we have $s_{\ell }=0$ and $t_{\ell }=%
\underline{t}$ in $\left\{ \hat{\sigma},\hat{\mu},\hat{\tau},\hat{m}\right\}
$ due to Theorem 6 in Han, Sam, and Shin (2024).

\section{Proof of Proposition \protect\ref{prop_unbounded_design}}

For any given $z_{\ell }\in \lbrack \underline{z},\overline{z}),$ we show
the existence of a solution satisfying (\ref{lem1e}) and (\ref{lem2e}).
Putting (\ref{lem1e}) and (\ref{lem2e}) together gives
\begin{equation}
g\left( t,n\left( z_{\ell }\right) ,s,z_{\ell }\right) +u(t,s,z_{\ell })=0.
\label{lemmaA-30pd}
\end{equation}
Assumption A.(v) and B.(v) imply that the left hand side of (\ref%
{lemmaA-30pd}) is continuously differentiable. The left hand side of (\ref%
{lemmaA-30pd}) is nonnegative at $s=0$ (Assumption E.(ii)). It is also
strictly negative as $s\rightarrow \infty $ because of Assumption C.
Therefore, by the intermediate value theorem there exists $s(t)\in \mathbb{R}
_{++}$ that satisfies (\ref{lemmaA-30pd}). Furthermore, $s(t)$ is
continuous. To show that $s(t)$ is continuous, suppose towards a
contradiction that $s(t)$ is not continuous. Then there is some $\tilde{t}$
such that $s(\tilde{t})\neq \lim_{t\rightarrow \tilde{t}}s(t):=\tilde{s}.$
This leads to a contradiction since $0=\lim_{t\rightarrow \tilde{t}}\left[
g\left( t,n\left( z_{\ell }\right) ,s(t),z_{\ell }\right) +u(t,s(t),z_{\ell
})\right] =g\left( \tilde{t},n\left( z_{\ell }\right) ,\tilde{s},z_{\ell
}\right) +u(\tilde{t},\tilde{s},z_{\ell }))\neq 0.$ The first equality holds
because $s(t)$ is a solution to (\ref{lemmaA-30pd}). The second equality
holds because $g\left( t,n\left( z_{\ell }\right) ,s,z_{\ell }\right)
+u(t,s(t),z_{\ell })$ is continuous by Assumption A.(v) and B.(v). Let $%
\Lambda (t):=g\left( t,n\left( z_{\ell }\right) ,s(t),z_{\ell }\right) $.
Since $g$ is continuous and $n(\cdot )$ and $s(t)$ are continuous, $\Lambda
(t)$ is continuous. Because (i) $\Lambda (0)\geq 0$, (ii) $%
\lim_{t\rightarrow \infty }\Lambda (t)=-\infty ,$ and (iii) $\Lambda (t)$ is
decreasing in $t,$ there exists a unique $t_{\ell }$ such that $\Lambda
(t_{\ell })=0$ by the intermediate value theorem. Because of (\ref%
{lemmaA-30pd}), $\Lambda (t_{\ell })=0$ also implies that $u(t_{\ell
},s(t_{\ell }),z_{\ell })=0,$ so that both (\ref{lem1e}) and (\ref{lem2e})
holds with equality at $s_{\ell }=s(t_{\ell })$ and $t_{\ell }.$

Now let us prove the final statement of the proposition.
Fix $z_\ell \in [\underline{z}, \bar{z})$. We aim to show that a pair
$(s_\ell, t_\ell) \in \mathbb{R}_+ \times [t^*(\underline{z}), \hat{t})$ satisfies
equations \eqref{lem1e} and \eqref{lem2e} if and only if the pair
$(z_\ell, s_\ell)$ is a solution of the same system.
	
Suppose $(s_\ell, t_\ell) \in \mathbb{R}_+ \times [t^*(\underline{z}), \hat{t})$
satisfies equations \eqref{lem1e} and \eqref{lem2e}. Then, in
particular, the triple $(t_\ell, s_\ell, z_\ell)$ satisfies both equations.
Hence, for this fixed $z_\ell$, the pair $(z_\ell, s_\ell)$ admits a value
$t_\ell$ in $[t^*(\underline{z}), \hat{t})$ such that both \eqref{lem1e} and
\eqref{lem2e} hold. Therefore, $(z_\ell, s_\ell)$ is a solution.
	
Conversely, suppose $(z_\ell, s_\ell) \in [\underline{z}, \bar{z}) \times \mathbb{R}_+$
is such that there exists $t_\ell \in [t^*(\underline{z}), \hat{t})$ satisfying
\eqref{lem1e} and \eqref{lem2e}. We claim that this value $t_\ell$ is uniquely
determined by $(z_\ell, s_\ell)$, and that the resulting pair $(s_\ell, t_\ell)$ is a solution.
To see this, define the function
	\[
	\Gamma(t) := g(t, n(z_\ell), s_\ell, z_\ell) + u(t, s_\ell, z_\ell),
	\]
and note that \eqref{lem1e} and \eqref{lem2e} imply $\Gamma(t_\ell) = 0$
and $u(t_\ell, s_\ell, z_\ell) = 0$. Therefore,
	\[
	\Gamma(t_\ell) = g(t_\ell, n(z_\ell), s_\ell, z_\ell).
	\]
	
By Assumption B.(i), $g$ is strictly decreasing in $t$. Therefore,
for $t_{\ell}\geq t^*(\underline{z})$, we have $\Gamma(t^*(\underline{z})):=g(t^*(\underline{z}), n(z_\ell), s_\ell, z_\ell) > g(t_\ell, n(z_\ell), s_\ell, z_\ell)=0$, and for $t_{\ell}< \hat{t}$, we have $\Gamma(\hat{t}):=g(\hat{t}, n(z_\ell), s_\ell, z_\ell) < g(t_\ell, n(z_\ell), s_\ell, z_\ell)=0$.
Hence, by continuity of $g$, there exists a unique $t_\ell \in [t^*(\underline{z}), \hat{t})$ such that $g(t_\ell, n(z_\ell), s_\ell, z_\ell) = 0$.
This complete the proof.

\section{Proof of Proposition \protect\ref{prop_bounded_design}}

For any given $z_{h}\in (z_{\ell },\overline{z}),$ we show the existence of
a solution satisfying (\ref{jumping_seller_e}) and (\ref{jumping_buyer_e}).
Putting (\ref{jumping_seller_e}) and (\ref{jumping_buyer_e}) together gives
\begin{align}
u(t,s,z_{h})+\mathbb{E}[g(t,s,z^{\prime },n\left( z_{h}\right) )|z^{\prime
}& \geq z_{h}]-u(\tilde{\tau}\left( \tilde{\sigma}\left( z_{h}\right)
\right) ,\tilde{\sigma}\left( z_{h}\right) ,z_{h})  \notag \\
-g(\tilde{\tau}\left( \tilde{\sigma}\left( z_{h}\right) \right) ,\tilde{
\sigma}\left( z_{h}\right) ,z_{h},n\left( z_{h}\right) )& =0.
\label{lemmaAd-30pd}
\end{align}
Assumption A.(v) and B.(v) imply that the left hand side of (\ref%
{lemmaAd-30pd}) is continuously differentiable. The left hand side of (\ref%
{lemmaAd-30pd}) is strictly negative as $s\rightarrow \infty $ because of
Assumption C, and also because $u(\tilde{\tau}\left( \tilde{\sigma}\left(
z_{h}\right) \right) ,\tilde{\sigma}\left( z_{h}\right) ,z_{h})+g(\tilde{\tau%
}\left( \tilde{\sigma}\left( z_{h}\right) \right) , \tilde{\sigma}\left(
z_{h}\right) ,z_{h},n\left( z_{h}\right) )>0.$ Note that $t_{h}$ that
satisfies (\ref{jumping_seller_e}) and (\ref{jumping_buyer_e}) must satisfy $%
t_{h}>\tilde{\tau}\left( \tilde{\sigma} \left( z_{h}\right) \right) $ (Lemma %
\ref{lemma2_e}). At $s=\tilde{\sigma} \left( z_{h}\right) ,$ we have that
\begin{multline*}
u(t,\tilde{\sigma}\left( z_{h}\right) ,z_{h})+\mathbb{E}[g(t,\tilde{\sigma}
\left( z_{h}\right) ,z^{\prime },n\left( z_{h}\right) )|z^{\prime }\geq
z_{h}]> \\
u(\tilde{\tau}\left( \tilde{\sigma}\left( z_{h}\right) \right) ,\tilde{%
\sigma }\left( z_{h}\right) ,z_{h})+g(\tilde{\tau}\left( \tilde{\sigma}%
\left( z_{h}\right) \right) ,\tilde{\sigma}\left( z_{h}\right)
,z_{h},n\left( z_{h}\right) )
\end{multline*}
for $t>\tilde{\tau}\left( \tilde{\sigma}\left( z_{h}\right) \right) .$
Therefore, by the intermediate value theorem there a continuous exists $%
s(t)\in \mathbb{R}_{++}$ that satisfies (\ref{lemmaAd-30pd}). The continuity
of $s(t)$ comes from the fact that the left hand side of (\ref{lemmaAd-30pd}
) is continuous. Let $\Lambda (t):=u(t,s(t),z_{h})-u(\tilde{\tau}\left(
\tilde{\sigma}\left( z_{h}\right) \right) ,\tilde{\sigma}\left( z_{h}\right)
,z_{h})$. Since $u$ is continuous and $n(\cdot )$ and $s(t)$ are continuous,
$\Lambda (t)$ is continuous. Because (i) $\Lambda (\tilde{\tau}\left( \tilde{
\sigma}\left( z_{h}\right) \right) )<0$, (ii) $\lim_{t\rightarrow \infty
}\Lambda (t)>0$, and (iii) $\Lambda (t)$ is increasing in $t,$ there exists
a unique $t_{h}$ such that $\Lambda (t_{h})=0$ by the intermediate value
theorem. Because of (\ref{lemmaAd-30pd}), $\Lambda (t_{h})=0$ also implies
that (\ref{jumping_buyer_e}) holds with equality at $s_{h}=s(t_{h})$ and $%
t_{h}.$

Now let us prove the final statement in the proposition. Fix $z_h \in (z_\ell, \bar{z})$.
We aim to show that a pair $(s_h, t_h) \in \mathbb{R}_+ \times (t_\ell, t_h^*(t_\ell))$
satisfies equations \eqref{jumping_seller_e} and \eqref{jumping_buyer_e}
if and only if the pair $(z_h, s_h)$ is a solution of the same system.
		
Suppose $(s_h, t_h) \in \mathbb{R}_+ \times (t_\ell, t_h^*(t_\ell))$
satisfies \eqref{jumping_seller_e} and \eqref{jumping_buyer_e}.
Then, for this fixed $z_h$, the pair $(z_h, s_h)$ admits a
value $t_h$ such that both constraints are satisfied.
Hence, $(z_h, s_h)$ is a solution.
		
Conversely, suppose $(z_h, s_h) \in (z_\ell, \bar{z}) \times \mathbb{R}_+$ is
such that there exists a value $t_h \in (t_\ell, t_h^*(t_\ell))$ for
which equations \eqref{jumping_seller_e} and \eqref{jumping_buyer_e}
hold. We claim that this value $t_h$ is uniquely determined by $(z_h, s_h)$,
and that the resulting pair $(s_h, t_h)$ satisfies both constraints.
Define the function:
		\[
		\Gamma(t) := \mathbb{E}[g(t, s_h, z', n(z_h)) \mid z' \geq z_h] -
		g(\tilde{\tau}(\tilde{\sigma}(z_h)), \tilde{\sigma}(z_h), z_h, n(z_h)).
		\]
From equation \eqref{jumping_buyer_e}, we know that $\Gamma(t_h) = 0$.
By Assumption~B(i), $g$ is strictly decreasing in $t$ for all $z'$,
and by Assumptions~B(iv) and B(v), the conditional
expectation inherits strict monotonicity and continuity.
Thus, $\Gamma(t)$ is strictly
decreasing and continuous on $(t_\ell, t_h^*(t_\ell))$.
We now characterize the limits of $\Gamma(t)$ near the interval boundaries.
Because $t_h^*(t_\ell)$ is the lowest maximum wage
that induces no pooled education chosen at the top, by
continuity of $g$, it must be that for $t < t_h^*(t_\ell)$
sufficiently close, 	$\mathbb{E}[g(t, s_h, z', n(z_h)) \mid z' \geq z_h] -
g(\tilde{\tau}(\tilde{\sigma}(z_h)), \tilde{\sigma}(z_h), z_h, n(z_h))=0$.
That is such $t < t_h^*(t_\ell)$ must satisfy \eqref{jumping_buyer_e}.
Because $g$ is strictly decreasing in $t$,
for $t < t_h^*(t_\ell)$ sufficiently close, $\mathbb{E}[g(t_h^*(t_\ell), s, z', n(z))
\mid z' \geq z] - g(\tilde{\tau}(\tilde{\sigma}(z)), \tilde{\sigma}(z), z, n(z))<
\mathbb{E}[g(t, s_h, z', n(z_h)) \mid z' \geq z_h] -
g(\tilde{\tau}(\tilde{\sigma}(z_h)), \tilde{\sigma}(z_h), z_h, n(z_h))=0$.
This implies that $\Gamma(t_h^*(t_\ell)) <0$.
		
By the definition of $\mathcal{T}^{\ast}$, the wage $t_{h}$ lies
tightly within the closed interval $[t_{\ell}, t_h^*(t_\ell)]$. Thus,
for $t > t_{\ell}$ sufficiently close, the continuity of $g$ implies
that $\mathbb{E}[g(t, s_h, z', n(z_h)) \mid z' \geq z_h] -
g(\tilde{\tau}(\tilde{\sigma}(z_h)), \tilde{\sigma}(z_h), z_h, n(z_h))=0$.
That is such $t > t_\ell$ must satisfy \eqref{jumping_buyer_e}.
Because $g$ is strictly decreasing in $t$,
for $t >t_{\ell}$ sufficiently close, $\mathbb{E}[g(t_{\ell}, s, z', n(z)) \mid
z' \geq z] - g(\tilde{\tau}(\tilde{\sigma}(z)), \tilde{\sigma}(z), z, n(z))>
\mathbb{E}[g(t, s_h, z', n(z_h)) \mid z' \geq z_h] -
g(\tilde{\tau}(\tilde{\sigma}(z_h)), \tilde{\sigma}(z_h), z_h, n(z_h))=0$.
This implies that $\Gamma(t_\ell) >0$.
		
Therefore, by the Intermediate Value Theorem, there exists
a unique $t_h \in (t_\ell, t_h^*(t_\ell))$ such that $\Gamma(t_h) =
0$. Since the constraint \eqref{jumping_seller_e} is also
satisfied by construction, this uniquely determines $(s_h, t_h)$.
We conclude that $(s_h, t_h)$ is a solution to equations
\eqref{jumping_seller_e} and \eqref{jumping_buyer_e} if
and only if $(z_h, s_h)$ is a solution. This completes the proof.

\FloatBarrier

\section{Additional  Figures}

\begin{figure}[tbh]
    \centering
    \caption{Additional Figures: Baseline}
    \label{figure: more baseline}

    \begin{minipage}[t]{0.48\textwidth}
        \centering
        \includegraphics[width=\linewidth]{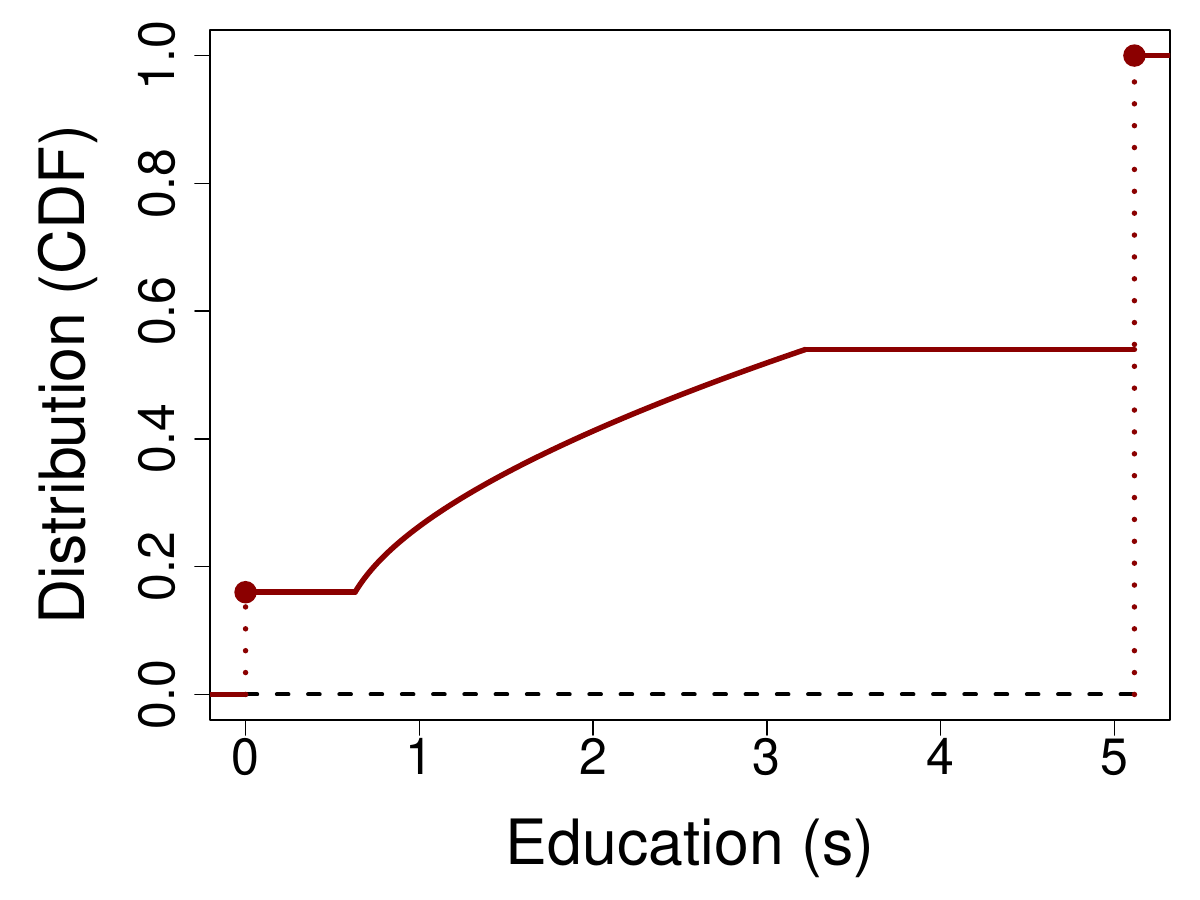}
    \end{minipage}
    \hfill
    \begin{minipage}[t]{0.48\textwidth}
        \centering
        \includegraphics[width=\linewidth]{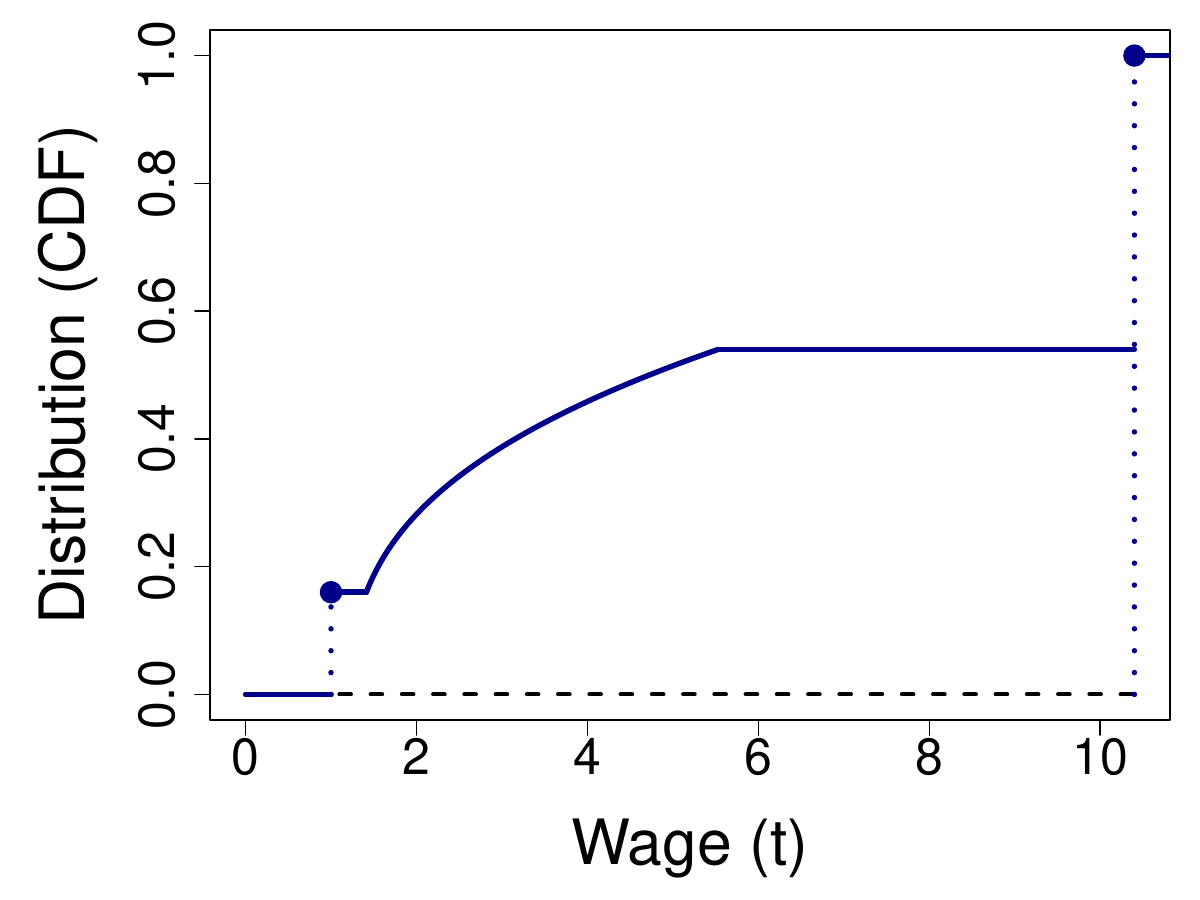}
    \end{minipage}
\end{figure}

\newpage

\begin{figure}[tbh]
    \centering
    \caption{Additional Comparative Statics: $a$}
    \label{figure: more comp statics a}

    \begin{minipage}[t]{0.48\textwidth}
        \centering
        \includegraphics[width=\linewidth]{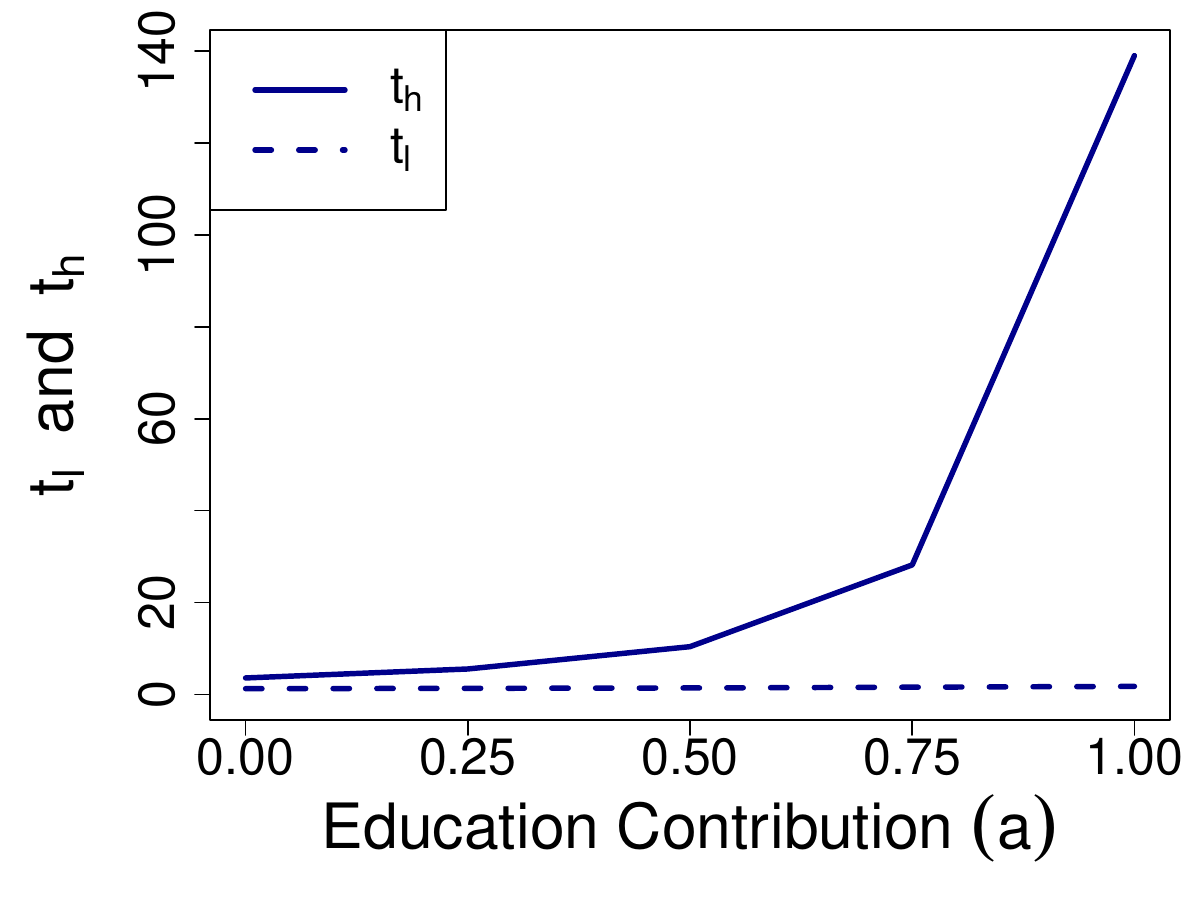}
    \end{minipage}
    \hfill
    \begin{minipage}[t]{0.48\textwidth}
        \centering
        \includegraphics[width=\linewidth]{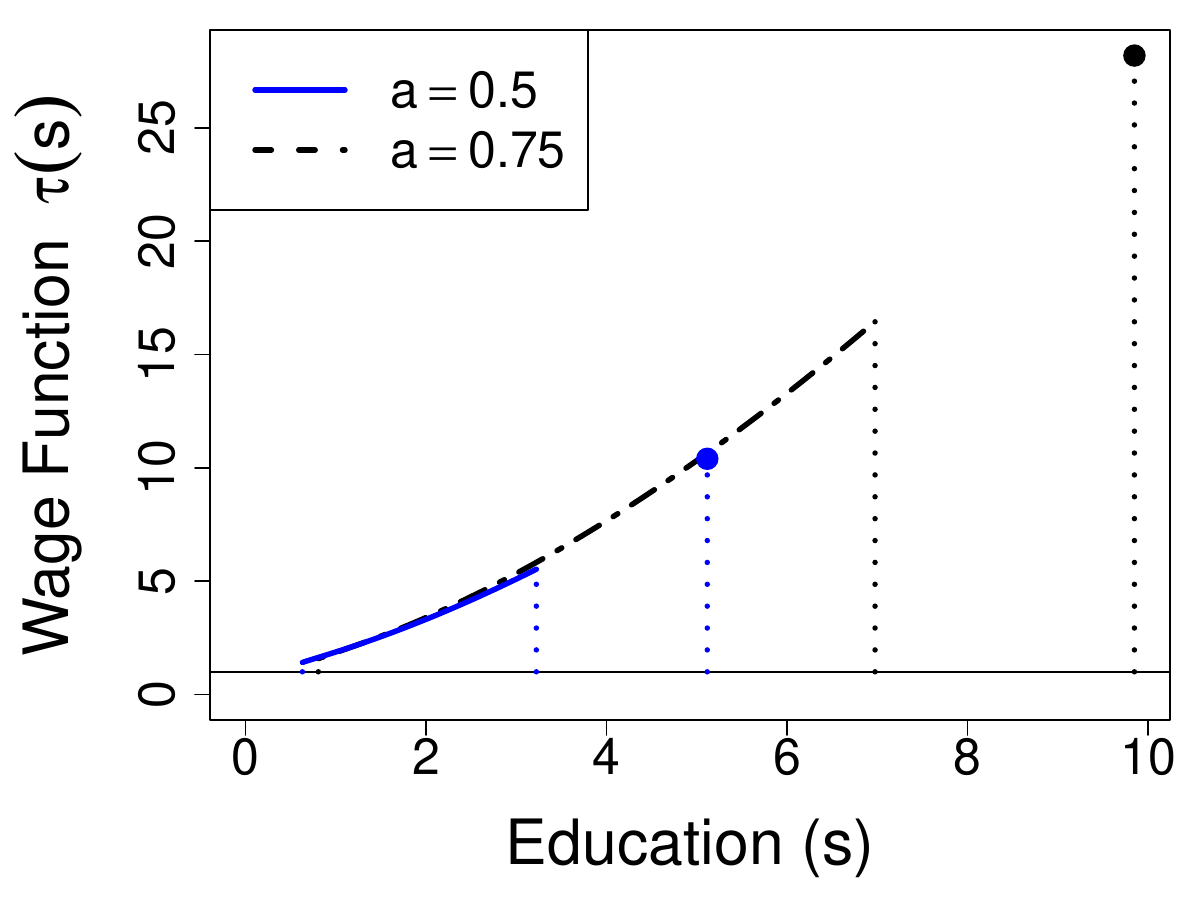}
    \end{minipage}
\end{figure}

\begin{figure}[tbh]
    \centering
    \caption{Additional Comparative Statics: $q$}
    \label{figure: more comp statics q}

    \begin{minipage}[t]{0.48\textwidth}
        \centering
        \includegraphics[width=\linewidth]{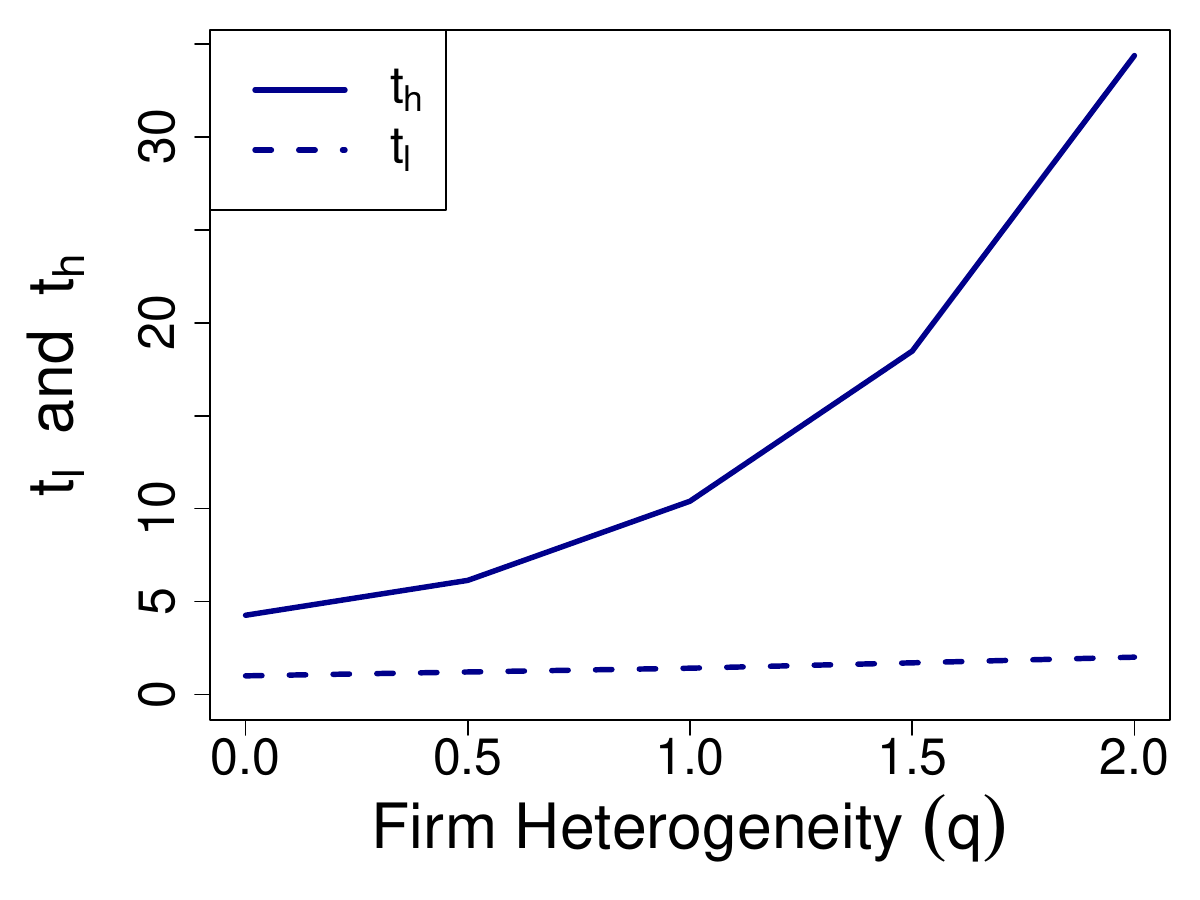}
    \end{minipage}
    \hfill
    \begin{minipage}[t]{0.48\textwidth}
        \centering
        \includegraphics[width=\linewidth]{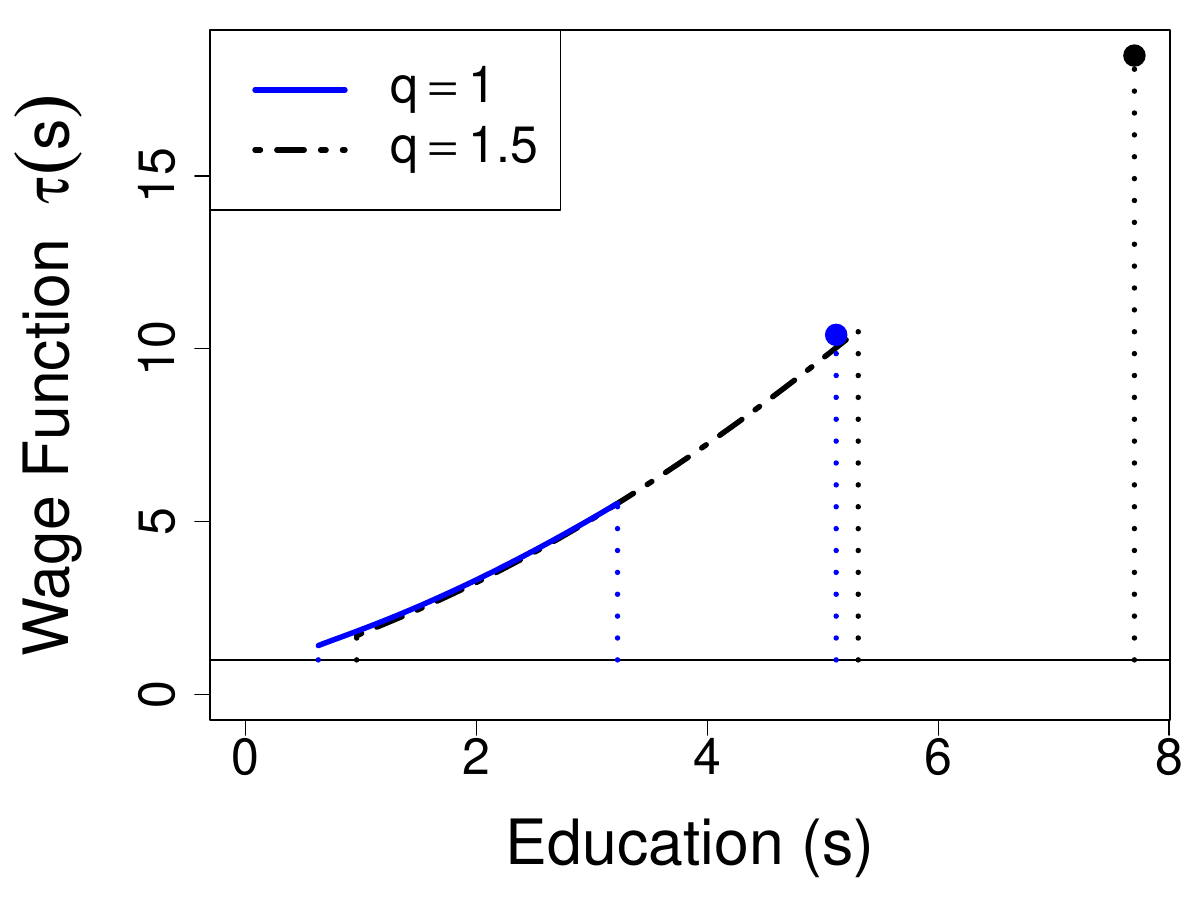}
    \end{minipage}
\end{figure}

\begin{figure}[tbh]
    \centering
    \caption{Additional Comparative Statics: $\rho$}
    \label{figure: more comp statics rho}

    \begin{minipage}[t]{0.48\textwidth}
        \centering
        \includegraphics[width=\linewidth]{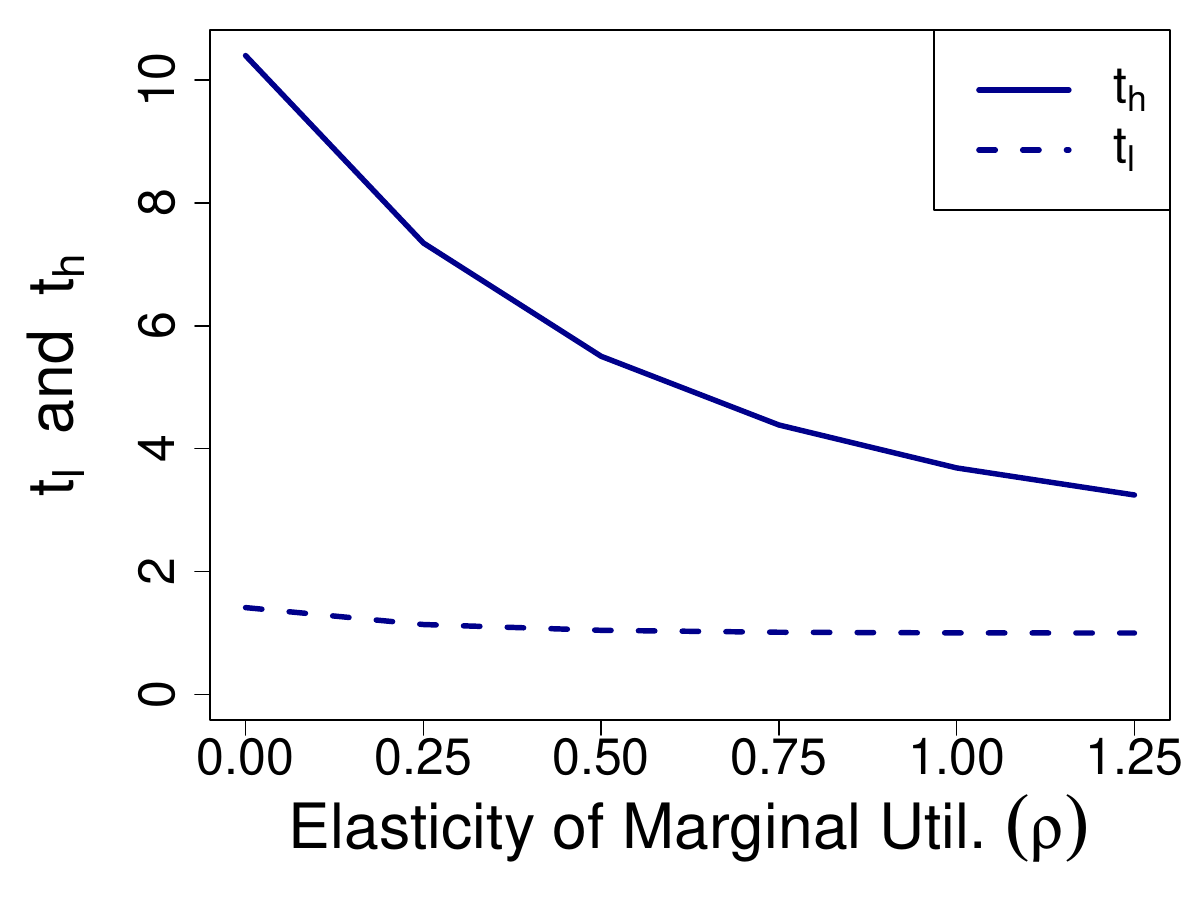}
    \end{minipage}
    \hfill
    \begin{minipage}[t]{0.48\textwidth}
        \centering
        \includegraphics[width=\linewidth]{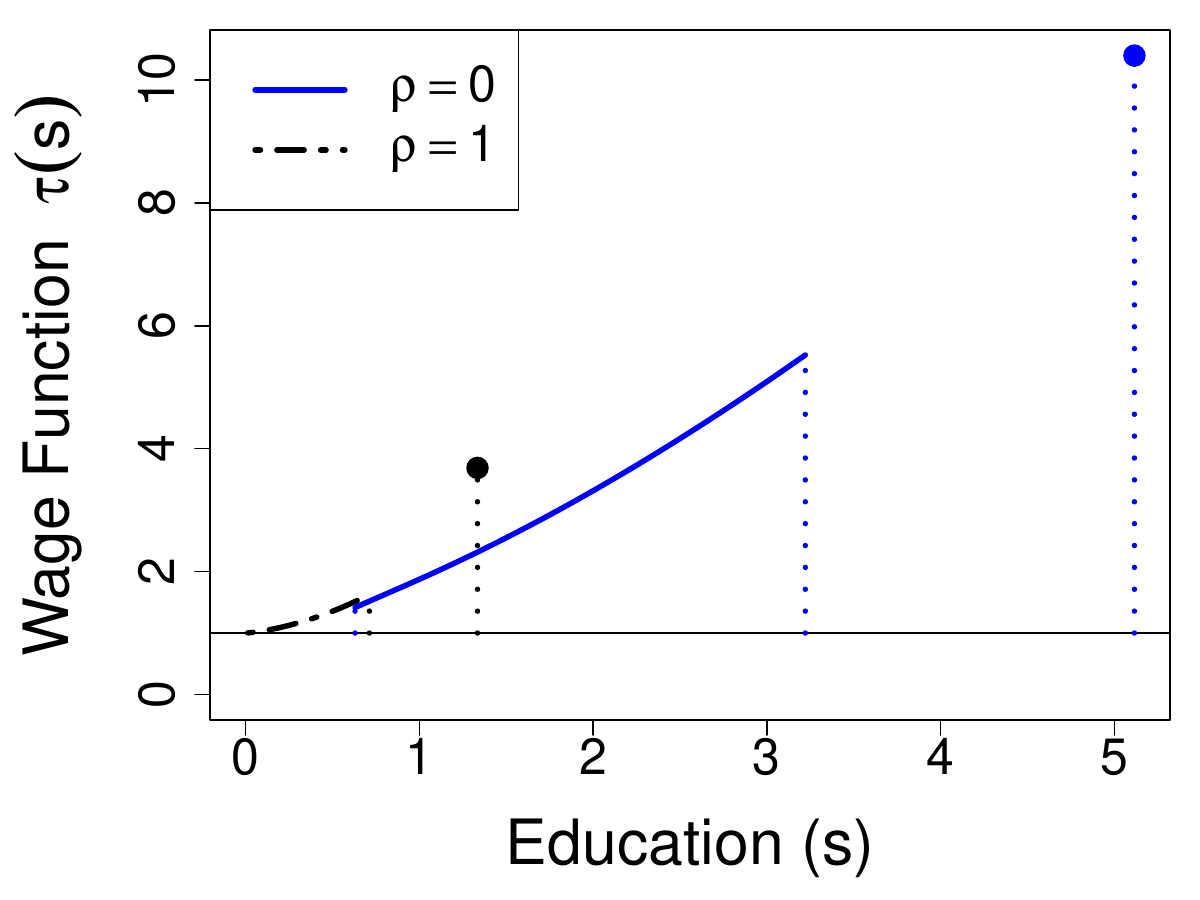}
    \end{minipage}
\end{figure}

\begin{figure}[tbh]
    \centering
    \caption{Additional Comparative Statics: $b$}
    \label{figure: more comp statics b}

    \begin{minipage}[t]{0.48\textwidth}
        \centering
        \includegraphics[width=\linewidth]{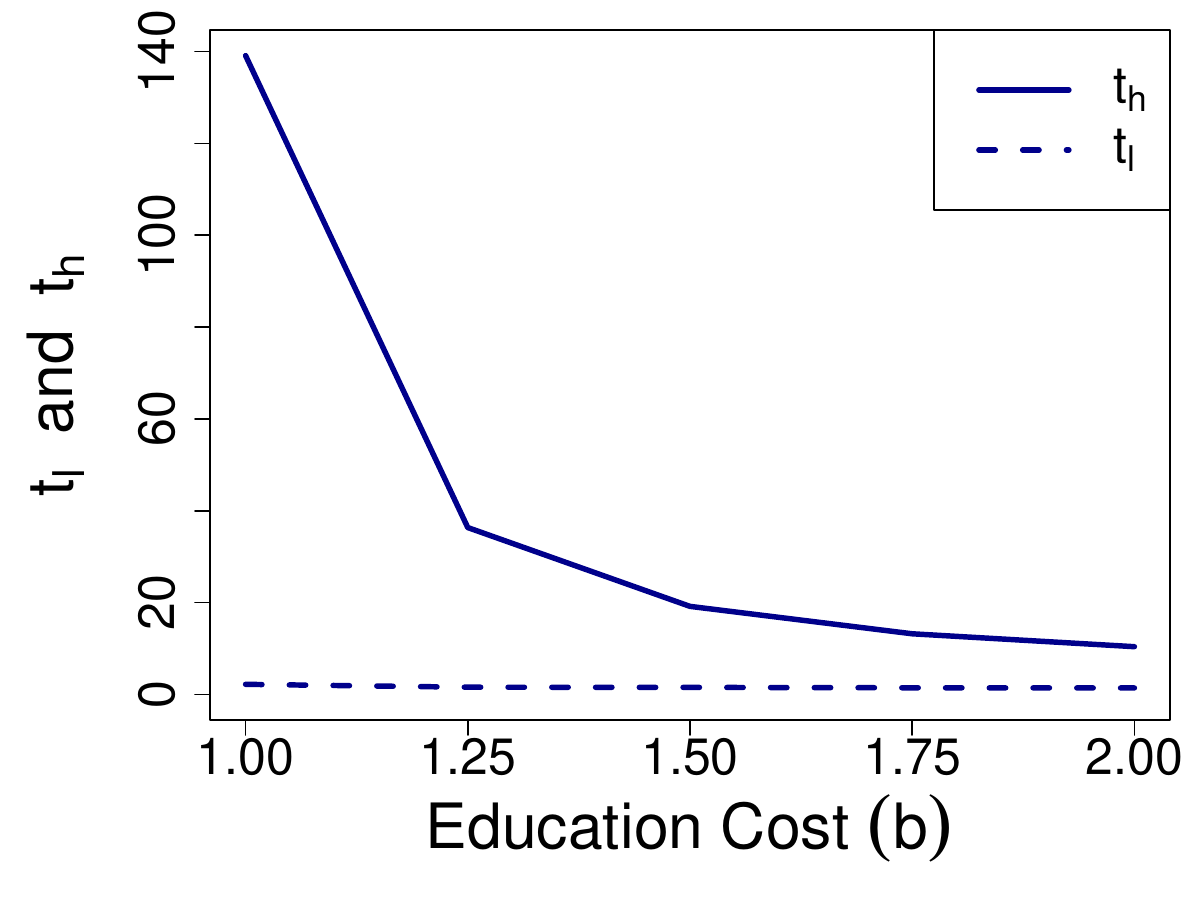}
    \end{minipage}
    \hfill
    \begin{minipage}[t]{0.48\textwidth}
        \centering
        \includegraphics[width=\linewidth]{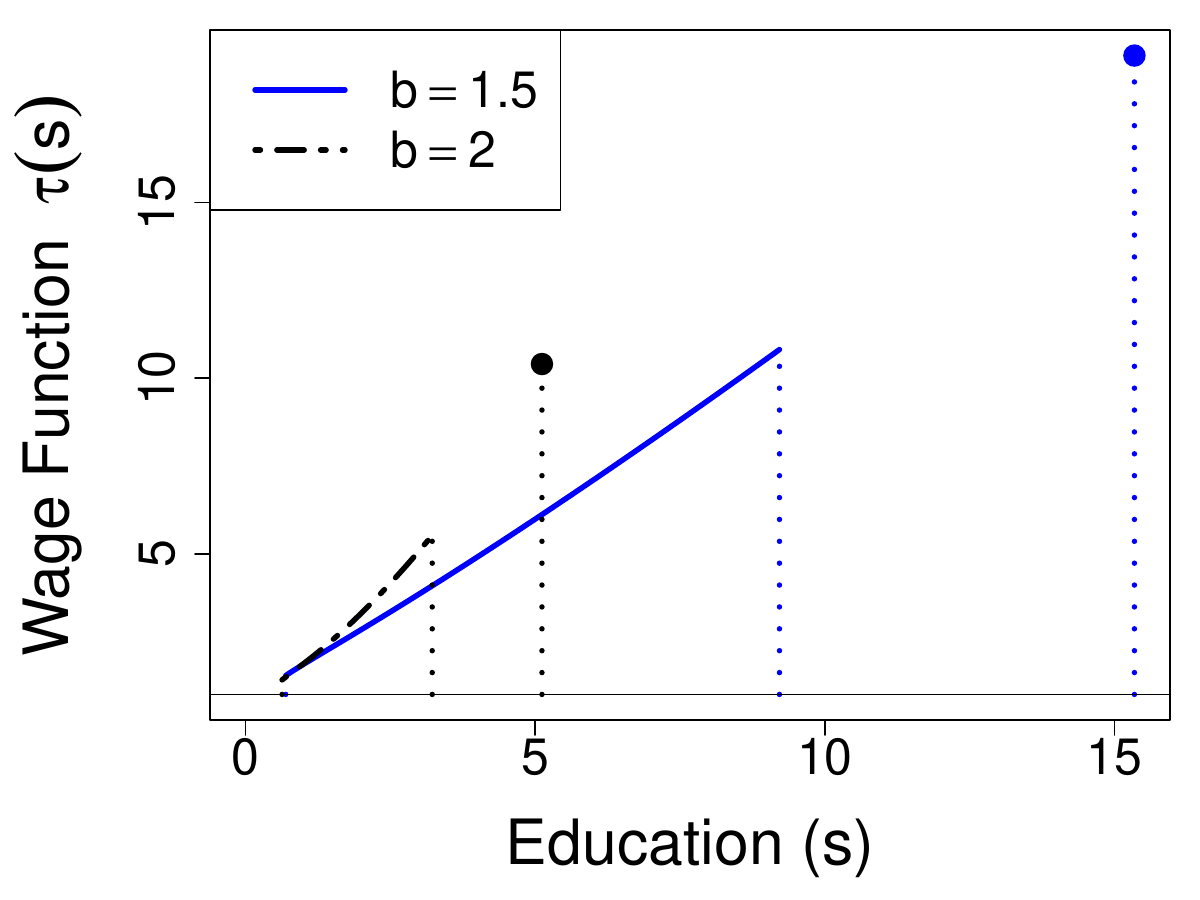}
    \end{minipage}
\end{figure}

\end{document}